\documentclass[%
 reprint,
superscriptaddress,
 amsmath,amssymb,
 aps,
prb,
]{revtex4-2}
\usepackage{algorithm}
\usepackage{algpseudocode}
\usepackage{graphicx}% Include figure files
\usepackage{dcolumn}% Align table columns on decimal point
\usepackage{bm}% bold math
\usepackage{multirow}
\usepackage{booktabs}

\usepackage[margin=1in]{geometry}

\usepackage{xcolor}
\usepackage{textcomp}
\usepackage[normalem]{ulem}
\usepackage{makecell}

\usepackage{subcaption}
\usepackage{comment}
\newcommand{\rd}[1]{ {{\color{blue} [RD: #1 ]}}}
\newcommand{\yl}[1]{ {{\color{red} [YL: #1 ]}}}
\definecolor{darkorange}{rgb}{1.0, 0.45, 0.0}

\begin{document}

%\title{Mendeleev materials: (ultra entropy alloys are) the final frontier of materials chemistry}

%\title{Strategies for simulation and exploration of Mendeleev materials}

\title{Exploring the extremes: atomic basis for multi-elemental materials science under complex thermodynamic conditions}

%\title{Strategies for simulation and exploration of Mendeleev materials}

\author{Anton Bochkarev}
\affiliation{Interdisciplinary Centre for Advanced Materials Simulation (ICAMS), Ruhr-University Bochum, 44780  Bochum, Germany}

\author{Yury Lysogorskiy}
\affiliation{Interdisciplinary Centre for Advanced Materials Simulation (ICAMS), Ruhr-University Bochum, 44780  Bochum, Germany}

\author{Aparna Subramanyam}
\affiliation{%
 Theoretical Division T-1, Los Alamos National Laboratory, Los Alamos, NM, 87545
}%
\author{Ralf Drautz}
\affiliation{Interdisciplinary Centre for Advanced Materials Simulation (ICAMS), Ruhr-University Bochum, 44780  Bochum, Germany}

\author{Danny Perez}
\affiliation{%
 X Computational Physics XCP-AI4ND, Los Alamos National Laboratory, Los Alamos, NM, 87545
}%

\date{\today}% It is always \today, today,

\begin{abstract}
\section*{Abstract}
Modern materials science has historically been founded on combining restricted subsets of the periodic table, favoring high-purity, few-element systems. However, the demands of an emerging circular economy, together with the need to understand materials behavior under planetary and industrial extremes, increasingly require mastering Mendeleev materials - chemically and structurally complex systems that span large portions of the periodic table. In these regimes, current universal machine-learning interatomic potentials often fail, largely due to systematic gaps in traditional training datasets that heavily emphasize low-energy, near-equilibrium structures.
We address this limitation by introducing a chemistry-agnostic, information-entropy–maximization protocol for data generation. By decoupling structural sampling from thermodynamic bias, our approach provides a robust physical prior for atomic interactions across the entire periodic table, including regimes far from equilibrium and under extreme conditions. Training a Graph Atomic Cluster Expansion (GRACE) model on the resulting statistically maximized entropy (SMAX) dataset yields markedly improved robustness across a range of stringent benchmarks. These include large-strain phase transformations in tin, defect evolution in tungsten-based alloys, and catalytic reaction barrier prediction.
More broadly, our approach establishes a scalable and principled methodology for navigating the vast chemical and configurational space relevant to future materials design. It enables a paradigm of discovery by simulation in which unbiased sampling protocols autonomously resolve emergent structures in multi-elemental mixtures-such as systems containing the nine most abundant elements in the Earth’s crust-without reliance on a priori chemical assumptions.
\end{abstract}

\maketitle

%\rd{A few general things:
%\begin{itemize}
%    \item Should we change 'Validation' to 'Application' or 'Demonstration' everywhere?
%    \item 'Discovery by simulation', does this make sense or can we come up with a better term?
%    \item Need to discuss the description of the examples, their detail and length
%    \item let's think about the title, it should give direction for writing
%    \item I think that we need some low dimensional map to demonstrate completeness compared to other databases that have holes
%\end{itemize}
%}

\section{Introduction}

Modern materials science-spanning high-performance steels, superalloys, and advanced polymers-relies predominantly on compounds composed of a remarkably limited subset of the periodic table. Although such materials may incorporate on the order of ten distinct elements, only three or four are typically present in substantial concentrations \cite{bhadeshiaSTEELSStructureProperties2024,reedSuperalloysFundamentalsApplications2006,youngIntroductionPolymers2011}. Even the paradigm shift toward high-entropy alloys, defined by five or more elements in near-equimolar ratios \cite{miracleCriticalReviewHigh2017,georgeHighentropyAlloys2019}, largely preserves this restriction in practice. %Compositional complexity is furthermore often paired with an implicit preference for structurally simple or well-characterized states, such as single-phase solid solutions or equilibrium multiphase microstructures.

This narrow exploration of chemical and structural space mirrors biological matter, which is founded on four bulk organic elements (H, C, N, O) augmented by around seven mineral and fifteen trace elements \cite{urryCampbellBiology2017}. In engineered materials, however, this constrained palette is not dictated by biological necessity but by history. Incremental design philosophies, coupled with the unequal abundance of elements in the Earth’s crust-where just eight elements account for about 98\% of its mass \cite{rudnickCompositionContinentalCrust2003,hanswedepohlCompositionContinentalCrust1995,Chapter5Composition1986}-have guided materials development toward familiar chemistries and well-charted regions of structure–property space.

These historical constraints have, in turn, limited the ambition of the field. By emphasizing small element counts and canonical structural motifs, materials science has left the overwhelming majority of chemical permutations, atomic configurations, and defect topologies accessible within the periodic table effectively unexplored. At the same time, research has disproportionately focused on behavior near ambient conditions, neglecting large swaths of phase space defined by extreme temperatures, pressures, irradiation environments, chemical potentials, and far-from-equilibrium defect populations. Yet it is precisely under such extreme thermodynamic and kinetic conditions that novel phases, unconventional bonding motifs, and emergent structural hierarchies are most likely to arise.

The challenge becomes especially acute in multi-elemental systems exposed to extreme environments, where intuition derived from dilute alloys, equilibrium processing, or simple crystal chemistry often breaks down. As a result, the fundamental chemistry, physics, and processing science required to intentionally design and control genuinely multi-elemental, multi-structural materials-particularly under extreme conditions-remains in its infancy.

A compelling motivation for confronting this challenge arises from the intrinsic complexity of the industrial and planetary material system itself. Modern household, hazardous, and radioactive waste streams comprise large numbers of chemical elements \cite{weibelChemicalAssociationsMobilization2017,panChemicalCharacteristicsRisk2013,ngulimiRadioactiveWasteManagement2025}, typically distributed across heterogeneous, metastable, and structurally complex phases; electronic waste alone is estimated to contain 60 or more elements \cite{WEEEManagementCircular2018,bressanelliCircularEconomyWEEE2020}. Realizing a functional circular economy therefore demands a materials science capable of operating far beyond the confines of simple compositions, ordered structures, and near-ambient processing conditions. It requires embracing chemical diversity, structural disorder, and thermodynamic extremes as central features rather than peripheral complications.

To address the widening gap between the complexity of real-world multi-elemental mixtures and the limited systematic exploration of chemically complex materials, we establish an atomic-scale foundation for multi-elemental materials science under complex and extreme conditions. In doing so, we simultaneously push the boundaries of both chemical diversity and accessible thermodynamic space.
We frame this emerging frontier through the concept of Mendeleev materials. These are materials or mixtures that, in principle, span the full periodic table, containing at least ten-and potentially all-chemical elements in arbitrary and highly non-uniform concentrations. In contrast to high-entropy alloys, which typically involve five elements in near-equimolar ratios, Mendeleev materials embrace maximal chemical heterogeneity and compositional hierarchy. They may be viewed as ultra-entropy materials that, in principle, encompass the complete atom-based materials universe, providing a unifying framework for exploring matter at the limits of chemical, structural, and thermodynamic complexity.

The concept of Mendeleev materials immediately raises a fundamental question of computational tractability. Of the 118 elements in the periodic table, roughly 95 can, in principle, participate in multi-component compounds and mixtures. A na\"ive enumeration of all possible compositions-even at coarse 1\% concentration increments-would require evaluating on the order of $\sim10^{190}$ distinct combinations, a number that renders brute-force exploration inconceivable.

Crucially, however, the intrinsic structure of matter dramatically reduces this apparent dimensionality. The chemical properties of elements are highly correlated, a fact encoded in the very organization of the periodic table. Structure maps and recent analyses demonstrate that the essential chemical characteristics of all elements can be embedded in low dimensional space, yielding an enormous compression of chemical complexity \cite{pettiforStructuresBinaryCompounds1986, onwuliElementSimilarityHighdimensional2023,zhouLearningAtomsMaterials2018,cerqueiraNonorthogonalRepresentationChemical2024}. At the atomic scale, interactions further constrain configurational freedom: bond lengths vary within narrow ranges, interatomic interactions are smooth and dominated by low-body-order contributions, and they decay rapidly with distance. Strong short-range electron overlap enforces repulsion at small separations, sharply limiting the number of physically realizable atomic configurations. Taken together, these fundamental insights-that elemental properties are strongly correlated, bond-length variations are bounded, and atomic interactions are smooth and low-order-transform the problem of exhaustive enumeration into one of systematic and physically informed sampling. They suggest that comprehensive exploration of elemental and configurational space, while challenging, is in principle achievable.

Despite these well-established correlations, a truly systematic sampling of atomic configurations across chemical space has yet to be realized. Existing computational materials databases have traditionally been anchored either in experimentally reported structures or in the targeted discovery of low-energy configurations. Only recently have these databases begun to expand by incorporating off-equilibrium structures generated through molecular dynamics, rattling, or controlled distortions, as well as by extending chemical coverage to include organic molecules alongside inorganic materials \cite{jainCommentaryMaterialsProject2013,schmidt2023machine,barroso2024open,levineOpenMolecules20252025,kuner2025mp,kaplan2025foundational,mazitov2025massive}. Nevertheless, none of the existing efforts treat the periodic table itself as a chemical space that must be sampled systematically. As we show below, current databases exhibit substantial and consequential gaps within the broader, unattended Mendeleev materials configuration space. This omission is particularly striking given the rapid emergence of universal machine-learning interatomic potentials designed explicitly to model chemical interactions across the entire periodic table within a single computational framework.

Establishing a computational foundation for Mendeleev materials therefore hinges on demonstrating robust, transferable simulations across arbitrarily complex chemical spaces. We address this challenge through a two-step strategy. First, we introduce a formal framework for exploring and sampling the full chemical interaction space, achieved by training machine-learning interatomic potentials (MLIPs) on databases constructed to systematically cover atomic configurations spanning the periodic table. Second, we deploy these models to simulate chemically complex materials under extreme thermodynamic conditions, thereby demonstrating their capability to operate far beyond the narrow regimes traditionally accessible to atomistic simulation.

%\rd{I need to check if we need a short outline of the paper here, i.e., what we do, or not.}

%Here, we demonstrate robust simulations in arbitrarily complex chemical spaces. This is achieved in two steps. First, we devise a strategy to formally explore and sample in principle all chemical interactions by training MLIPs to databases that sample all atomic configurations.

%\rd{TODO: add critic of current databases, add references including links to MLPs, add references to sampling}

%\rd{far reaching consequences for materials discovery and design if one can sample the space of all possible atomic configurations systematically, in molecules and compounds.}

\section{Results and discussion}

\subsection{Sampling the space of all atomic configurations}

The space of atomic configurations was sampled by extending the maximum-entropy structure generation framework \cite{karabin2020entropy,montes2022training,pa2025information} to multi-component systems of arbitrary chemical complexity. The method is explicitly agnostic to chemical identity and interaction models, relying only on physically-motivated topological feature diversity augmented by simple boundary conditions such as minimum interatomic distances to exclude unrealistically high-energy configurations and minimum atomic densities to prevent the generation of non-interacting fragments.
Within these constraints, the approach enables the principled and fully automated construction of large-scale datasets with unprecedented topological and chemical diversity. New atomic configurations are generated iteratively to increase an approximation of the information entropy of the dataset, evaluated in a chemically meaningful feature space. Conceptually, this strategy is closely related to classical optimal design-of-experiments methodologies, such as D-optimality \cite{podryabinkin2017active}, which aim to maximize the information content available for parameter estimation. Full methodological details are provided in the Methods section.

Using this protocol, we generated a large database of atomic structures spanning the periodic table and characterized them using density functional theory. The dataset was subsequently enriched with relaxed structures and low-energy configurations drawn from existing materials and organic databases, thereby increasing data density in thermodynamically relevant regions of configuration space. In total, the generation process was terminated after computing 1,693,386 distinct structures comprising 23,071,117 atoms, covering nearly the entire periodic table (excluding actinide elements).

\begin{figure*}
    \centering
    \includegraphics[width=1\textwidth]{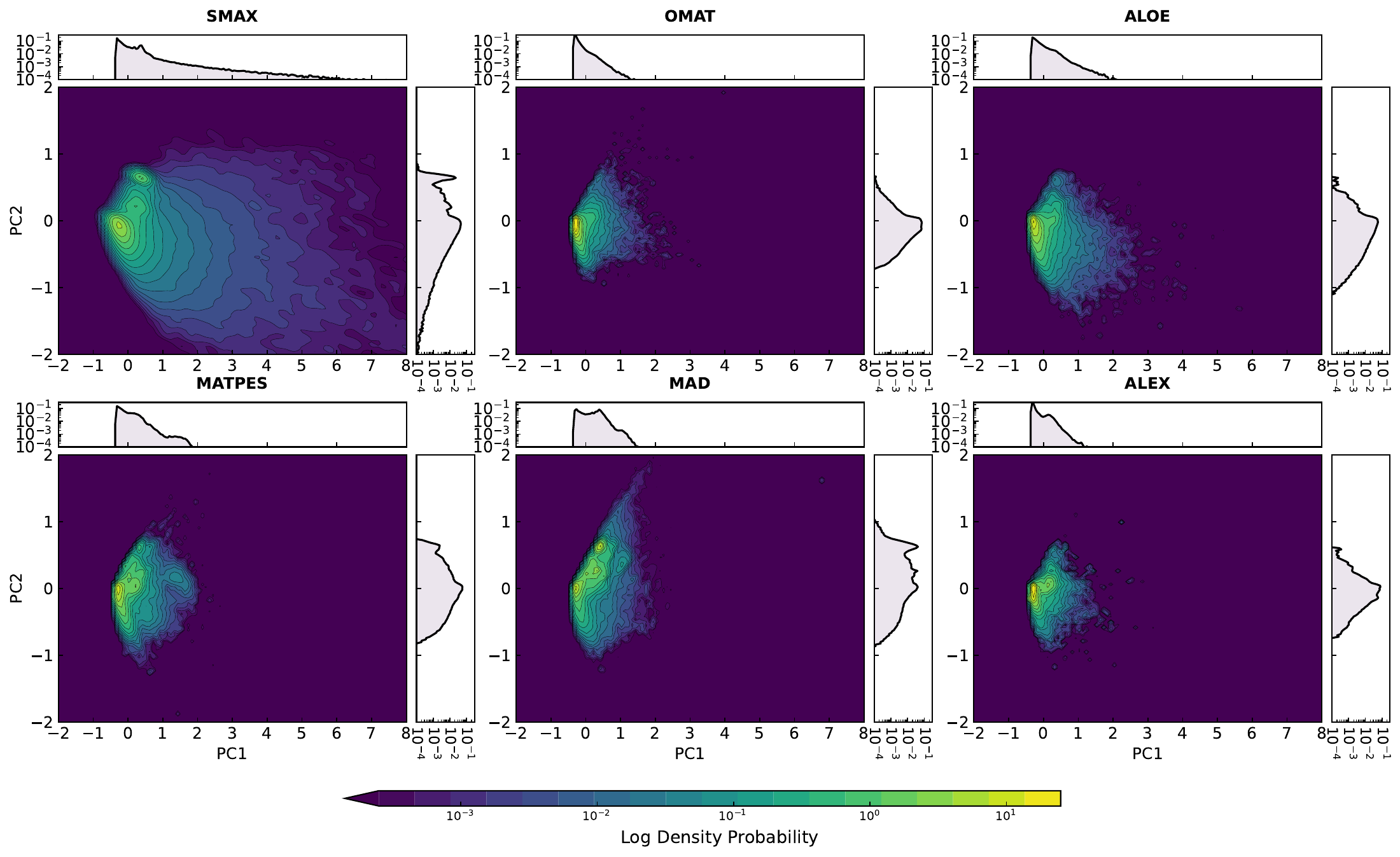}
    \caption{First two PCA components in basis function space. The maximum entropy generated SMAX database shows a significantly more comprehensive sampling than other databases.}
    \label{fig:PCAUEA}
\end{figure*}
The resulting maximum-entropy (SMAX) dataset forms the foundation for all subsequent machine-learning interatomic potential (MLIP) training. Figure~\ref{fig:PCAUEA} demonstrates the effectiveness of the sampling strategy, showing that SMAX achieves substantially broader coverage of the atomic feature space than existing datasets. In particular, SMAX spans the largest region in the plane defined by the first two principal components of the chosen chemically informed feature representation, exceeding the coverage of datasets such as Massive Atomic Diversity (MAD) \cite{mazitov2025massive}, which explicitly targets broad chemical and environmental diversity. The contrast is even more pronounced when compared to more conventional datasets, including MATPES \cite{kaplan2025foundational} and Alexandria~\cite{schmidt2023machine,wang2023symmetry}, as well as diversification approaches based on molecular dynamics sampling, such as OMAT~\cite{barroso2024open} and ALOE~\cite{kuner2025mp}.

This enhanced diversity is not limited to the leading principal components. As detailed in the Supplementary Material, SMAX consistently exhibits broader coverage across higher-order PCA components, indicating that the total volume of feature space captured by SMAX is dramatically larger than that of existing datasets. As shown in Fig.~\ref{fig:dataset_statistics_6_panel_robust}, this topological and chemical diversity translates directly into a substantially wider distribution of reference energies, forces, and stresses-an essential ingredient for achieving robust and transferable performance under extreme thermodynamic and mechanical conditions.

% \rd{Would be great to include Alexandria here}
% http://localhost:8888/lab/tree/acefit/GRACE/UEA/FOR_PAPER_Feature_map_1L-ALOE-MATPES-UEA-OMAT-global-per-element-resolution.ipynb

% \begin{figure}
%     \centering
%     \includegraphics[width=0.5\textwidth]{figures/PCA_1_2_All_Species.pdf}
%     \caption{First two PCA components in basis function space. The maximum entropy generated \ylchanged{SMAX dataset} shows a significantly more comprehensive sampling than other databases.}
%     \label{fig:PCAUEA}
% \end{figure}

\subsection{Probing the limits of massively trained MLIPs}

State-of-the-art universal machine-learning interatomic potentials (MLIPs) rely on the curation of extremely large and diverse training datasets. Among these, the Open Materials (OMAT24) dataset~\cite{barroso2024open} stands out due to its unprecedented scale, comprising approximately 118 million computed atomic configurations, an order of magnitude larger than most other foundational datasets, which typically contain only a few million structures. As a result, OMAT has rapidly become a primary training resource for the latest generation of foundational MLIPs.

Using the OMAT dataset, we trained a Graph Atomic Cluster Expansion (GRACE) potential \cite{bochkarev2024,drautzAtomicClusterExpansion2019}. GRACE employs a complete, systematically improvable set of graph-based body-order functions and represents the current state of the art in balancing accuracy, transferability, and computational efficiency for universal interatomic potentials \cite{lysogorskiyGraphAtomicCluster2025a}.

\begin{figure}
    \centering
    \includegraphics[width=0.5\textwidth]{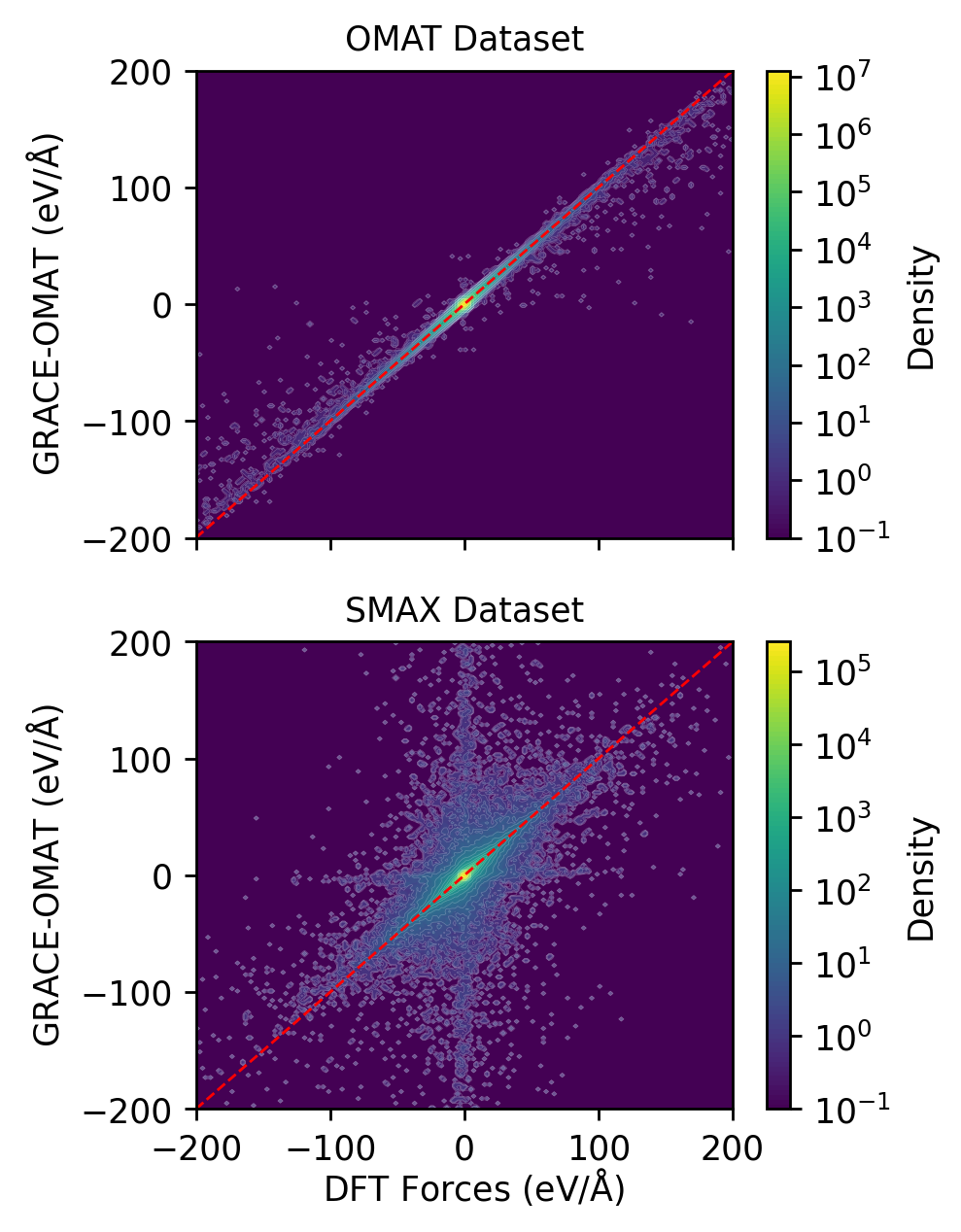}
    \caption{Parity plot of force components predicted by GRACE-OMAT model evaluated on the OMAT (top) test set and SMAX (bottom) test set. }
    \label{fig:parityOMATUEA}
\end{figure}

Figure~\ref{fig:parityOMATUEA} (top) shows the parity plot obtained by evaluating the resulting GRACE model on a standard OMAT hold-out test set. As expected, the model exhibits the high predictive accuracy characteristic of leading universal MLIPs when assessed on data drawn from the same distribution as the training set. A markedly different behavior, however, is observed when the OMAT hold-out set is replaced by configurations drawn from the SMAX dataset (Fig.~\ref{fig:parityOMATUEA}, bottom).

In this case, the apparent accuracy inferred from OMAT validation deteriorates substantially. This divergence reveals that, despite its sheer size, the OMAT dataset leaves significant regions of chemically and structurally relevant configuration space effectively unsampled. These regions correspond to configurations that are statistically rare or absent in conventional dataset construction workflows, yet are systematically explored by the entropy-maximization strategy underlying SMAX. The inability of the OMAT-trained GRACE model to extrapolate reliably in these regimes demonstrates that dataset scale alone is insufficient to ensure universal coverage.

Taken together, these results establish SMAX as a powerful benchmark for assessing the true transferability of universal MLIPs. By probing configurations that lie outside the implicit support of existing large-scale datasets, SMAX exposes failure modes that remain hidden under conventional validation protocols and provides a more stringent test of uniform accuracy across the periodic table. In the following, we show that the increased diversity of SMAX also translates into marked improvement of the description of a broad range of properties of practical relevance to applications.

\subsection{Foundational GRACE from SMAX dataset}

To quantify the impact of systematic, entropy-maximizing sampling on model robustness and transferability, we compare three Graph Atomic Cluster Expansion (GRACE) models trained on distinct datasets:
\begin{itemize}
\item \textbf{SMAX Model}: GRACE trained exclusively on the Maximum Entropy (SMAX) dataset.
\item \textbf{OMAT Model}: baseline GRACE trained on the conventional large-scale public OMAT dataset \cite{barroso2024open}, representative of the datasets underpinning current state-of-the-art foundational MLIPs.
\item \textbf{SMAX–OMAT Model}: GRACE trained on the combined SMAX and OMAT datasets.
\end{itemize}
All models share the same architecture and training protocol; details are provided in the Methods section.

The consequences of dataset construction become immediately apparent when examining cross-dataset performance. As discussed in the previous section, a GRACE model trained on OMAT exhibits excellent accuracy when evaluated on an OMAT hold-out test set, yet can fail catastrophically when exposed to configurations drawn from SMAX (Fig.\ref{fig:parityOMATUEA}). Table~\ref{tab:model_performance} quantifies this effect. While the OMAT-trained model achieves low mean absolute errors (MAEs) on OMAT test data (0.065 eV/\AA\ for forces and 0.013 eV/atom for energies), its performance deteriorates by more than two orders of magnitude when evaluated on SMAX, with force errors approaching 3eV/\AA, stress errors exceeding 9,000 GPa, and energy errors rising to several eV per atom.

These extreme error levels reflect the presence of large, systematically unattended regions in the OMAT training distribution. Despite its unprecedented size, OMAT underconstrains substantial portions of the chemical and structural space spanned by SMAX.
The resulting failures demonstrates that reported hold-out test errors on large-scale datasets can severely underestimate true generalization error, and should therefore be interpreted with considerable caution.

In stark contrast, the GRACE model trained exclusively on SMAX exhibits balanced and consistent performance against holdout data from both datasets. As shown in Table~\ref{tab:model_performance}, the SMAX-trained model achieves comparable accuracy on SMAX and OMAT test sets, with force MAEs of 0.140 eV/\AA\ and 0.108 eV/\AA, respectively, and energy MAEs near 0.02~eV/atom in both cases. Although slightly less optimized for the narrow OMAT distribution, the SMAX-trained model avoids catastrophic extrapolation failures entirely. This symmetry in cross-dataset performance indicates that SMAX provides a far more uniform and constraining coverage of atomic configuration space.

These results establish a central conclusion: dataset size alone is insufficient to produce a truly foundational MLIP. What matters instead is the completeness of the sampled chemical and configurational spaces. By systematically maximizing information entropy, SMAX eliminates large extrapolative gaps and enforces physically meaningful coverage across chemical and structural degrees of freedom. Models trained on SMAX therefore generalize not by interpolation within a narrow distribution, but by construction across a much broader domain. Of course, the same caution should be exercised when interpreting the results of this cross-validation on the SMAX dataset, as the test errors of GRACE-OMAT on SMAX should be seen as a worst-case scenario that are obviously not typical of the vast majority of atomistic studies of materials.

In the following sections, we build on this foundation by examining increasingly complex material scenarios. Through a series of application-driven benchmarks, we demonstrate that the enhanced diversity and uniformity of SMAX translate directly into reliable predictions for material properties across a wide range of chemical compositions, structural motifs, and thermodynamic conditions.

%\ab{omat trained model:
%F MAE on omat test (789248 structures) - 0.065 eV/A$^3$, on %smax test (28955 structures) - 2.9 eV/A$^3$.
%S MAE on omat test - 0.5 GPa, on smax test - 9500 GPa
%E MAE on omat test - 0.013 eV/atom, on smax test - 7.3 eV/atom

%smax trained model:
%F MAE on omat test - 0.108 eV/A$^3$, on smax test - 0.140 %eV/A$^3$.
%S MAE on omat test - 0.6 GPa, on smax test - 1.5 GPa
%E MAE on omat test - 0.023 eV/atom, on smax test - 0.014 %eV/atom}]

\begin{table}
    \centering
    \caption{Comparison of Mean Absolute Errors (MAE) for OMAT and SMAX trained models on respective test sets.}
    \label{tab:model_performance}
    \begin{tabular}{llccc}
        \hline
        Training Set & {Test Set} & {Force MAE} & {Stress MAE} & {Energy MAE} \\
        & & (eV/\AA) & (GPa) & (eV/atom) \\
        \hline
        OMAT & OMAT & 0.065 & 0.5 & 0.013 \\
                              & SMAX  & 2.9 & 9,500 & 7.3 \\
        \hline
        SMAX & OMAT & 0.11 & 0.60 & 0.023 \\
                              & SMAX  & 0.14 & 1.50 & 0.014 \\
        \hline
    \end{tabular}
\end{table}

\subsection{Tin large deformation maps}
\label{sec:tin}
Predicting materials behavior under large mechanical deformation constitutes a stringent test for universal machine-learning interatomic potentials (MLIPs). Most large, general-purpose datasets are heavily biased toward near-equilibrium configurations and modest strains, leaving strongly deformed states poorly constrained. In contrast, the SMAX dataset is constructed to deliberately sample broad and diverse regions of configurational space, including those accessed under extreme mechanical conditions.

Tin provides a particularly demanding benchmark in this regard. It exhibits a rich sequence of temperature- and pressure-driven allotropes that reflect a subtle balance between covalent and metallic bonding. 
%At ambient pressure, $\alpha$-Sn (diamond cubic) is stable below 13.2 °C and transforms to $\beta$-Sn (body-centered tetragonal, bct) at higher temperatures. Under compression, Sn undergoes further transitions through a bct $\gamma$-Sn phase to a bcc structure (often denoted $\delta$-Sn), and finally to an hcp $\delta$-Sn phase at high pressures \cite{Freville2024}. 
These phase transformations are highly sensitive to stress state, making Sn a canonical test system for pressure-induced structural transitions in condensed-matter physics, planetary science, and materials design, as well as a stringent benchmark for interatomic models \cite{cusentino2023molecular}.

\begin{figure}
    \centering
    \includegraphics[scale=0.25]{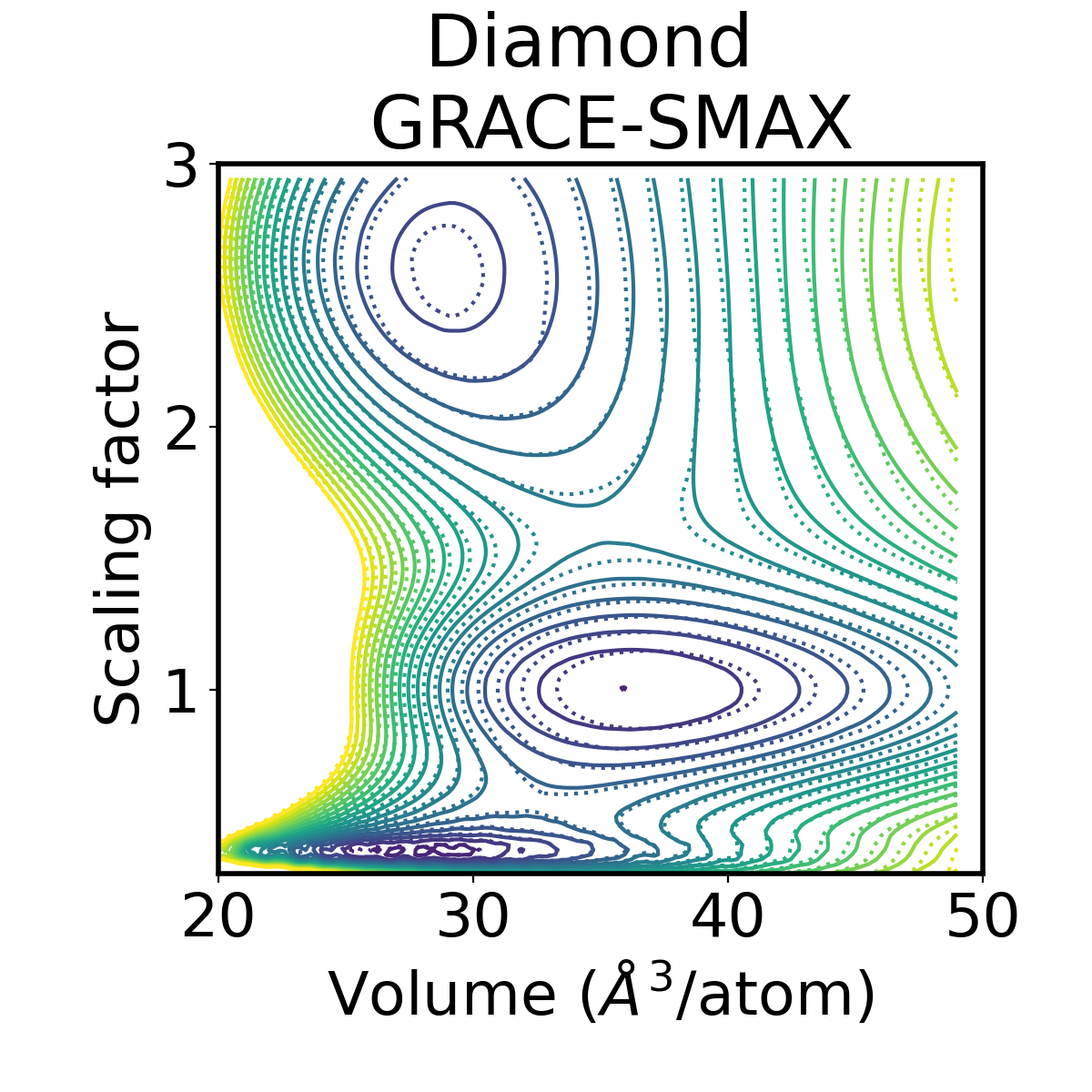}
    \includegraphics[scale=0.25]{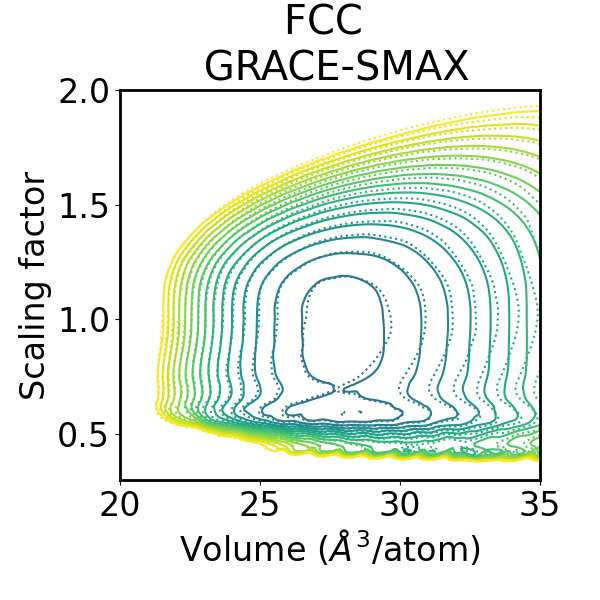}
    \includegraphics[scale=0.25]{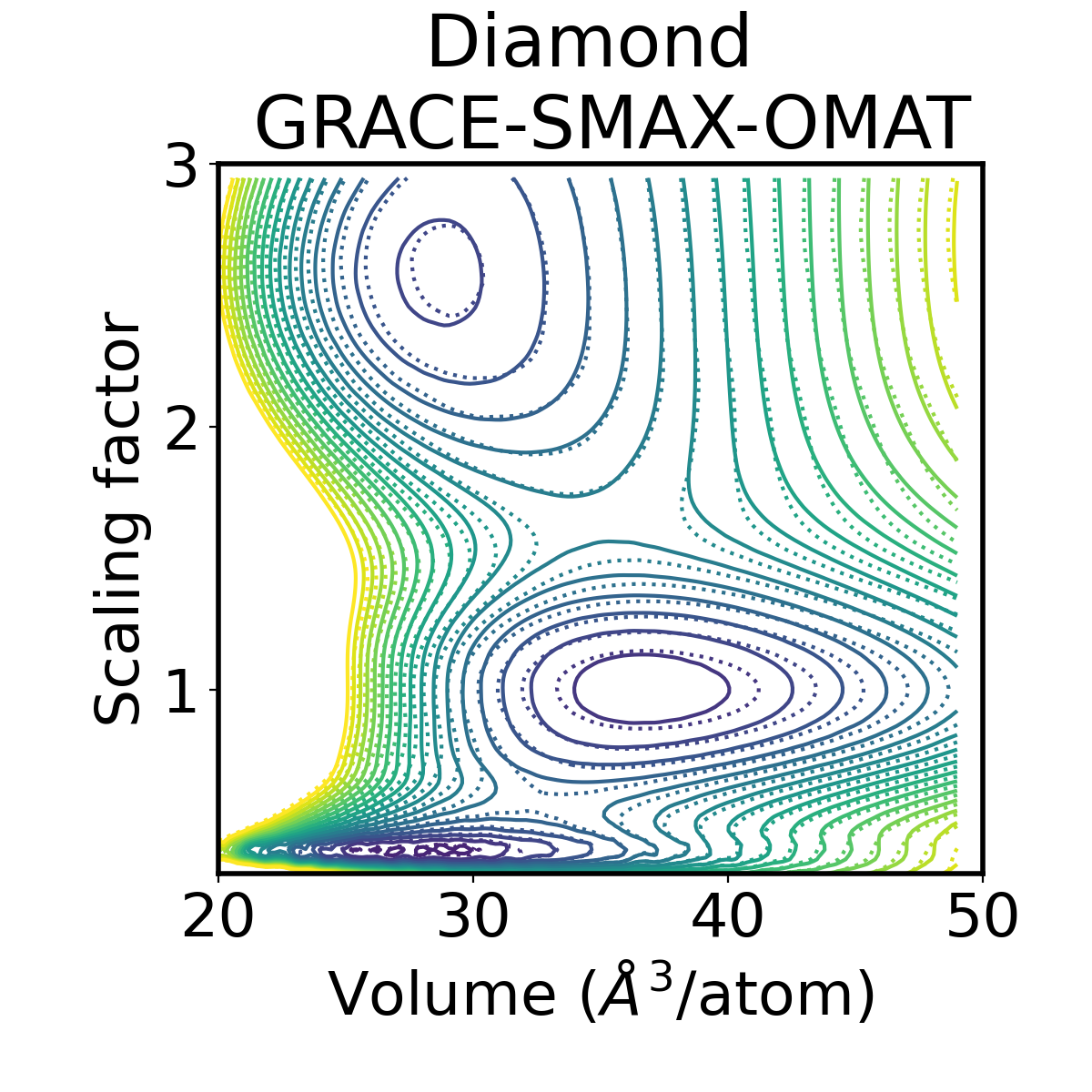}
    \includegraphics[scale=0.25]{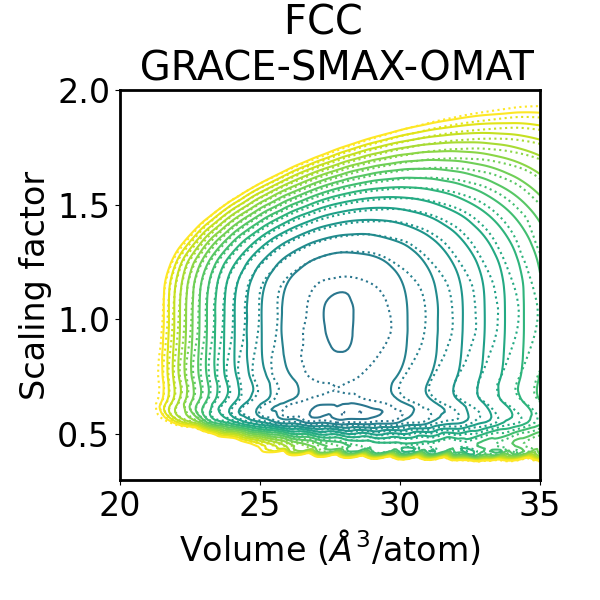}
    \includegraphics[scale=0.25]{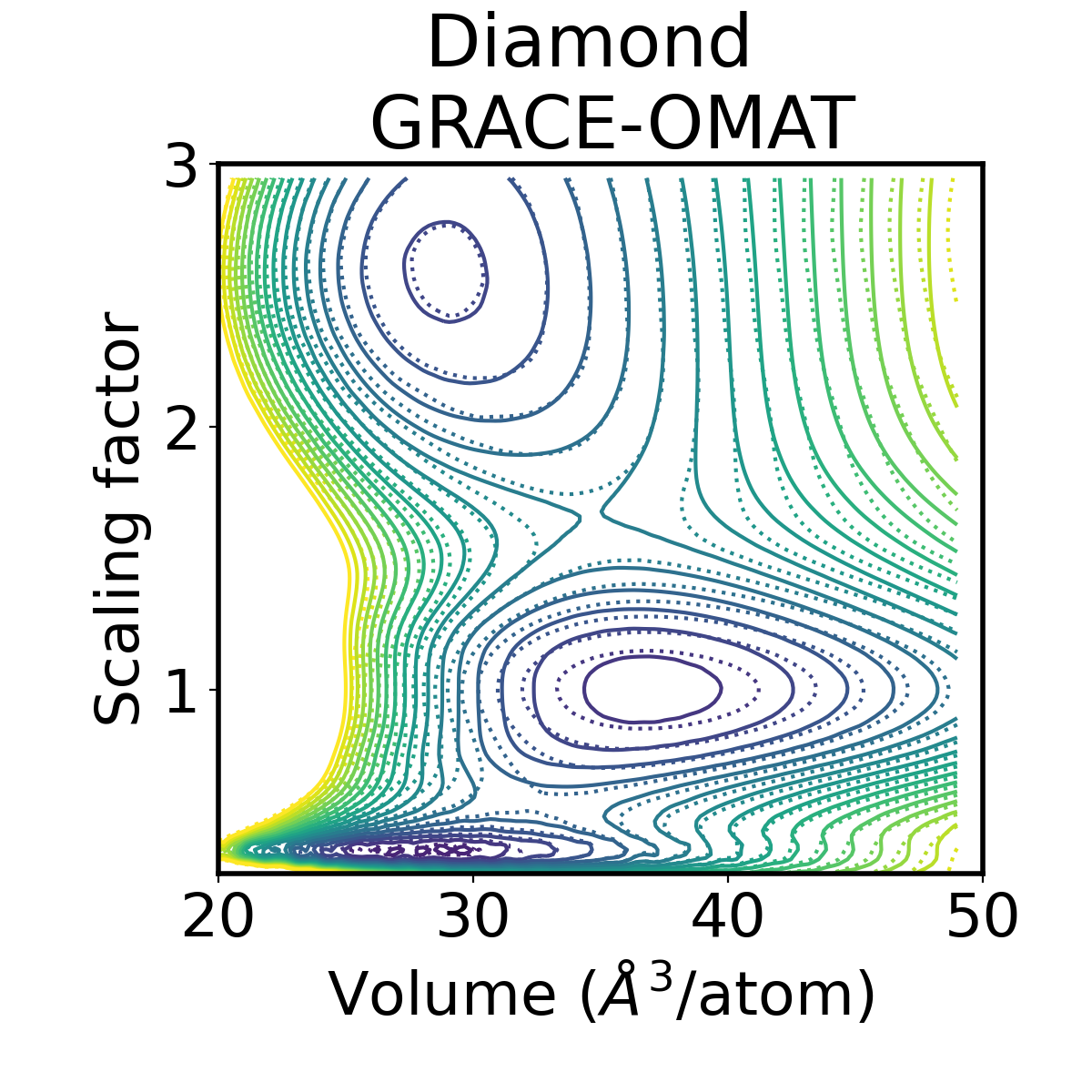}
    \includegraphics[scale=0.25]{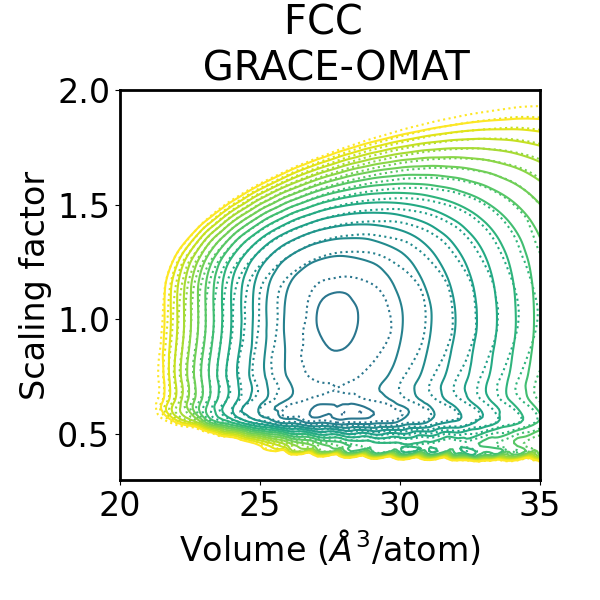}
    \caption{Deformation map for tin under uniaxial compression. Left column: deformation around the diamond structure; Right column: deformation around the FCC structure. The dotted lines correspond to the DFT reference, while the filled lines are the GRACE predictions. Contours are plotted every 0.05 eV/atom and are matched across figures. All models are variants of GRACE-2L-L. }
    \label{fig:tin}
\end{figure}

Figure~\ref{fig:tin} compares deformation energy landscapes obtained from density functional theory (DFT) with predictions from the three GRACE models. The left panels show energy maps centered on the $\alpha$-Sn diamond-cubic structure (scaling factor = 1). Upon uniaxial compression (scaling $<$ 1), a secondary minimum corresponding to the $\beta$-Sn/bct polymorph emerges. In this near-ground-state region, all three models closely reproduce the DFT reference over a wide range of volumes and distortions, with the SMAX-trained model exhibiting marginally improved fidelity.

The right panels present an analogous deformation map constructed around an fcc reference lattice. Although fcc Sn is not a stable phase at positive pressure, it serves as a convenient parent structure for exploring the Bain path (fcc → bct → bcc) through tetragonal distortions. In this representation, the bcc basin appears near a uniaxial compression of $1/\sqrt{2} \simeq 0.707$, while the bct minimum emerges near a scaling factor of approximately 0.6. In this more challenging regime, the SMAX-trained model reproduces the DFT energy surface with near-quantitative accuracy across basins and saddle regions alike. By contrast, the OMAT and SMAX–OMAT models display small but systematic deviations, particularly around the fcc basin and the bcc saddle point.

Accurately reproducing these deformation maps has historically been difficult for both empirical interatomic potentials and MLIPs, unless highly specialized and tightly curated training datasets are employed \cite{cusentino2023molecular,ChenPRM2023}. In contrast, the fine features of the Sn deformation landscape emerge naturally from the SMAX-trained model without any explicit targeting, refinement, or prior knowledge of Sn allotropy. Notably, the SMAX data generation procedure is entirely agnostic to the existence of stable or metastable Sn phases, underscoring the generality of the approach.

Finally, we emphasize that the relative stability and properties of Sn phases are known to depend sensitively on the choice of exchange–correlation functional used in DFT \cite{Aguado2003}. Accordingly, agreement with DFT reference data should not be interpreted as a guarantee of quantitative agreement with experiment. Rather, this example demonstrates the ability of SMAX-trained models to robustly capture complex, large-deformation energy landscapes within a consistent first-principles framework.

\subsection{Defects in Tungsten-Based Materials}

We next assess the performance of the GRACE potentials for a family of tungsten-based materials relevant to operation under extreme conditions, particularly as candidate first-wall or structural materials in fusion reactors. In such environments, materials are exposed to intense neutron irradiation and plasma fluxes, making the accurate description of radiation-induced defects essential. Defect formation energies and migration barriers govern long-term microstructural evolution, swelling, embrittlement, and fuel retention, and thus represent a critical benchmark for any interatomic potential intended for extreme-condition simulations.

Configurations associated with point defects, interstitial clusters, and strongly distorted local environments are, however, sparsely represented in conventional large-scale datasets such as OMAT, which predominantly sample near-equilibrium crystalline configurations. By construction, the SMAX dataset instead provides systematic coverage of far-from-equilibrium atomic arrangements, including a wide range of bonding topologies, coordination environments, and local densities. As shown below, this difference translates directly into improved robustness and accuracy for defect energetics.

\subsubsection{Interdiffusion in the W/Be System}
\label{subsec:wbe}

The W/Be system serves as a canonical mixed-material platform for fusion first-wall and divertor research. In experimental fusion devices, beryllium has historically been used in the main chamber due to its low atomic number, which minimizes radiative plasma losses, while tungsten has been employed in the divertor because of its high melting point and low sputtering yield. The W–Be combination therefore provides a prototypical system for studying plasma–surface interactions, including erosion, migration, and fuel retention under reactor-relevant conditions \cite{Matthews2013JETILW,Brezinsek2015BeW_PSI,Okamoto1986BeW,Wiltner2021WBeInterface,Doerner2005BeWmixed}. Recent experiments further show that Be deposition on W, followed by annealing or plasma exposure, can lead to the formation of W–Be intermetallic compounds (e.g., WBe$2$, WBe${12}$) near the interface, substantially altering local thermo-mechanical properties and melting resistance.

To assess whether the different GRACE models correctly capture the energetics governing W/Be interdiffusion and compound formation, we relaxed 1,224 prototype structures using DFT and compared GRACE predictions under three increasingly demanding conditions: (i) single-point energy evaluation on DFT-relaxed geometries, (ii) internal relaxation with fixed cell parameters, and (iii) full relaxation of both atomic positions and cell degrees of freedom.

As summarized in Tables~\ref{table:prototypes_WBe_rms} and ~\ref{table:prototypes_WBe_mae}, the GRACE-2L-OMAT-L model-trained on a dataset dominated by crystalline structures-achieves the lowest root-mean-square (RMS) errors for single-point evaluations (0.022eV/atom) and internal relaxations (0.022 eV/atom). However, when full relaxation is permitted, the GRACE-2L-SMAX-L model becomes the most accurate, with an RMS error of 0.080 eV/atom, compared to 0.081 eV/atom for SMAX–OMAT and 0.084~eV/atom for OMAT. This trend is consistent with the SMAX model’s exposure to a much broader distribution of bonding environments and densities, which stabilizes the optimization trajectory through configuration space when constraints are removed. 

The dominant contributions to the remaining error under full relaxation arise from a small subset of structures that become mechanically unstable once cell constraints are lifted, collapsing toward lower-energy configurations. Importantly, these cases occur at positive formation energies and therefore do not affect predicted phase stability. Indeed, all three models correctly reproduce the DFT convex hull, identifying Be, Be$_{12}$W, Be$_2$W, and W as the stable phases. This result demonstrates that the extreme topological diversity of the SMAX dataset does not compromise accuracy for low-energy crystalline states, despite being trained on a substantially broader configuration space. Perhaps unsurprisingly, models trained to OMAT (which is extremely rich in low-energy crystal structures) perform extremely well in a metric that is less sensitive to outliers, such as MAE (c.f., Tab. \ref{table:prototypes_WBe_mae}). In this case, both GRACE-OMAT and GRACE-SMAX-OMAT outperform GRACE-SMAX.

Under fusion-relevant conditions, continuous neutron bombardment and hydrogen/helium implantation generate high densities of point defects and small defect clusters. Accurate prediction of defect formation energies is therefore indispensable for modeling microstructural evolution and fuel retention. Table~\ref{table:defects_WBe} reports absolute errors for selected vacancy and interstitial defects in Be, WBe$_{12}$, and WBe$_2$. The GRACE-2L-SMAX-L model yields the lowest error in 16 of the 21 cases considered, with particularly large improvements for configurations involving strongly distorted local coordination. Notable examples include tungsten interstitials in Be (e.g., BC: 0.432eV for SMAX versus 1.691eV for OMAT) and beryllium interstitials in W (110: 0.272eV versus 0.463eV for OMAT).

While the OMAT-trained model remains competitive in a small number of cases, and the SMAX–OMAT model typically yields intermediate performance, the overall pattern is clear: SMAX’s systematic coverage of chemically and topologically diverse environments confers a decisive advantage whenever local atomic coordination deviates strongly from ideal crystalline motifs. In these regimes, models trained on conventional datasets can exhibit errors that far exceed those suggested by hold-out validation on their own training distributions. This behavior underscores the presence of substantial underrepresented regions in conventional datasets and highlights the advantage of entropy-maximizing sampling strategies for achieving reliable predictions under extreme conditions.

%%%%%%%%%%%%%%%%%%%%%%%%%%%%%%%
\begin{table}
{\small
\setlength{\tabcolsep}{1pt}% tighter columns
\begin{tabular}{l|ccc}
\hline
% Relaxation protocol & \makecell{SMAX \\ (eV/atom)}  & \makecell{SMAX-OMAT \\ (eV/atom)}  & \makecell{OMAT \\ (eV/atom)}  \\
Relaxation protocol & \makecell{SMAX}  & \makecell{SMAX-OMAT}  & \makecell{OMAT}  \\
\hline
Single point & 0.033 & 0.023 & \textbf{0.022} \\
Internal relaxation & 0.032 & 0.024 & \textbf{0.022} \\
Full relaxation & \textbf{0.080} & 0.081 & 0.084 \\
\hline
\end{tabular}
}
\caption{RMS test errors (eV/atom) on 1224 different W/Be  inter-metallic structures. The lowest error in each row is highlighted in bold. 
%All models are variants of GRACE-2L-L. 
See text for details of the different relaxation protocols.}
\label{table:prototypes_WBe_rms}
\end{table}

%%%%%%%%%%%%%%%%%%%%%%%%%%%%%%%

\begin{table}
{\small
\setlength{\tabcolsep}{4pt}% tighter columns
\begin{tabular}{l|ccc}
\hline
% Relaxation protocol & \makecell{SMAX \\ (eV/atom)}  & \makecell{SMAX-OMAT \\ (eV/atom)}  & \makecell{OMAT \\ (eV/atom)}  \\
Relaxation protocol & \makecell{SMAX}  & \makecell{SMAX-OMAT}  & \makecell{OMAT}  \\
\hline
Single point & 0.024 & \textbf{0.017} & \textbf{0.017} \\
Internal relaxation & 0.023 & \textbf{0.017} & \textbf{0.017} \\
Full relaxation & 0.031 & \textbf{0.025} & \textbf{0.025} \\
\hline
\end{tabular}
}
\caption{MAE test errors (eV/atom) on 1224 different W/Be  inter-metallic structures. The lowest error in each row is highlighted in bold. 
%All models are variants of GRACE-2L-L. 
See text for details of the different relaxation protocols.}
\label{table:prototypes_WBe_mae}
\end{table}
%%%%%%%%%%%%%%%%%%%%%%%%%%%%%%%

\begin{table}
{\small
\setlength{\tabcolsep}{4pt}% tighter columns
\begin{tabular}{llccc}
\hline
Family & Defect & \makecell{SMAX \\ (eV)}  & \makecell{SMAX-OMAT \\ (eV)}  & \makecell{OMAT \\ (eV)}  \\
\hline
Be IA in W & 110 & \textbf{0.272} & 0.363 & 0.463 \\
% & 110 & \textbf{0.272} & 0.363 & 0.463 \\
 & 111 & \textbf{0.336} & 0.364 & 0.371 \\
 & OS & \textbf{0.272} & 0.363 & 0.463 \\
 & TS & 0.460 & 0.369 & \textbf{0.355} \\
 W IA in Be & BC & \textbf{0.432} & 0.978 & 1.691 \\
  & BS & 0.917 & 0.578 & \textbf{0.134} \\
  & BT & \textbf{0.432} & 0.978 & 1.691 \\
  & O & \textbf{0.418} & 0.462 & 1.049 \\
  & S & \textbf{0.371} & 0.830 & 1.359 \\
  & T & \textbf{0.432} & 0.978 & 1.691 \\
V in Be   & V & \textbf{0.085} & 0.172 & 0.213 \\
 & V+V (1NN) & \textbf{0.187} & 0.240 & 0.413 \\
% & Be\_mono\_1NN & \textbf{0.087} & 0.216 & 0.331 \\
% & Be\_mono\_2NN & 0.154 & \textbf{0.112} & 0.221 \\
%WBe_{12} & W & \textbf{1.139} & 1.445 & 2.828 \\
%WBe12 & bulk & \textbf{1.216} & 1.304 & 2.698 \\
%WBe12 & Be & \textbf{1.050} & 1.361 & 2.721 \\
%WBe12 & WBe & \textbf{1.110} & 1.578 & 2.938 \\
V in WBe$_{12}$ & W V  & \textbf{0.078} & 0.146 & 0.138 \\
 & Be V  & 0.164 & 0.079 & \textbf{0.058} \\
 & W+Be V  & \textbf{0.104} & 0.301 & 0.283 \\
%\#\# WBe2 & W & \textbf{0.023} & 0.481 & 2.055 \\
%\#\# WBe2 & Bulk & \textbf{0.132} & 0.356 & 2.136 \\
%\#\# WBe2 & Be & \textbf{0.480} & 0.773 & 2.475 \\
%\#\# WBe2 & WBe & \textbf{0.205} & 0.801 & 2.249 \\
WBe$_2$ & W V  & 0.156 & 0.130 & \textbf{0.072} \\
  & Be V   & \textbf{0.350} & 0.439 & 0.375 \\
  & W+Be V & \textbf{0.075} & 0.471 & 0.157 \\
%\#\# WBe22 & W & \textbf{0.978} & 1.322 & 1.337 \\
%\#\# WBe22 & Bulk & \textbf{0.058} & 0.239 & 0.326 \\
%\#\# WBe22 & Be & \textbf{0.088} & 0.108 & 0.298 \\
%\#\# WBe22 & WBe & \textbf{0.578} & 0.958 & 1.124 \\
WBe$_{22}$ & W V  & \textbf{0.919} & 1.088 & 1.019 \\
 & Be V  & 0.143 & 0.109 & \textbf{0.007} \\
 & W+Be V  & \textbf{0.522} & 0.745 & 0.842 \\
\hline
\end{tabular}
}
\caption{Absolute errors on defect formation energies in W/Be. The lowest error for any defect is highlighted in bold. 
%All models are variants of the GRACE large 2 layers model. 
The defect structure reflects the initial state before any relaxation, but will not reflect the post-relaxation structure if the initial structure was unstable.}
\label{table:defects_WBe}
\end{table}

\subsubsection{Vacancies in W/Ta/Cr/V high-entropy alloys}
Fusion-relevant tungsten surfaces undergo severe degradation under helium and neutron bombardment, manifesting as surface blistering, the growth of fiber-like “fuzz,” and irradiation-induced hardening and embrittlement. These damage mechanisms pose significant risks to both component lifetime and plasma performance in fusion reactors. Tungsten-based high-entropy alloys (HEAs) have therefore emerged as a promising mitigation strategy. By distributing multiple refractory elements on a common bcc lattice, these alloys introduce pronounced chemical and elastic disorder that can slow defect migration, enhance defect recombination, and suppress helium-induced morphological instabilities. Experimental studies of W–Ta–Cr–V class alloys report substantially improved radiation tolerance-evidenced by reduced defect loop densities, suppressed hardening, and delayed fuzz formation relative to pure tungsten, with further improvements observed in related quinary systems \cite{el2019outstanding,qin2024influence,el2023quinary}.

To evaluate how the GRACE models handle this level of chemical and configurational complexity, we created vacancies across 78 distinct local environments in a W$_{20}$Ta$_{20}$Cr$_6$V$_8$ alloy. As in the previous analyses, we considered three increasingly demanding evaluation modes: single-point energy prediction on DFT-relaxed geometries, internal relaxation with fixed cell parameters, and full relaxation of both atomic positions and the simulation cell. The resulting errors are summarized in Tables~\ref{table:vac_hea_mae} and \ref{table:vac_hea_rms}.

Across all three modes, the SMAX-trained GRACE model consistently achieves the lowest root-mean-square (RMS) error-0.11eV for single-point, internal, and full relaxations-outperforming both the SMAX–OMAT model (0.13/0.14/0.14eV) and the OMAT-trained model (0.15/0.16/0.16~eV). The systematic nature of this improvement indicates that training on the SMAX dataset, with its intentionally broad coverage of bonding topologies, coordination environments, and local densities, yields more reliable defect energetics in chemically-disordered, multi-component alloys.

Taken together with the results for W/Be intermetallics, these findings reinforce a consistent conclusion: models trained on entropy-maximized datasets are better equipped to handle the complex, far-from-equilibrium local environments that dominate radiation damage in chemically complex materials. In contrast, models trained on conventional datasets can be vulnerable encountering  underconstrained regions of configuration space, leading to degraded accuracy precisely in the regimes most relevant for extreme-environment applications.

\begin{table}
{\small
\setlength{\tabcolsep}{4pt}% tighter columns
\begin{tabular}{lccc}
\hline
Method & \makecell{SMAX \\ (eV)} & \makecell{SMAX-OMAT \\ (eV)}  & \makecell{OMAT \\ (eV)}  \\
\hline
Single point & \textbf{0.11} & 0.13 & 0.15 \\
Internal relaxation & \textbf{0.11} & 0.14 & 0.16 \\
Full relaxation & \textbf{0.11} & 0.14 & 0.16 \\
\hline
\end{tabular}
}
\caption{RMS test errors (eV) for energy change upon vacancy creation ($E_{\mathrm{bulk}} - E_{\mathrm{bulk+V}}$) in a W$_{20}$Ta$_{20}$Cr$_{6}$V$_{8}$ high-entropy alloy. 
The lowest error in each row is highlighted in bold. 
%All models are variants of GRACE-2L-L. 
Relaxation protocols are as in Sec.\ \ref{subsec:wbe}.}
\label{table:vac_hea_rms}
\end{table}

\begin{table}
{\small
\setlength{\tabcolsep}{4pt}% tighter columns
\begin{tabular}{lccc}
\hline
Method & \makecell{SMAX \\ (eV)} & \makecell{SMAX-OMAT \\ (eV)}  & \makecell{OMAT \\ (eV)}  \\
\hline
Single point & \textbf{0.08} & 0.10 & 0.13 \\
Internal relaxation & \textbf{0.09} & 0.11 & 0.14 \\
Full relaxation & \textbf{0.09} & 0.11 & 0.14 \\
\hline
\end{tabular}
}
\caption{MAE test errors (eV) for energy change upon vacancy creation ($E_{\mathrm{bulk}} - E_{\mathrm{bulk+V}}$) in a W$_{20}$Ta$_{20}$Cr$_{6}$V$_{8}$ high-entropy alloy. 
The lowest error in each row is highlighted in bold. 
%All models are variants of GRACE-2L-L. 
Relaxation protocols are as in Sec.\ \ref{subsec:wbe}.}
\label{table:vac_hea_mae}
\end{table}

\subsubsection{Segregation in W/Nb/Mo/Ta/V alloy}

%\yl{I restarted/continued them, but they look already converged ...}
%\rd{Thank you. Actualy, which GB do we have here? And both GBs should be equivalent? Because the concentrations of the elements are not symmetric. Can the GBs be highlighted in the graph? Possibly overlaying the atomic simulated structure with the graph?}
%\rd{Can we rerun everything with 2L}

\begin{figure}
    \centering
    \includegraphics[width=\linewidth]{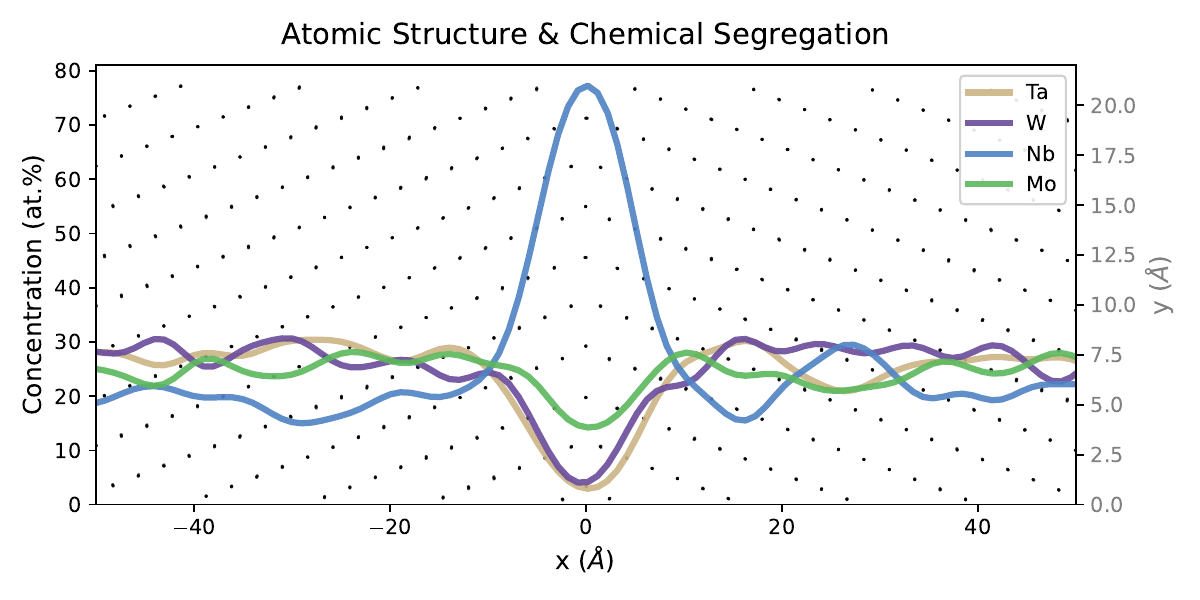}
    
    \caption{Segregation in refractory BCC-HEA (equimolar MoNbTaW) with $\Sigma$7 $[111]$ $\{123\}$ GB
    %GRACE-1L-SMAX-OMAT-large
    % \yl{TODO-SIM: increase bulk between GB, show only one GB, show atomic layers.}
    }
    \label{fig:BCC_HEA_4el_GB_seg_profile_1L}
\end{figure}

High-entropy alloys (HEAs) based on refractory elements, such as W-Nb-Mo-Ta and W-Nb-Mo-Ta-V, were among the first refractory HEAs synthesized and have since been extensively studied for their exceptional high-temperature stability \cite{senkovRefractoryHighentropyAlloys2010,kormannPhononBroadeningHigh2017,liuRecentProgressBCCStructured2022}. 
While the matrix remains stable and homogeneous down to low temperatures, the presence of extended defects, such as grain boundaries (GBs), can induce local segregation and demixing of the metallic elements.
% \rd{Which GB?}\yl{Sigma 7 - oriented x:1-32, y: 5-1-4, z:111}
We first analyzed segregation in the quaternary refractory alloy W-Nb-Mo-Ta, previously studied in Ref.~\cite{li2020complex}, using a representative BCC $\Sigma$7 $[111]$ $\{123\}$ symmetric tilt grain boundary. 
These simulations were performed using a hybrid MD-MC approach which included atom swaps between all pairs of elements~\cite{LAMMPS} and GRACE-1L-SMAX-OMAT-L potential.
The elemental concentration profiles, shown in Figure~\ref{fig:BCC_HEA_4el_GB_seg_profile_1L}, reveal distinct segregation.
A significant enrichment of niobium ($c(\text{Nb}) \approx 70$~at.\%) is observed at the grain boundary (GB) regions located at $x = 0$~\AA{}.
Contrary, Ta and W exhibit depletion at the GB with a corresponding enrichment in the bulk lattice. Molybdenum displays a relatively uniform distribution ($\approx 25$~at.\%) throughout the entire sample. 
%http://localhost:8888/lab/tree/acefit/GRACE/UEA/GB-many-impurities/Analyze_GB_segregation-MoNbTaW-GB.ipynb

This segregation pattern aligns closely with observations in Ref.~\cite{li2020complex}, where the Nb concentration was reported to increase to 57~at.\% at the GB and decrease to 15.5~at.\% in the bulk. 
Similarly, Li et al.\ observed that the Mo concentration remained constant at $\approx 25$~at.\% across both regions, while Ta and W were depleted from the GB and enriched in the bulk. 
Thus, the GRACE model effectively captures segregation in this compositionally complex alloy.

\begin{figure}
    \centering
    \includegraphics[width=\linewidth]{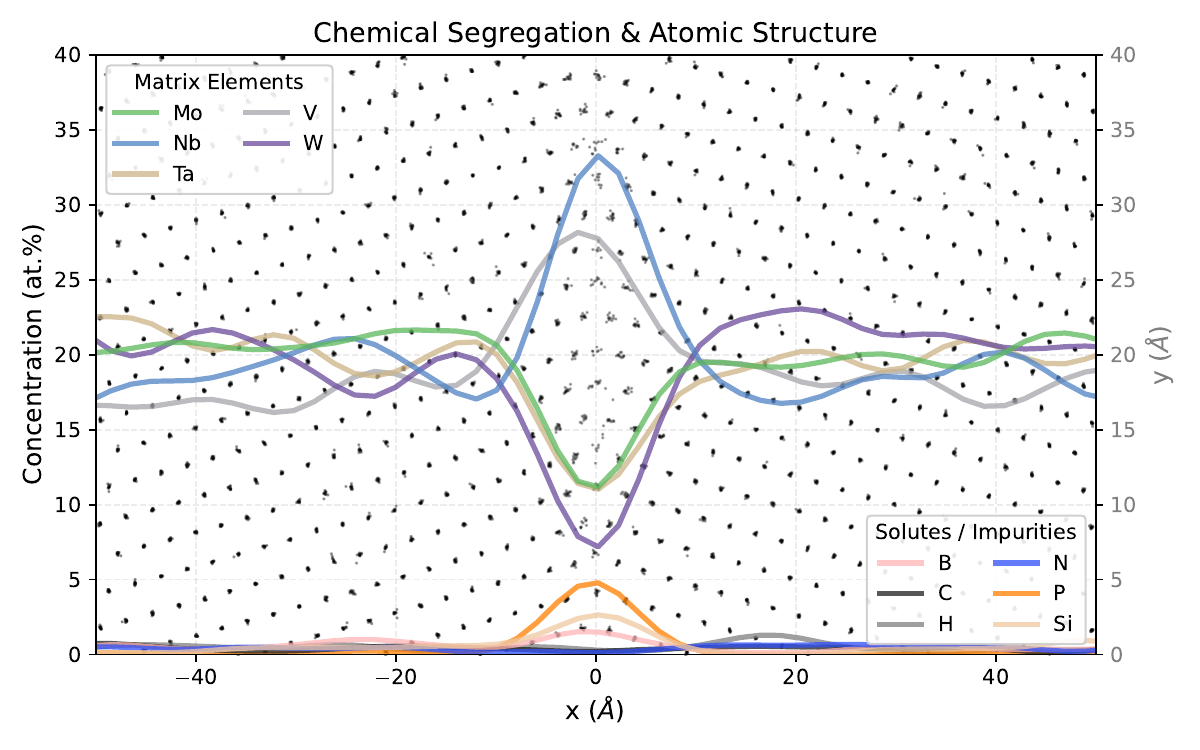}
    \caption{Segregation in BCC  W-Nb-Mo-Ta-V $\Sigma$7 $[111]$ $\{123\}$ GB with impurity elements (H,B,C,N,Si and P). 
    %as computed with GRACE-1L-SMAX-OMAT-large
    }
    \label{fig:BCC_HEA_GB_seg_profile_1L}
\end{figure}

%\yl{TODO:check}
Building upon the quaternary system, we now extend our analysis to the quinary W-Nb-Mo-Ta-V alloy.
In technological high entropy alloys, interstitial and light alloying elements are frequently present in  non-negligible concentrations.
To model this critical complexity, we further introduced interstitial impurities (H, B, C, and N) and substitutional impurities (Si and P).
These elements often exhibit a strong tendency for GB segregation, resulting in a complex competition between light element partitioning and local transition metal demixing. 
A grain boundary saturated with multiple alloying and interstitial elements may undergo fundamental property changes, potentially strengthening or weakening inter-atomic bonding, or inducing the formation of brittle defect phases.
Fig.~\ref{fig:BCC_HEA_GB_seg_profile_1L} presents the elemental profiles for both matrix elements and impurities across this specific $\Sigma$7 boundary.
These profiles were computed using MD-MC with atom swaps between all metallic pairs and additional displacement moves for the interstitial species (H, B, C, N) \cite{LAMMPS}. 
 At 300 K, we observe a pronounced segregation of Nb and V to the GB. 
 Notably, this strong metallic partitioning is accompanied by the concurrent segregation of P, Si, and B, whereas N and H maintain a uniform distribution across the bulk alloy. 
 These results emerge from a complex interplay between chemical interactions (cohesive energies) and steric effects-specifically the atomic size mismatch between bulk and GB sites.
%As indicated in the figure, the P, Si, and B impurities show a strong tendency to segregate to the GB, whereas N, H, and C show no significant segregation preference.

%While further detailed simulations are required to decouple these effects, such analysis is beyond the scope of the current work.

% http://localhost:8888/lab/tree/acefit/GRACE/UEA/GB-many-impurities/Analyze_GB_segregation-8k.ipynb

The ability to capture these multi-element segregation patterns highlights the robustness of the GRACE potential in describing the complex local environments found in refractory alloys.

\subsection{Reaction Barriers in Catalytic Systems}

The OC20NEB-OOD dataset \cite{wander2025cattsunami} comprises 460 nudged elastic band (NEB) calculations for heterogeneous catalytic reactions and serves as a challenging out-of-distribution benchmark for universal MLIPs which are typically not trained to reaction barriers. The dataset spans three reaction classes-desorption, bond dissociation/association, and intra- or intermolecular transfer-and considers reactions occurring on solid slabs randomly derived from bulk structures. The slabs expose surfaces with Miller indices $\leq 2$ and contain up to three chemical elements. 

The results, summarized in Table~\ref{table:ocneb}, show that the SMAX–OMAT model achieves the lowest overall errors, closely followed by the SMAX-only model. Both models substantially outperform the OMAT-trained baseline, despite the absence of explicitly catalytic or transition-state-like configurations in the SMAX training data. While the absolute accuracy remains insufficient for quantitative kinetics given the exponential sensitivity of reaction rates to barrier heights, the relative improvement is consistent and significant.

These results reinforce a central theme of this work: although the entropy-maximization strategy underlying SMAX is chemistry-agnostic, it effectively constrains chemically and physically relevant regions of the potential energy surface. This includes rare, high-energy configurations such as transition states, which are typically difficult to capture through conventional dataset construction strategies unless large numbers of expensive explicit barrier calculations are carried out.

\begin{table*}
\centering
\begin{tabular}{l|cccc|cccc|ccc}
\hline
 &  \multicolumn{4}{c}{SMAX}  &   \multicolumn{4}{c}{SMAX-OMAT}  & \multicolumn{3}{c}{OMAT}      \\
 \hline
 & \makecell{\%  $>$ 0.1 eV}  & \makecell{MAE \\ (eV)} & \makecell{RMSE \\ (eV)} &  & \makecell{\%  $>$ 0.1 eV}  & \makecell{MAE \\ (eV)} & \makecell{RMSE \\ (eV)}  & & \makecell{\%  $>$ 0.1 eV}  & \makecell{MAE \\ (eV)} & \makecell{RMSE \\ (eV)}   \\
\hline
Dissociation & 66.2 & 0.225 & 0.320 &  & \textbf{62.4} & \textbf{0.214} & \textbf{0.296}  & & 66.8 & 0.286 & 0.467  \\
Transfer     & 62.8 & 0.207 & 0.283 &  & 69.7 & 0.205 & \textbf{0.275}  & & \textbf{59.4} & \textbf{0.184} & 0.316  \\
Desorption   & 87.4 & \textbf{0.393} & \textbf{0.479} &  & \textbf{85.0} & 0.423 & 0.516  & & 91.3 & 0.718 & 1.25  \\
 &  &  &  &  &  &  &  &  &  &  &     \\
 Overall &  \textbf{70.8} & \textbf{0.264} & \textbf{0.359} & &  71.5 & 0.269 & 0.364 &  & \textbf{70.8} & 0.367 & 0.738   \\
\hline
\end{tabular}
\caption{Performance on the LAMBench/OC20NEB benchmark of activation energies of reactions relevant to heterogeneous catalysis. 
All models are variants of the GRACE large 2 layers model. See text for details.}
\label{table:ocneb}
\end{table*}

\subsection{Structure formation in multi-elemental mixtures}
Conventional atomistic simulations are typically driven by specific hypotheses or intended outcomes. For instance, the study of phases in tin (Sec. \ref{sec:tin}) relies on the targeted analysis of known deformation maps. However, such a prescriptive approach is largely unfeasible for Mendeleev materials, as the governing mechanisms and emergent properties are often impossible to pinpoint a priori.
To address this, we propose a new paradigm for the exploration of chemically complex space: ``discovery by simulation''. This approach relies on unbiased simulation protocols that allow for the autonomous generation of atomic arrangements, which are only subsequently analyzed for their structural and functional properties.

\begin{figure}
    \centering
    \includegraphics[width=\linewidth]{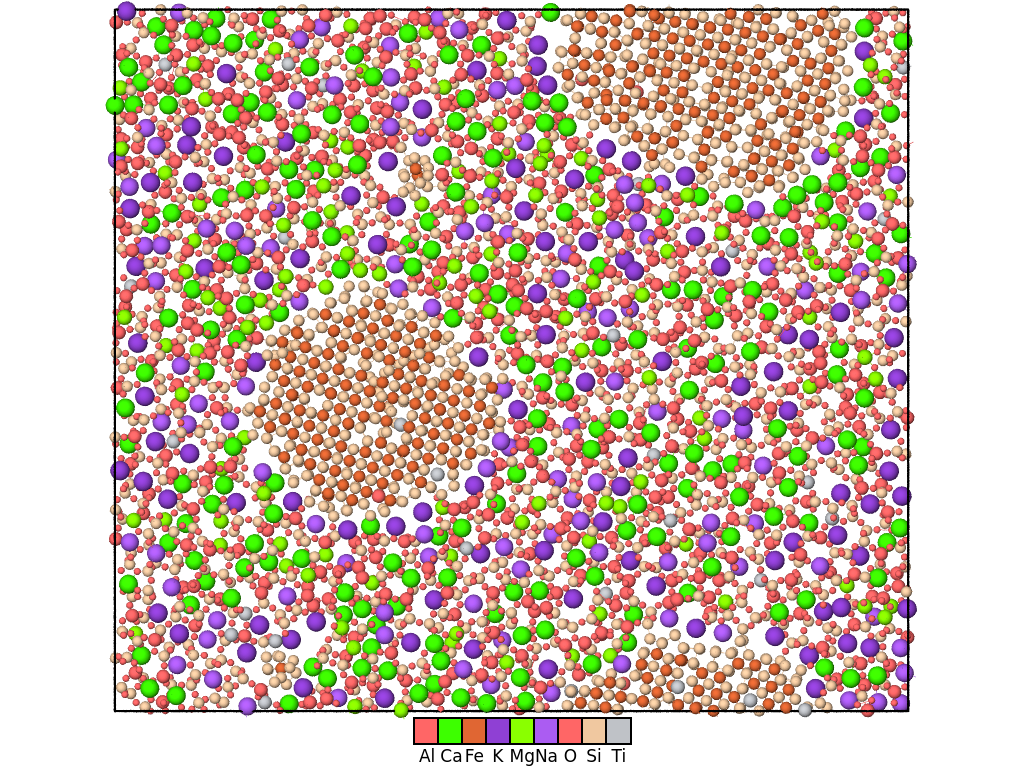}
    \caption{
    %\yl{TODO: clustering phase separation}
    Final snapshot of the ``Lava'' simulation after 4ns. The spontaneous phase separation and nucleation of Fe-Si clusters embedded into oxide matrix are clearly visible.
    }
    \label{fig:lava_snapshot}
\end{figure}

%\yl{Lava-sims (put before Mendeleev); little bit more info}
To demonstrate the power of discovery-driven simulation, we first prepared a system containing the nine most abundant elements in the Earth’s crust - O, Si, Al, Fe, Ca, Na, Mg, K, and Ti-mixed in their relative concentrations as displayed in Table~\ref{tab:lava_composition}. 
This ``Lava'' model serves as a benchmark for the ability of the SMAX-trained GRACE potential to navigate an arbitrarily complex potential energy surface and spontaneously form physically meaningful structures from a disordered state.
We employed the GRACE SMAX-OMAT model and quenched the temperature linearly from 5000\,K to 300\,K while the pressure was simultaneously reduced from 1\,GPa to 0\,GPa using a Nose-Hoover thermostat and an anisotropic barostat. 
The simulation ran for a total of 4 million steps with a 1 fs timestep.
The simulation results, illustrated in Fig.~\ref{fig:lava_snapshot}, demonstrate the emergence of distinct chemical environments. 
In this oxygen-lean mixture, we observe the nucleation of Fe-Si metallic clusters within the broader oxide matrix, indicating that the potential correctly captures the competitive bonding preferences between transition metals and the silicates.
To validate these observations, several clusters were randomly extracted from the final structure and recalculated using DFT.
The resulting comparison between DFT and various GRACE model predictions shows excellent agreement and is detailed in the Supplementary Information. 
Further analysis of the Fe-Si phase, including a comparison of its radial distribution function with known iron silicide phases, is also provided in the Supplementary Information.

\begin{table}[htbp]
    \centering
    \caption{Atomic Composition for ``Lava'' simulations}
    \vspace{0.2cm}
    \begin{tabular}{lcc}
        \hline
        \textbf{Element} & \textbf{Count} & \textbf{Concentration (\%)} \\
        \hline
        O & 2305 & 46.26 \\
        Si & 1410 & 28.30 \\
        Al & 412 & 8.27 \\
        Fe & 282 & 5.66 \\
        Ca & 208 & 4.17 \\
        Na & 118 & 2.37 \\
        Mg & 116 & 2.33 \\
        K & 104 & 2.09 \\
        Ti & 28 & 0.56 \\
        \hline
    \end{tabular}
    \label{tab:lava_composition}
\end{table}

% /home/users/lysogy36/acefit/GRACE/UEA/MENDELEEV_ALLOY/EARTH_ABUNDANT_NON_OPT_1L-UEA-OMAT-large-NPT/dump.npt-equil.dump

\begin{figure*}[htbp]
    \centering
    % Adjust the width fraction (0.19) to fit 5 images + spacing (5 * 0.19 = 0.95)
    
    \begin{subfigure}[b]{0.19\textwidth}
        \centering
        \includegraphics[width=1.05\textwidth]{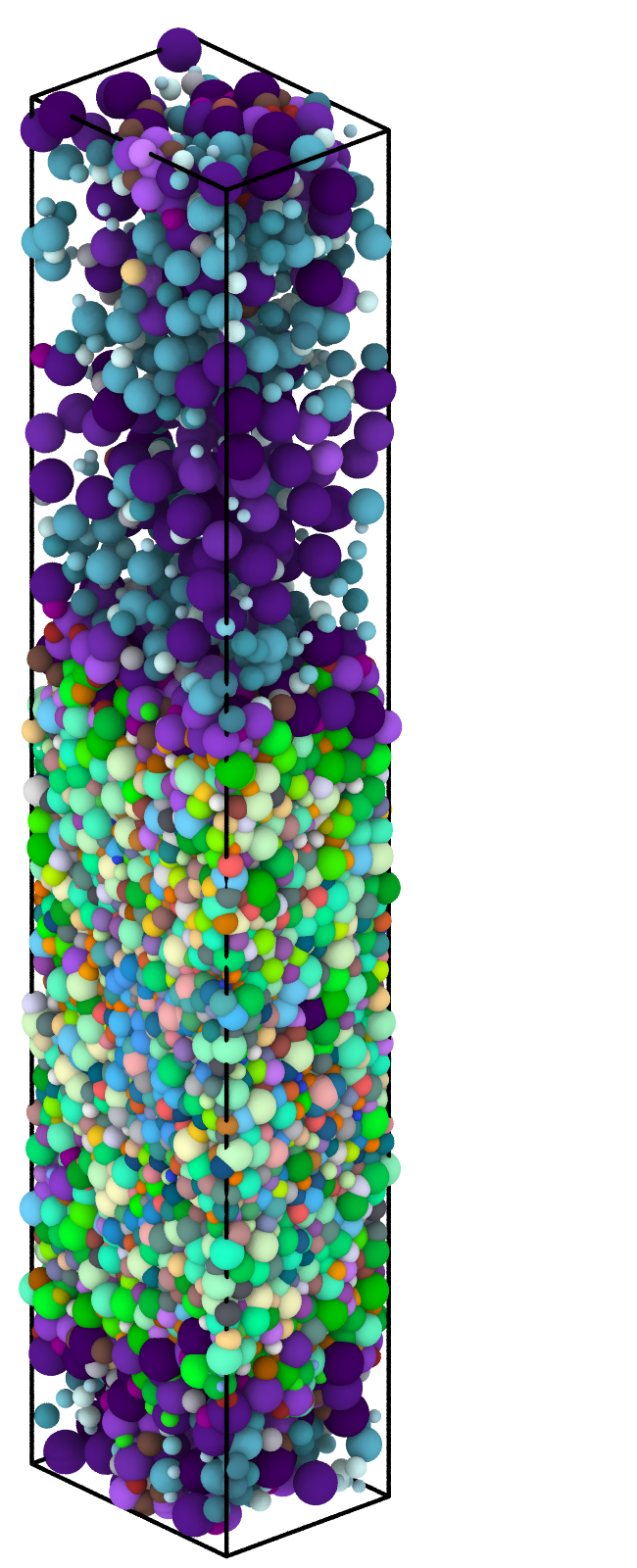}
        \caption{Mendeleev Alloy}
        \label{fig:snap0}
    \end{subfigure}
    \hfill % Adds flexible space between images
    \begin{subfigure}[b]{0.19\textwidth}
        \centering
        \includegraphics[width=1.05\textwidth]{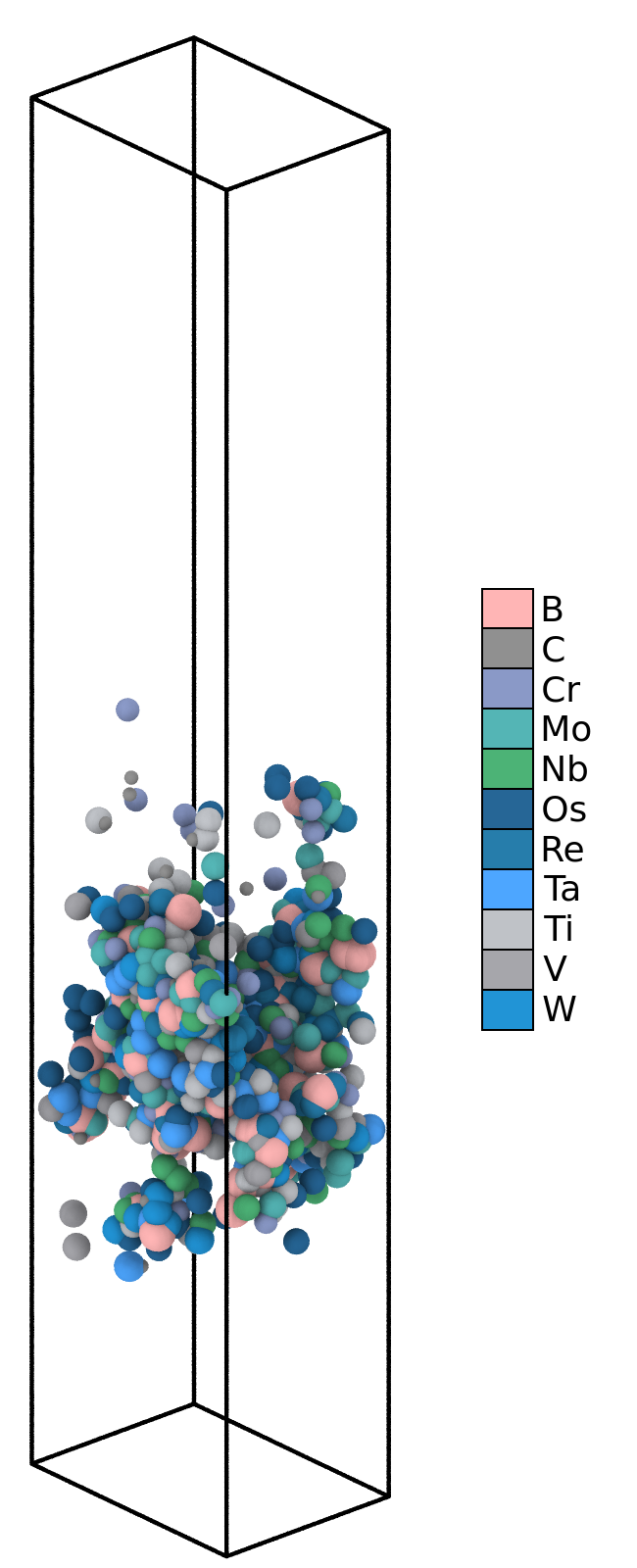}
         \caption{Refractory}
        
        \label{fig:snap1}
    \end{subfigure}
    \hfill
    \begin{subfigure}[b]{0.19\textwidth}
        \centering
        \includegraphics[width=1.05\textwidth]{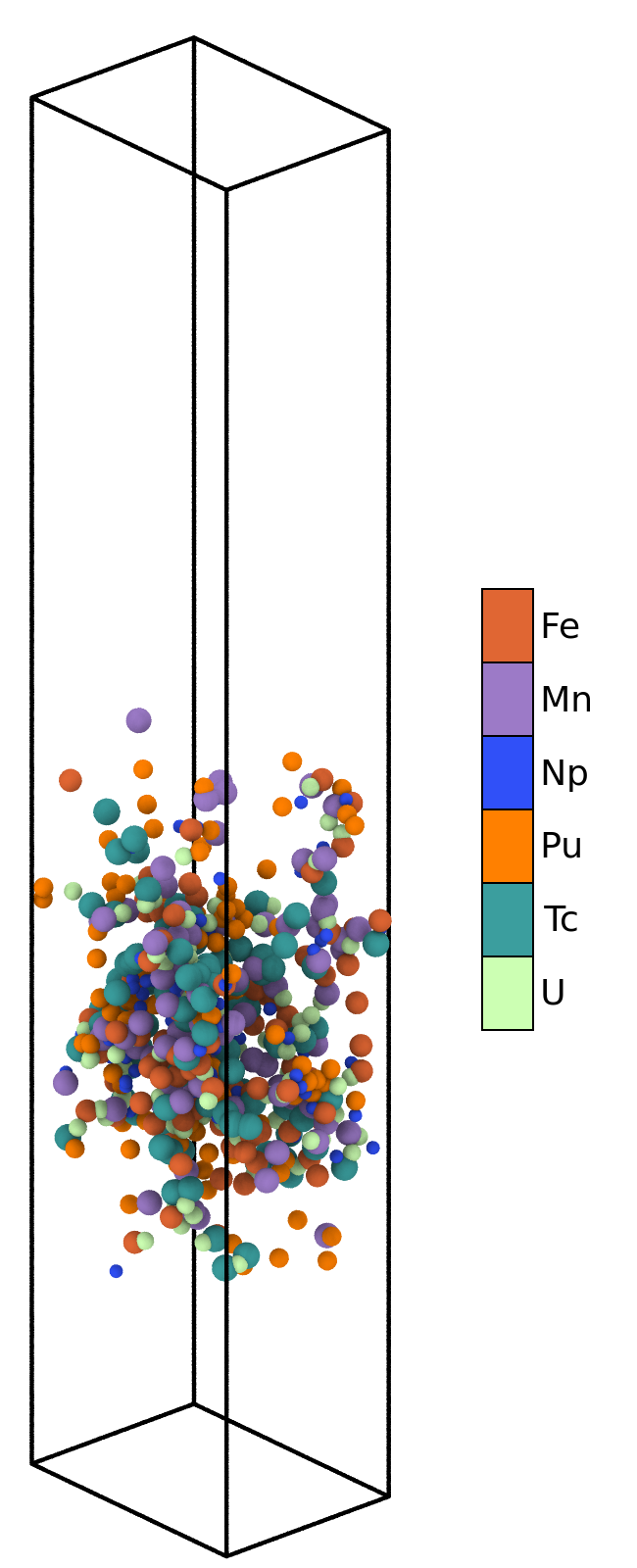}
        \caption{Actinide-Fe}
        
        \label{fig:snap2}
    \end{subfigure}
    \hfill
    \begin{subfigure}[b]{0.19\textwidth}
        \centering
        \includegraphics[width=1.05\textwidth]{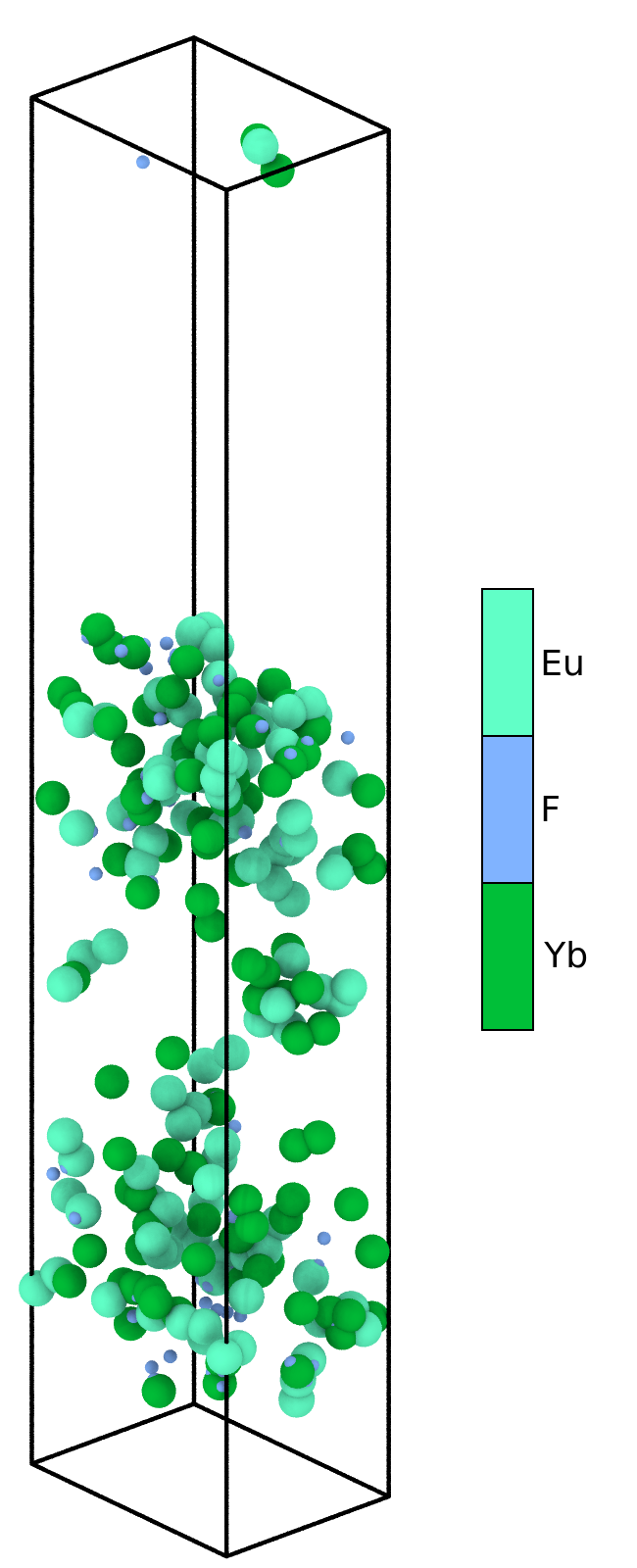}
        \caption{Fluorides}
        \label{fig:snap3}
    \end{subfigure}
    \hfill
    \begin{subfigure}[b]{0.19\textwidth}
        \centering
        \includegraphics[width=1.05\textwidth]{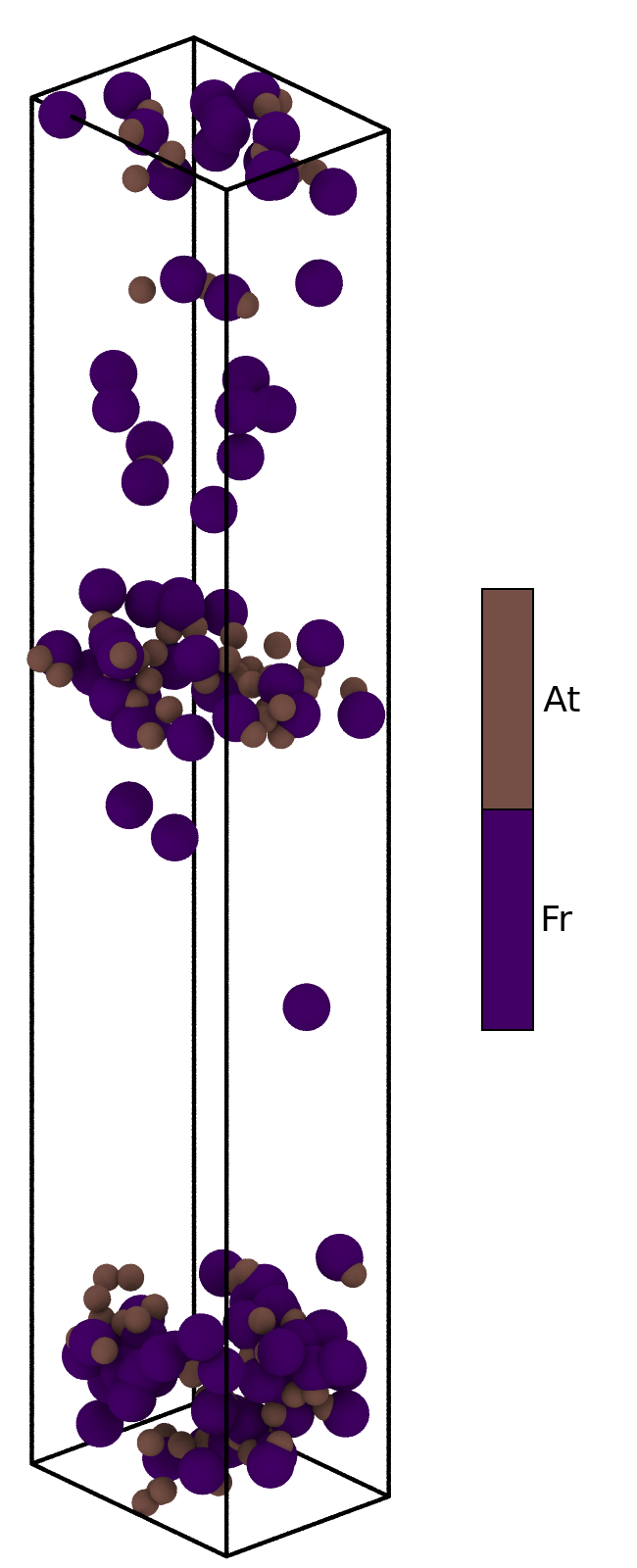}
        \caption{At-Fr Salt}
        \label{fig:snap4}
    \end{subfigure}
    
    \caption{Evolution of chemical clusters in the Mendeleev material. From left to right: (a) Complete structure, (b) Refractory Carbides/borides, (c) Actinide-Iron Eutectic, (d) Stable Fluorides and (e) Radioactive Salts.}
    \label{fig:mendeleev_five_snapshots}
\end{figure*}

%\yl{TODO: Put Mendeleev material sims, see Fig.~\ref{fig:mendeleev_five_snapshots}}
To test the limits of chemical variety, we performed a series of simulations on a ``Mendeleev'' material using the GRACE-2L-SMAX-OMAT-medium model. In this setup, we randomly placed 94 different elements - with 106 atoms of each species-into a cubic box. We followed the same quenching procedure as before, cooling the system from 5000 K to 300 K while reducing the pressure from 2 GPa to 0 GPa. 
Across five independent runs, we observed consistent and reproducible segregation trends.
Fig.~\ref{fig:mendeleev_five_snapshots}(a) displays the final structure from a representative simulation.
The system evolved into an elongated morphology, driven by the minimization of surface energy between the separating phases.
To decode the complex chemistry of this system, we performed an automated graph-theoretical analysis of the bond network. 
By constructing a connectivity graph - where elements act as nodes and bond counts determine edge weights - we extracted the four most significant chemical clusters (see Supplementary Information).
Atoms of corresponding elements from these clusters, visualized in Fig.~\ref{fig:mendeleev_five_snapshots}(b--e), reveal distinct chemical segregation driven by the cooling kinetics. 
Refractory carbide and boride ceramics (Fig.~\ref{fig:mendeleev_five_snapshots}(b)) solidified first due to their extremely high melting points. 
In contrast, iron-actinide mixtures (Fig.~\ref{fig:mendeleev_five_snapshots}(c)) segregated into eutectic domains (particularly Fe-U and Fe-Pu), remaining liquid until lower temperatures.
We also identified highly stable europium and ytterbium fluorides (Fig.~\ref{fig:mendeleev_five_snapshots}(d)), which formed due to the unique stability of the divalent (+2) oxidation states of Eu and Yb. Finally, the segregation of astatine-francium salts (Fig.~\ref{fig:mendeleev_five_snapshots}(e)) represents an extreme limit of ionic bonding between the heaviest alkali and halogen elements.

% http://localhost:8888/lab/tree/acefit/GRACE/UEA/Mendeleev_alloy_simulation_setup-MANY_ELEMENTS-Many-2L.ipynb
% /home/users/lysogy36/acefit/GRACE/UEA/Mendeleev_clusters_slices/Mendeleev_clusters_slices.xyz

\begin{comment}
\subsection{Metal-Organic Frameworks}

\rd{I think that Danny suggested to add something on MOFs, if I remember correctly.}
\end{comment}

% \subsection{Cellulose pyrolysis}

% \rd{Should we also write something here? -> but still need to quantify reactions products, etc. ...  Current analysis uses DFT to identify compute reaction barrieres for specific pathways \cite{wangInitialPyrolysisMechanism2020}. Here, instead, we use our discovery by simulation approach that does not require an initial hypothesis...}
% \yl{Maybe too much...}

\section{Conclusions}

In this work, we have demonstrated that the primary bottleneck in the development of universal machine learning interatomic potentials (MLIPs) is not the expressivity of the models, but the inherent bias of the datasets upon which they are trained.
By moving beyond the traditional reliance on human-curated, low-energy configurations, we have established the Maximum Entropy (SMAX) dataset as a foundational framework for atomistic simulation.

The core of our approach lies in its chemistry-agnostic sampling. By utilizing information-entropy maximization, we have decoupled structural exploration from the underlying energetics of specific materials. This shift is transformative: it allows the dataset to systematically populate the chemical feature space independently of thermodynamic relevance, effectively filling critical data holes that lead to instabilities in traditional MLIPs. This provides a physical prior for the periodic table, enabling models to interpolate across the atomic universe with unprecedented reliability.

The enhanced robustness enables a significant leap in our simulation capabilities, specifically allowing us to tackle the vast complexity of Mendeleev materials. These systems, which embrace the full elemental diversity of the periodic table, have historically been inaccessible. By providing the necessary atomic basis to navigate these arbitrarily complex chemical spaces, we enable a new paradigm of discovery by simulation. In this approach, simulations are no longer restricted to testing narrow a priori hypotheses, but can instead autonomously explore the potential energy surface to resolve emergent structures and properties across chemistries and structures.

Ultimately, the SMAX framework provides the rigorous methodology required to bridge the gap between idealized laboratory materials and the chemical complexity of the industrial and planetary systems. Whether applied to the extreme environments of fusion reactors or the intricate recovery of elements from global waste streams, this chemistry-agnostic foundation opens the door to a nearly infinite chemical landscape, transforming the way we explore, understand, and design the materials of the future.

\section{Methods}

\subsection{Multicomponent maximum information entropy structure generation}

A large dataset of atomistic structures was generated using a generalization of the information-entropy maximization method introduced in Refs.\ \cite{karabin2020entropy,montes2022training} and extended in Ref.\ \cite{pa2025information}. In this approach, the process of curating a training set is reframed as an optimization procedure whose objective function is the information entropy of the entire dataset as measured in a feature space that characterizes local atomic environments. This approach is completely agnostic to the energetics of specific materials, as it is driven only by topological features, in contrast to common techniques like active learning that explicitly tailor datasets to specific materials, usually through dynamical samplers that favor thermodynamically or dynamically accessible structures. Instead, information entropy maximization attempts to cover as much of the feature space as possible independently of the potential thermodynamic relevance of the corresponding configurations. This philosophy implicitly values transferability and robustness over local accuracy in a subset of the configuration space. This can be expected to be particularly advantageous in applications where it is {\em a priori} difficult to delineate the local environments that are likely to be encountered in practice, e.g., for applications to extreme conditions, but also more generally to avoid the presence of deleterious holes in the dataset where the lack of constraining data can lead to catastrophic losses in accuracy, sometimes even compromising the stability of simulations \cite{montes2022training,fu2022forces}. In the case of relatively simple ML potentials like SNAP \cite{wood2018extending} and ACE \cite{drautz2019atomic}, this approach was previously shown to yield ultra-transferable models at the cost of a modest increase in test errors on human-curated data sets, as the extreme diversity of the data set severely strains the expressivity limits of simpler functional forms. In contrast, models trained to human-curated data sets dramatically failed when tested on high-diversity data, with observed errors increasing by multiple orders of magnitude compared to hold-out data from their own training set \cite{montes2022training,pa2025information}. In the context of extremely flexible functional forms such as GRACE, one can expect this local accuracy/transferability trade-off to be even more favorable, an intuition that is  confirmed in the present study.

The information entropy is defined as:
\begin{equation}
    S = -\int{p\left(B\right)\log{p\left(B\right)dB}} \,,
    \label{eq:entropy}
\end{equation}
where $B$ is an $m$-dimensional feature vector characterizing atomistic environments and $p(B)$ is the corresponding probability distribution over the corresponding dataset. At large $m$, numerical quadrature is in general intractable, so Eq.\ \ref{eq:entropy} is not a practical objective function. Instead, a parametric approximation is invoked where the $p(B)$ is approximated as an $m$-dimensional normal distribution with mean $\mu$ and covariance matrix $\Sigma$, i.e., 
\begin{equation}
p(B) \simeq N(\mu, \Sigma),
\end{equation}
and $\mu$ and $\Sigma$ are obtained using conventional sample estimators, as described below. 

Within this parametric space, the information entropy is analytically given by
\begin{equation}
S=\frac{m}{2}\log(2\pi e) + \frac{1}{2}\log \det(\Sigma),
\label{eq:entropy_normal}
\end{equation}
which, up to irrelevant constants, is proportional to the logarithm of the determinant of the feature covariance matrix; in an abuse of language, we refer to $\log \det(\Sigma)$ as the information entropy in the following. Within this approximation, the objective function admits a simple interpretation in terms of the sum of the logarithm of the variance of the feature distribution along all principal directions, with narrow distributions corresponding to low entropies and broad distributions to high entropies. Therefore, whether or not the parametric approximation provides a numerically accurate approximation to the true information entropy, this objective function naturally promotes diversity and minimizes redundancy, which is the key objective of this approach.

In practice, the dataset is incrementally constructed using a greedy optimization scheme. 
Denote the dataset at iteration $k$ by $\mathcal{S}_k$. At each iteration, a candidate configuration $\mathcal{C}_k$ is initialized at random, including cell size, shape, and atomic positions. The feature entropy of the augmented dataset $\mathcal{S}_k \cup \mathcal{C}_k$ is then locally maximized with respect to the atomic positions  in $\mathcal{C}_k$ only, using a local optimization scheme. If, upon convergence to a local maximum of the objective function, i.e., if $S(\mathcal{S}_k \cup \mathcal{C}_k) > S(\mathcal{S}_k)$, it is added to the dataset --- $\mathcal{S}_{k+1}=\mathcal{S}_k \cup \mathcal{C}_k$ --- otherwise the dataset remains unchanged and $\mathcal{S}_{k+1}=\mathcal{S}_k$.

In order to avoid the generation of non-physical structures with overlapping atoms, which would lead to diverging energies and forces, the objective function is augmented by a short-range  repulsive term that prevents atomic approaches closer than some user-defined core radius. The function that is maximized is of the form:

\begin{equation}
    F(\mathcal{C}_k | \mathcal{S}_k) = K*\log \det [\Sigma(\mathcal{S}_k \cup \mathcal{C}_k)]  - V_\mathrm{core}(\mathcal{C}_k) \,,
    \label{eq:objective}
\end{equation}

where $K$ is an adjustable constant that controls the relative strength of the information entropy and core repulsion terms. As stated above, the feature mean and covariance are estimated using sample estimators:

\begin{equation}
\begin{aligned}
\mu(\mathcal{S}_k \cup \mathcal{C}_k)  = &  \frac{1}{N+1} \left[ \sum_{\alpha \in \mathcal{S}_k} B_\alpha + \sum_{\beta \in \mathcal{C}_k} B_\beta \right] \,,\\
\Sigma(\mathcal{S}_k \cup \mathcal{C}_k)  = &  \frac{1}{N+1} \left[ \sum_{\alpha \in \mathcal{S}_k} (B_\alpha -\mu)(B_\alpha -\mu)^T \right.\\  
&+ \left. \sum_{\beta \in \mathcal{C}_k}(B_\beta-\mu)(B_{\beta}-\mu)^T\right] \,,
\end{aligned}
\end{equation} 
where $N$ is the number of items in the augmented dataset. As discussed in Ref.\ \cite{pa2025information}, we consider two variants, a so-called ``per-atom'' version, where the sum runs over atomic feature vectors and a ``per-configuration'' version, where the sum runs over configurations and the feature vectors are averaged over all atoms in the corresponding configuration. In principle, any set of descriptors can be used in conjunction with this approach. As in previous studies, we employed the first 55 bispectrum components \cite{wood2018extending}. Further details can be found in the original references \cite{karabin2020entropy,montes2022training,pa2025information}. 

\begin{figure}
    \centering
    \includegraphics[width=0.35\textwidth]{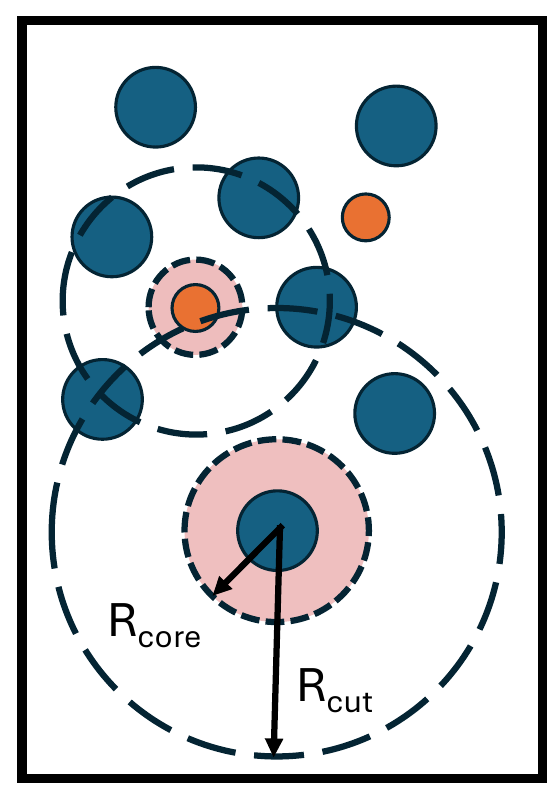}
    \caption{Schematic illustration of the multi-component featurization approach. Each atom is assigned a featurization cutoff radius $R_{cut}$ and core exclusion radius $R_{core}$ sampled from a chemical-element-dependent distribution. Chemical elements are indicated by different atom colors and sizes. From an initially random spatial arrangement, the feature information entropy of the current dataset augmented by the candidate configuration under consideration is then maximized.}
    \label{fig:cutoffs_cartoon}
\end{figure}

The methodological innovation in the present context is the treatment of multi-component systems that contain an arbitrary number of chemical elements, generalizing the approach described in Ref.\ \cite{pa2025information}. At each iteration, a new configuration is initialized randomly as described above. The number of distinct elements $N_e$ in the configuration is then randomly sampled according to a user-defined probability distribution. The identity of the specific element $\{e_1,..., e_{N_e}\}$ and the number of atoms of each type $\{n_1,..., n_{N_e}\}$  
are further randomly sampled according to additional user-specified probability distributions. 
To account for chemical diversity, the hyper-parameters controlling the computation of the atomistic features are set on a per-atom basis by drawing an effective radius $r_i$ from a Beta distribution $B(1.25,1.25)$ centered on the tabulated covalent radius of the corresponding chemical element, extending over a radius range of 30\% of the tabulated covalent radius; the cutoff radius of the atomic descriptor calculation is then set to $R_{\mathrm{cut}}=3r_i$ and the corresponding core exclusion radius to $R_{\mathrm{core}}=0.8r_i$ (c.f., Fig.\ \ref{fig:cutoffs_cartoon}). With these definitions in place, the greedy optimization loop proceeds as before. This mechanism introduces {\em configurational} complexity that reflects the {\em chemical} diversity within each structure even when the features that describe the local atomic density around each atom don't explicitly resolve atomic species. This simple approach avoids the combinatorial explosion that would result from the use of many-body features that explicitly resolve chemical species.

This procedure was used to generate a core dataset of 916,388 entropy-maximized structures that forms the basis of the SMAX dataset.

\subsection{Dataset enrichment}

The core entropy-maximized dataset was enriched through a number of mechanisms meant to balance the tradeoffs between extreme transferability and coverage of lower energy structures that are often emphasized in applications to materials discovery. We therefore also included additional configurations from various sources, as follows: 

\begin{itemize}
\item 492,006 low-force configuration were generated by optimizing structures using a preliminary GRACE potential trained to core set of entropy-maximized data. Relaxations were carried out either while holding the simulation cell fixed or allowing it to also relax. Note that these structures were not re-optimized using DFT, so they do not necessarily exhibit small reference forces.
\item 131,256 structures found on the convex hulls identified by the Materials Project
\item 132,745 structures of organic molecules extracted from the QM9 dataset \cite{ramakrishnan2014quantum}.
\item 20,991 structures of organic molecules extracted from the Transition1X-a dataset \cite{schreiner2022transition1x}.
\end{itemize}
%\ab{Is it possible that the rest are actinides? Because apart from the t1x dupes, I removed 3685 structures containing Cm, and 3570 by energy/force min/max. I don't know what's the rest}
%\dannyp{We can go with your numbers since that is what you trained to and release your cleaned up data.}

In all cases, the reference energies, forces, and stresses were recomputed with the method described in the next section to ensure consistency.

\subsection{Reference data generation}

Reference energies, forces, and stresses corresponding to these configurations were generated using VASP \cite{hafner2008ab} version 6.3.0 using a 500 eV plane wave cutoff, a 0.125 $/\text{\AA}$ k-point density, the PBE exchange-correlation functional, and using the VASP-recommended pseudo-potentials. The complete specification of the VASP options is provided in Supplementary Material \ref{supp:vasp}.
Calculations were executed on the Venado supercomputer at Los Alamos National Laboratory using custom high-throughput workflows based on the Flux resource management framework \cite{ahn2014flux}.

\subsection{Maximum Entropy (SMAX) dataset characteristics}

The SMAX dataset is composed of 1,693,386 configurations containing a total of 23,071,117 atoms. For training and evaluation purposes, we partitioned the data into a training set of 1,660,322 configurations and a testing set of 33,064 configurations. 
To evaluate its performance against the current state of the art, we compared the structural and chemical diversity of SMAX to several other foundational datasets, including OMAT \cite{barroso2024open}, ALOE \cite{kuner2025mp}, MATPES \cite{kaplan2025foundational}, MAD \cite{mazitov2025massive}, and sAlex~\cite{schmidt2023machine,wang2023symmetry}.

For these comparisons, we randomly sampled subsets from each dataset. 
These representative samples included 33,064 structures (449,659 atoms) from SMAX, 25,000 structures (677,677 atoms) from MAD, 45,489 structures (293,057 atoms) from MP-ALOE, 50,000 structures (447,300 atoms) from MATPES, 27,406 structures (547,588 atoms) from OMAT, and 34,578 structures (412,528 atoms) from sAlex.

\begin{figure*}
    \centering
    \includegraphics[width=\linewidth, height=0.9\textheight, keepaspectratio]{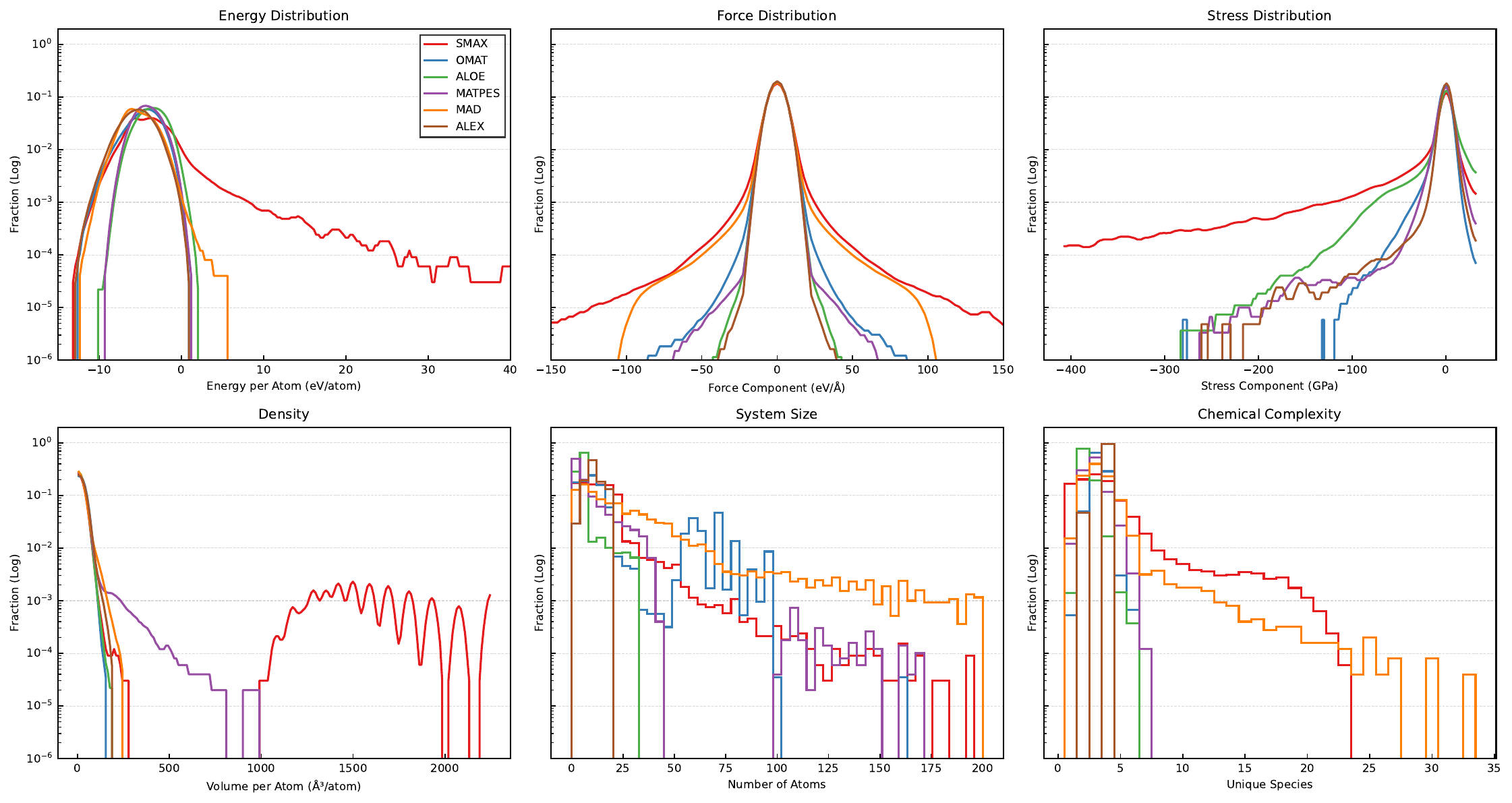}
    \caption{
        Statistical characteristics of the SMAX dataset compared to other foundational datasets, including OMAT \cite{barroso2024open},
        ALOE \cite{kuner2025mp}, MATPES \cite{kaplan2025foundational}, MAD \cite{mazitov2025massive}, and ALEX~\cite{schmidt2023machine,wang2023symmetry}.
    }
    \label{fig:dataset_statistics_6_panel_robust}
\end{figure*}

The comparisons illustrate the significantly increased configurational and energetic diversity of SMAX compared to other datasets. For example, as displayed in Fig.~\ref{fig:dataset_statistics_6_panel_robust}, the energy distribution in most datasets is confined to bound configurations. While SMAX's distribution is also focused on low energies, it exhibits a considerably longer tail that extends up to 200 eV/atom.
These configurations are usually not sampled in conventional materials science applications, but can be extremely relevant to simulations of extreme conditions, including irradiation, detonation, shock propagation, etc. 
Similar behavior is observed for the force and stress distributions, where SMAX is significantly broader than in other datasets, while remaining peaked in the close-to-ambient conditions that are the most common. In terms of chemical diversity, SMAX also contains significantly more configurations containing more than 10 chemical elements, ensuring broad configurational and chemical coverages.

%The distribution of number of atoms per cell in the core entropy-maximized dataset is roughly uniform between $N=1$ and 25, followed by a rapid decay for larger cells. Similarly, most configurations contain between 2 and 10 different atomic species per configuration, with the tail of the distribution extending up to 24 different species per cell.

Fig.~\ref{fig:chemical_distribution_pt} shows that the dataset is particularly rich in elements from the first three rows of the periodic table, whose chemistry is of particular interest in many applications, while the rest of the periodic table is roughly uniformly sampled. The typical number of configuration containing any given chemical element is on the order of 50,000 with a corresponding total number of atoms of a particular element of around 150,000. 

%\begin{figure}
    %\centering
    %\includegraphics[width=0.35\textwidth]{figures/atom-count-histogram.png}
    %    \includegraphics[width=0.35\textwidth]{figures/species-count-histogram.png}
    %\caption{Distributions of the number of atoms (top) and of the number of chemical species (bottom) per configuration across %the SMAX dataset.}
%    \label{fig:chemical_distribution}
%\end{figure}

\begin{figure*}
    \centering
    \includegraphics[width=\linewidth, height=0.9\textheight, keepaspectratio]{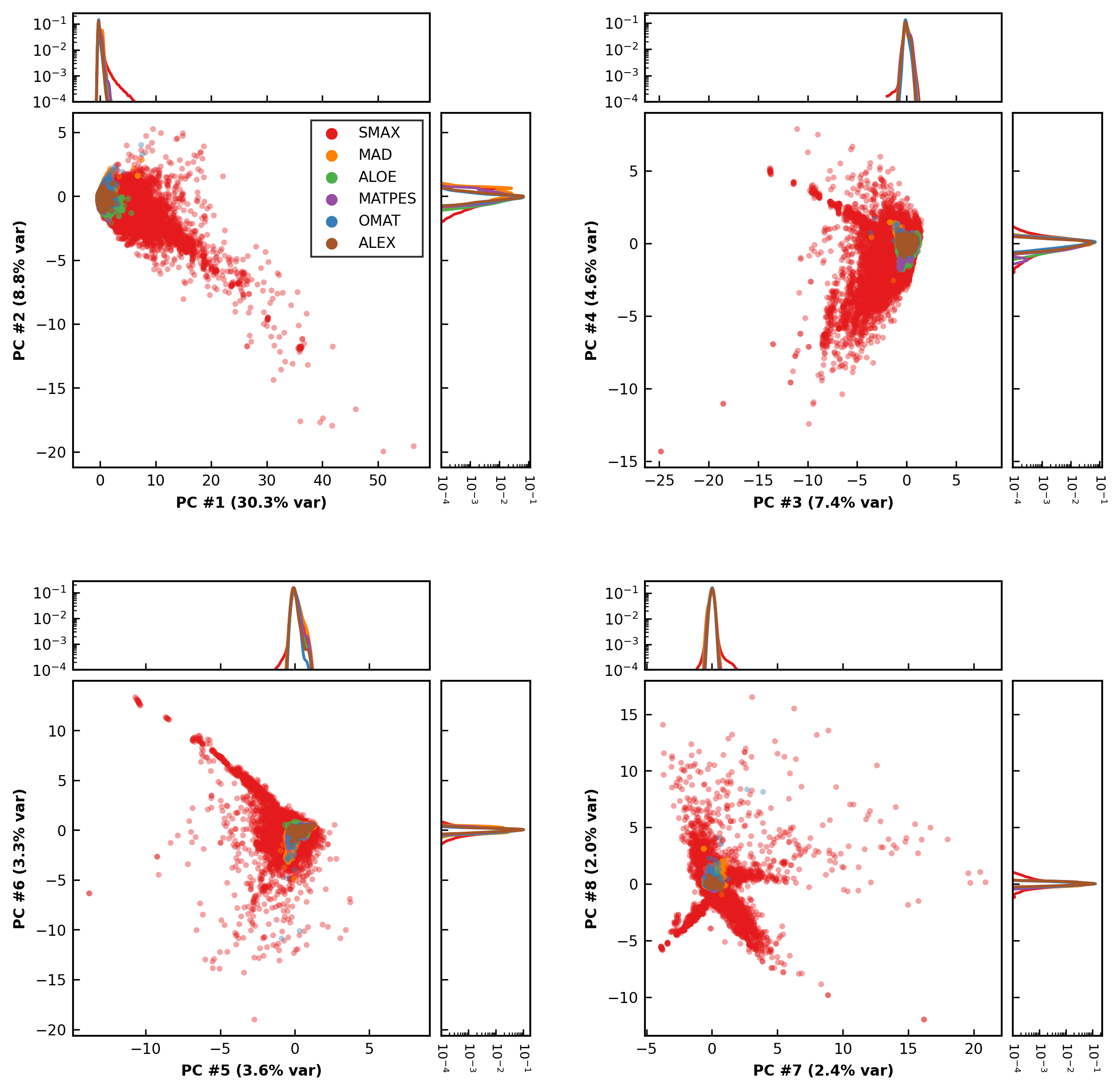}
    \caption{First eight PCA projections of the six foundational datasets, highlighting the expansive feature-space coverage of the SMAX dataset.}
    \label{fig:pca_pairs_6_ds}
\end{figure*}

While the energetic diversity of SMAX is evident, energies, forces and stresses are but an indirect proxy of topological diversity. 
This aspect can be probed more directly by comparing the distributions of the invariant features that are internally learned by GRACE during training.
To describe each atomic environment, we utilized 5,664 invariant features learned by the GRACE-1L-SMAX-OMAT-large model. We combined 5,000 structures sampled from each of the six foundational datasets into a single feature matrix comprising 436,726 atoms.
A Principal Component Analysis (PCA) was trained on this combined matrix to reduce the dimensionality of the feature space. 
We then applied on-the-fly PCA transformations to the larger subsets to visualize their distributions. 
As shown in Fig.~\ref{fig:pca_pairs_6_ds}, SMAX covers the GRACE feature space much more extensively than the other datasets. 
While the bulk of the distributions overlap, SMAX exhibits ``fat tails'' that span a significantly larger volume of the feature space. 
We hypothesize that this broad coverage acts as a regularizer, providing the model with sparse but critical information about the outer regions of the feature space that would otherwise be undefined.

\subsection{Graph atomic cluster expansion}

The Graph Atomic Cluster Expansion (GRACE) framework \cite{bochkarev2024} provides a rigorous and general mathematical foundation for constructing high-fidelity and computationally efficient MLIPs. GRACE can be viewed as a systematic generalization of the Atomic Cluster Expansion (ACE) formalism. While ACE furnishes a complete and systematically improvable basis for representing local atomic environments using star-graph topologies centered on a single atom, GRACE extends this concept to a complete basis of graph functions. This extension enables the explicit and systematic inclusion of semilocal and nonlocal interactions beyond strictly local atomic neighborhoods.

Through efficient recursive evaluation, the GRACE formalism naturally gives rise to equivariant message-passing graph neural networks. GRACE derives multi-ACE that inserts star-graph ACE for message passing on each layer \cite{batatiaDesignSpaceE3equivariant2025} as implemented in MACE \cite{NEURIPS2022_4a36c3c5}. In this sense, GRACE provides a unifying theoretical framework that encompasses both traditional local MLIPs based on ACE and modern graph-based architectures, offering to systematically improve model expressivity while retaining clear physical interpretability.

Chemical diversity is incorporated in GRACE via tensor decomposition and chemical embedding, allowing arbitrary chemical interactions to be represented within a single, unified parameterization. This enables the description of interactions spanning the entire periodic table within one model. In the present work, long-range electrostatic interactions extending beyond the product of the number of message-passing layers and the local cutoff radius are not explicitly treated. Likewise, variations in magnetic structure are not modeled through explicit magnetic degrees of freedom; instead, the potential is trained to reproduce spin-polarized density functional theory energies as a function of atomic positions alone.

\subsection{Training and model description}

We developed and parameterized local GRACE-1L-L and semi-local GRACE-2L-L architectures \cite{lysogorskiyGraphAtomicCluster2025a} utilizing the SMAX and OMAT datasets. To maintain a high-fidelity representation across diverse chemical environments, we implemented element-specific cutoff radii (see Table~\ref{tab:element_cut}), with bond cutoffs determined by the arithmetic mean of the constituent elemental values. This adaptive cutoff strategy facilitates simulation of compounds with large nearest-neighbor separations without incurring the prohibitive computational overhead associated with a global interaction range. Models were trained on both the standalone SMAX repository and an augmented SMAX+OMAT dataset to ensure robustness across a broad structural landscape. Comprehensive parameterization details are available in Supplementary Information Sec.~\ref{sec:mod_detail}.

\section{Data availability}

The SMAX Maximum Entropy dataset and the GRACE models will be made available when the present manuscript is accepted for publication.

\section{Acknowledgements}

AB and RD acknowledge funding from the Deutsche Forschungsgemeinschaft (DFG, German Research Foundation) through SFB 1394 (project number 409476157).
AB and YL acknowledge high-performance computing resources, provided by the Paderborn Center for Parallel Computing (PC$^2$) and the Elysium HPC cluster at Ruhr-University Bochum.
RD acknowledges the support of the Seaborg Institute at Los Alamos National Laboratory for their hospitality while this project was initiated.
DP acknowledges the support of the U.S.~Department of Energy (DOE) ARPA-E CHADWICK Program through Project DE‐AR0001988. 
This research used resources provided by the Los Alamos National Laboratory Institutional Computing Program, which is supported by the U.S. Department of Energy National Nuclear Security Administration under Contract No. 89233218CNA000001
Los Alamos National Laboratory is operated by Triad National Security, LLC, for the National Nuclear Security Administration of the U.S. Department of Energy (contract no.89233218CNA000001).

%\section{Author information}

\newpage
\clearpage

\section{Supplementary Materials}

\subsection{Maximum Entropy (SMAX) dataset characteristics}

\begin{figure*}
  \includegraphics[width=0.9\textwidth]{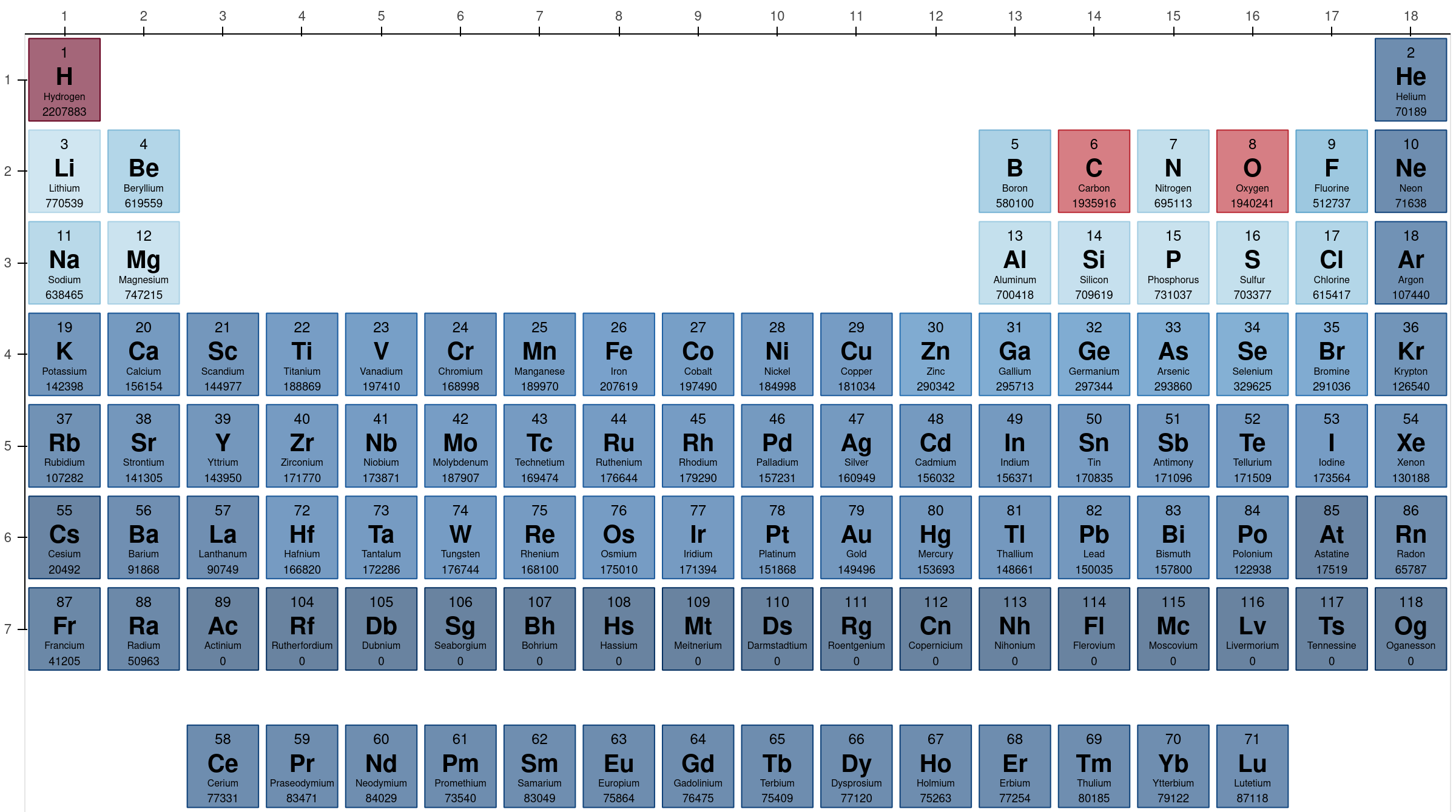}
    \includegraphics[width=0.9\textwidth]{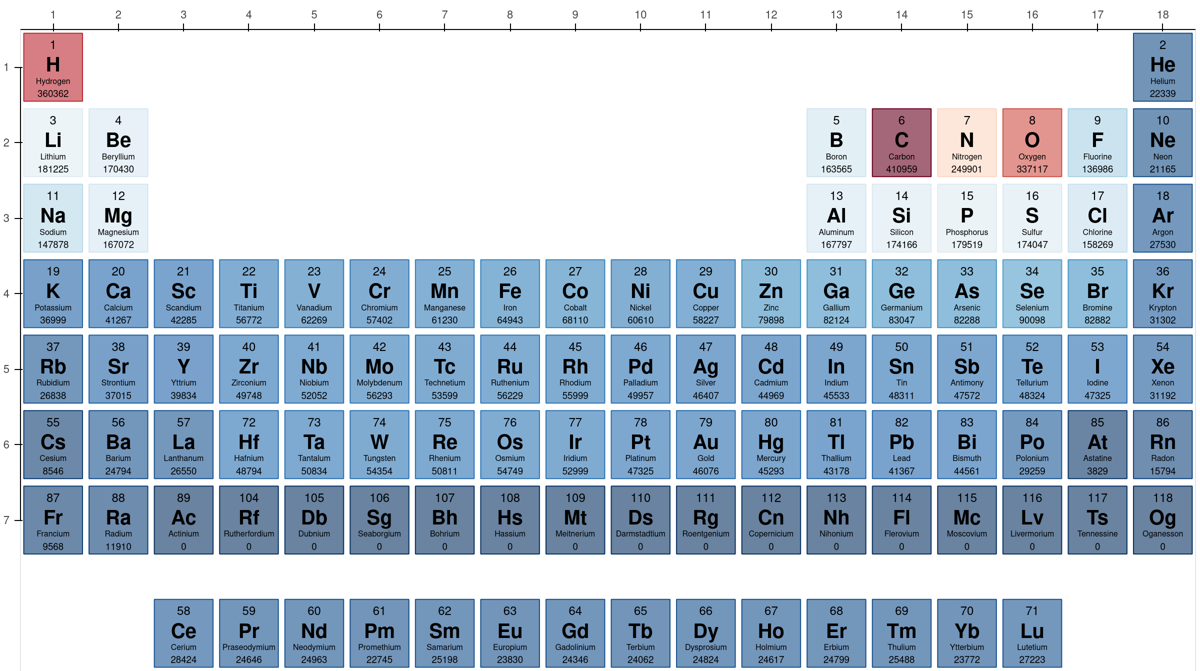}
 \caption{Number of atoms of each chemical species (top) and number of configurations that contain at least one atom of a given chemical species (bottom), across the SMAX dataset.}
      \label{fig:chemical_distribution_pt}
\end{figure*}

Fig.~\ref{fig:chemical_distribution_pt} shows the distribution of chemical species and the number of structures that contain each chemical species in the periodic table, excluding actinides.

\subsection{GRACE models characteristics and validation}

We have carried out validation assessment of the different GRACE parameterizations. We focus on training set dependence and comparison to a number of recent foundational MLIPs.

\subsubsection{Elastic moduli}
\begin{figure*}
    \centering
    \includegraphics[width=0.7\linewidth]{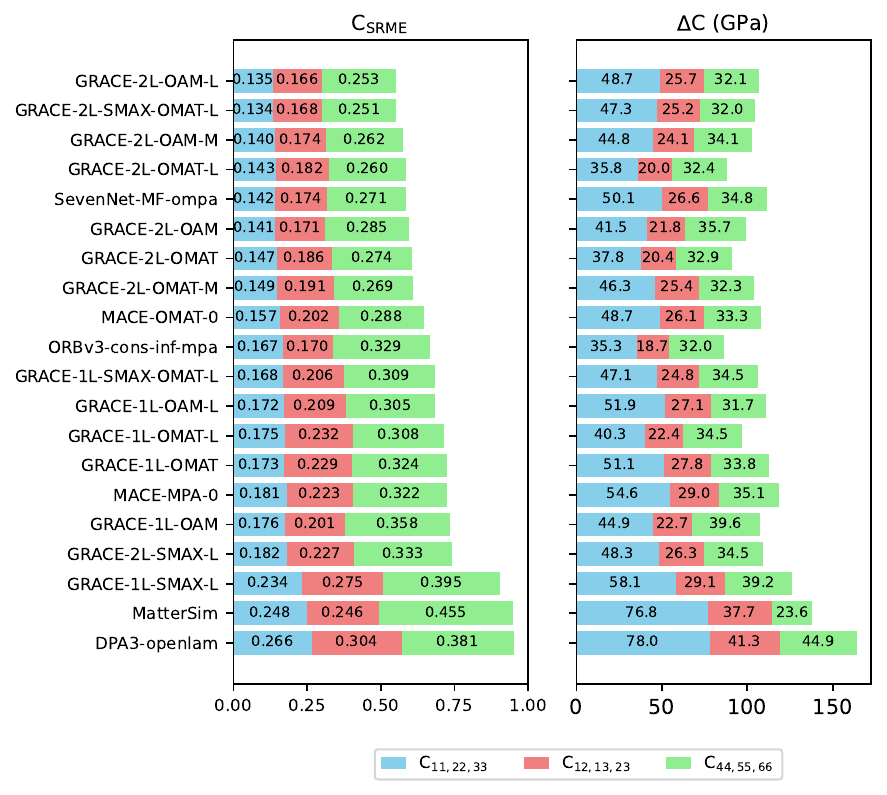}
    \caption{
    % \yl{Remove majority of older models, keep only best from GRACE-UFF, 
    % individually for each category}
    The symmetric relative mean error (SRME) and MAE ($\Delta C$ in GPa) for elastic constants, categorized into three subgroups:  longitudinal ($C_{11}, C_{22}, C_{33}$), Poisson's ratio-related ($C_{12}, C_{13}, C_{23}$), and shear ($C_{44}, C_{55}, C_{66}$). See text for more details.
    }
    \label{fig:elast_delta_overview}
\end{figure*}

For a comprehensive evaluation of the elastic properties, we followed the test protocols detailed in  Ref.~\cite{lysogorskiyGraphAtomicCluster2025a}.
In Fig.~\ref{fig:elast_delta_overview}, we present the symmetric relative mean error (SRME) and the absolute Mean Absolute Error (MAE) for a test set consisting of 7,962 structures. These structures were originally sourced from the Materials Project~\cite{de2015charting} and represent a common subset for which elastic tensors could be successfully computed by all evaluated potentials using an energy-based method \cite{golesorkhtabar2013elastic}. 
We see that GRACE models generally improve over other available models that were trained on OMAT. We also note that the models that were trained on the SMAX dataset only are slightly worse than models trained on OMAT. This is due to the abundance of structures in the OMAT dataset that are at or close to local minima, as elastic moduli specifically characterize the deviation from these equilibrium states.
% http://localhost:7777/notebooks/storage/UFF-tests/Elastic/Plot-elastic-new-models.ipynb

\subsubsection{Surface energies}
\begin{figure}
    \centering
    \includegraphics[width=\linewidth]{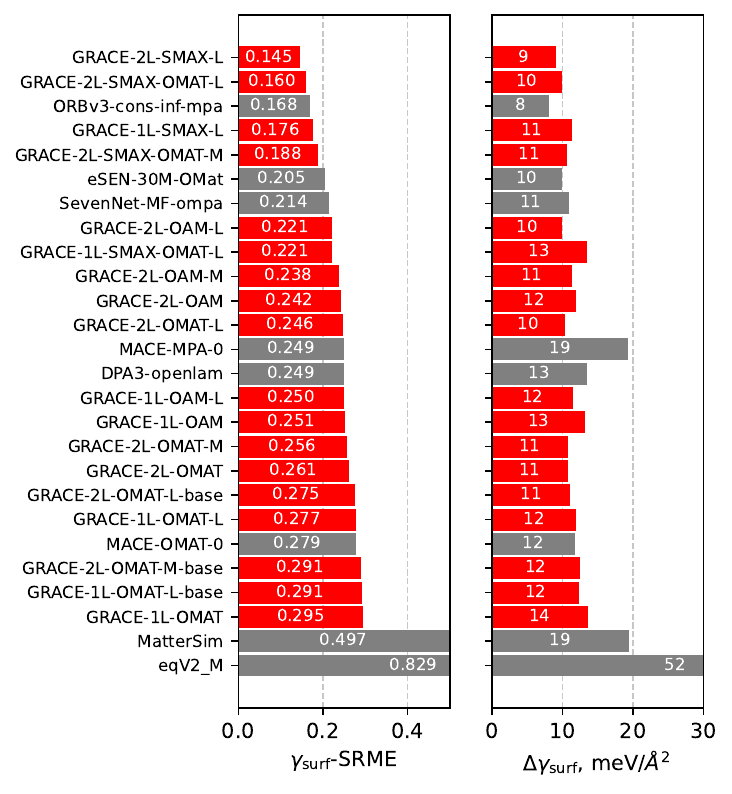}
    \caption{{Accuracy for surface formation energies of unary systems: symmetric relative mean error $\gamma_\mathrm{surf}$-SRME and mean absolute error $\Delta \gamma_\mathrm{surf}$.} GRACE models are highlighted in red.}
    \label{fig:surf_delta_overview}
\end{figure}

In Fig.~\ref{fig:surf_delta_overview} we compare surface energies for elemental materials, see Ref.~\cite{lysogorskiyGraphAtomicCluster2025a} for more details. Again, the GRACE models are of leading accuracy. Here, the formal sampling of the SMAX dataset shows its strength, models trained on the SMAX dataset generally perform better than models trained on OMAT alone.
% http://localhost:7777/notebooks/storage/UFF-tests/Surfaces/SURF-table-plot.ipynb

\subsubsection{Grain boundary energies}
\begin{figure}
    \centering
    \includegraphics[width=\linewidth]{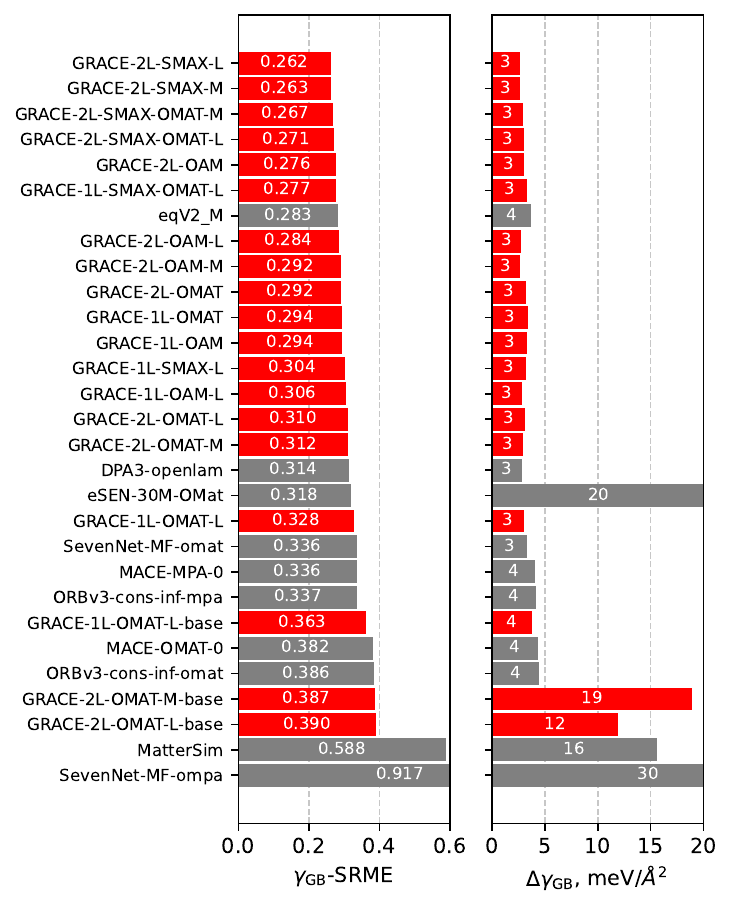}
    \caption{
        {Accuracy for grain boundary formation energies of unary systems: symmetric relative mean error $\gamma_\mathrm{GB}$-SRME and mean absolute error $\Delta \gamma_\mathrm{GB}$.} GRACE models are highlighted in red. 
    }
    \label{fig:gb_delta_overview}
\end{figure}

A similar trend can be observed in Fig.~\ref{fig:gb_delta_overview}. The prediction of grain boundary energies in elemental materials benefits from the diversity of the SMAX dataset and models trained on the SMAX dataset have leading predicitve accuracy. 
See Ref.~\cite{lysogorskiyGraphAtomicCluster2025a} for more details of the test.

% http://localhost:7777/notebooks/storage/UFF-tests/GBs/GB-table-plot.ipynb

\subsubsection{Self-interstitial energies}
\begin{figure}
    \centering
    \includegraphics[width=\linewidth]{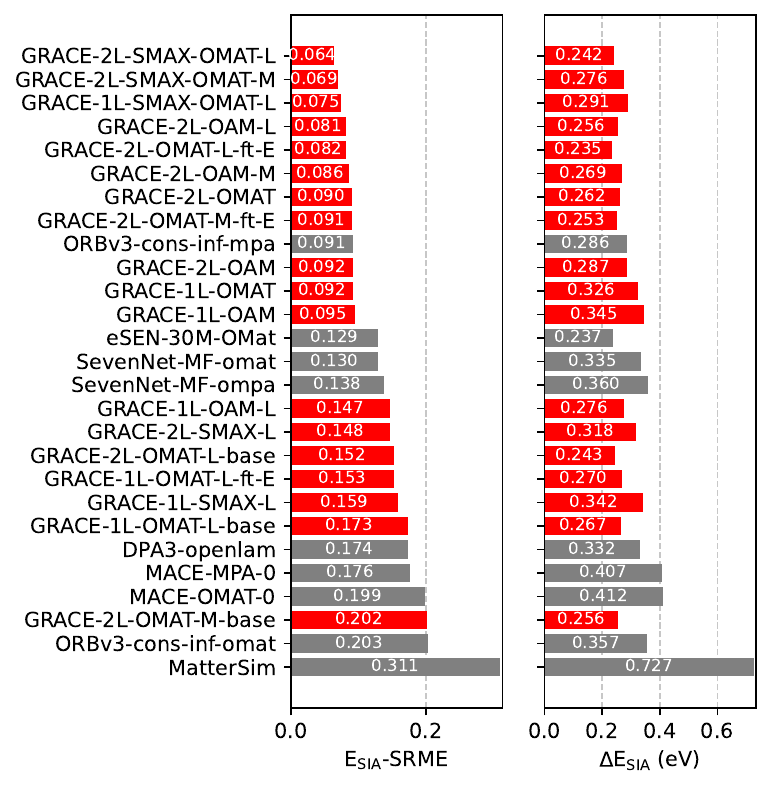}
    \caption{
        {Accuracy for self-interstitials formation energies of unary systems.} GRACE models are highlighted in red.
    }
    \label{fig:SIA_SRME_overview}
\end{figure}

The importance of the sampled SMAX training dataset also is visible for the prediction of self-interstitial energies in elemental materials displayed in Fig.~\ref{fig:SIA_SRME_overview}. The SMAX-trained models provide the most accurate predictions of self-interstitial energies. 
See Ref.~\cite{lysogorskiyGraphAtomicCluster2025a} for more details of the test.

% http://localhost:7777/notebooks/storage/UFF-tests/PointDefects/SIAs-table-plot.ipynb

\subsubsection{Vacancy formation energies}
\begin{figure}
    \centering
    \includegraphics[width=\linewidth]{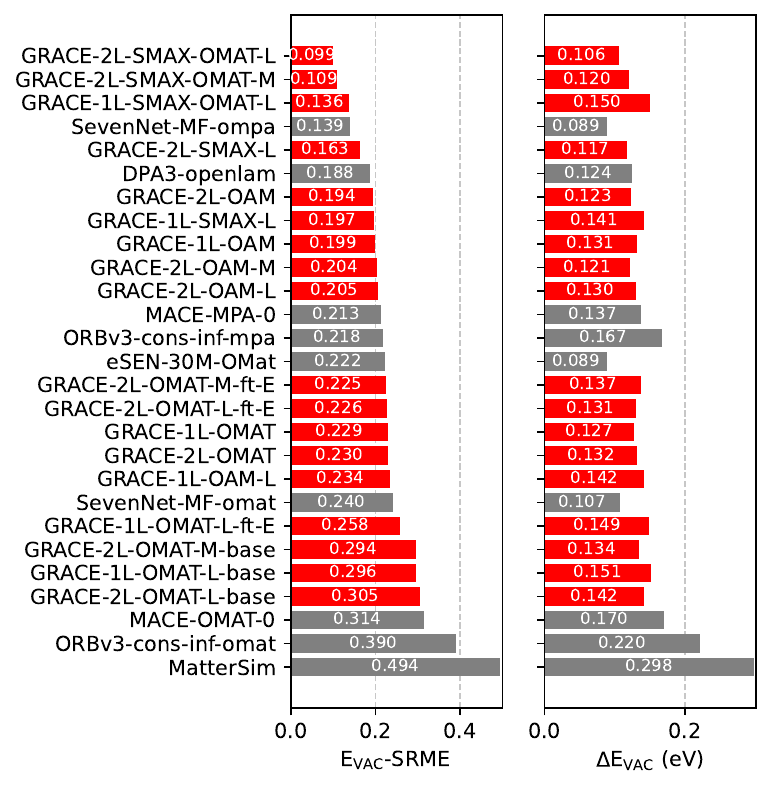}
    \caption{
        {Accuracy for vacancies formation energies of unary systems.} GRACE models are highlighted in red.
    }
    \label{fig:VAC_SRME_overview}
\end{figure}

An analogous conclusion is obtained for the prediction of vacancy formation energies displayed in Fig.~\ref{fig:VAC_SRME_overview}. The SMAX-trained models provide the most accurate predictions also for vacancy formation energies.
See Ref.~\cite{lysogorskiyGraphAtomicCluster2025a} for more details of the test.

% http://localhost:7777/notebooks/storage/UFF-tests/PointDefects/SIAs-table-plot.ipynb

\subsubsection{Ni-NiO defoliation}
\begin{figure*}
    \centering
    \includegraphics[width=\linewidth]{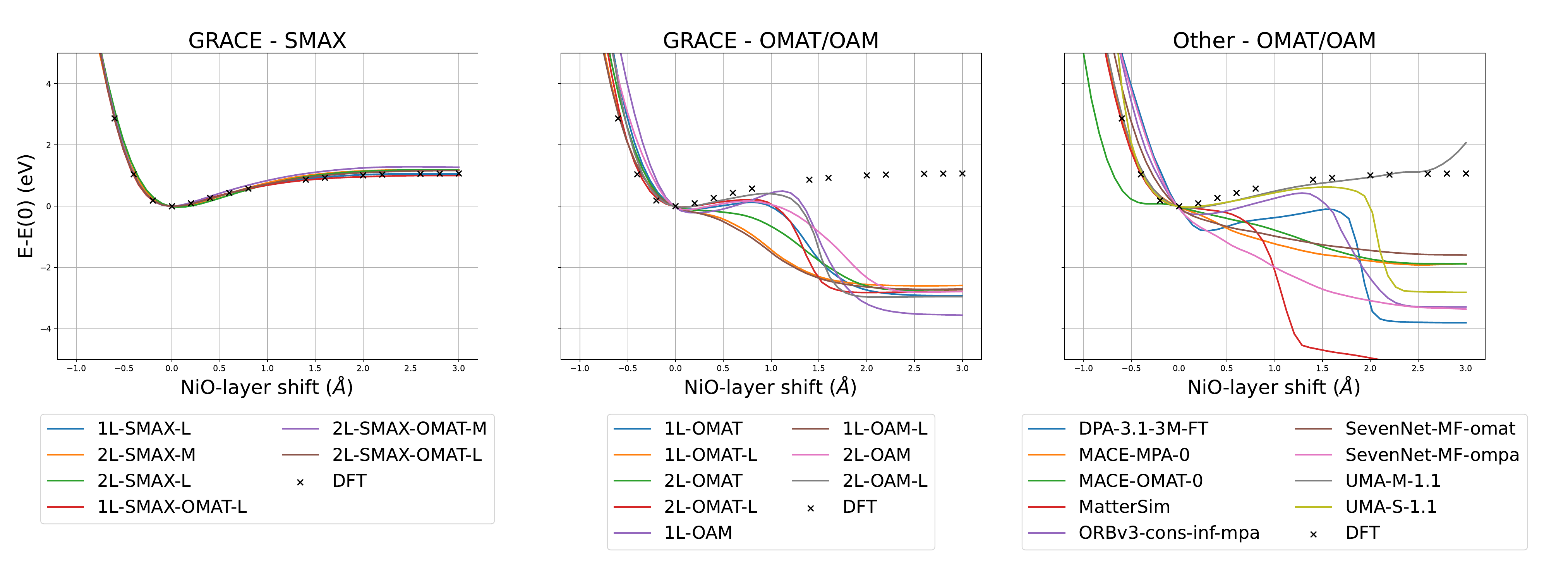}
    \caption{Ni-NiO-Defoliation energy as a function of distance between Ni nd NiO layer.}
    \label{fig:Ni_NiO_Defoliation_Plots}
\end{figure*}

%\yl{TODO: Double Check relaxation of NiO (111?) slab; try W-C decohesion}
Problems of state-of-the-art datasets become obvious also when bonds are stretched far away from their equilibrium distances. We carried out illustrative calculations for the defoliation of NiO from Ni. In the calculations we separate the NiO layer from Ni and plot the energy as a function of separation distance in Fig.~\ref{fig:Ni_NiO_Defoliation_Plots}. The models trained on SMAX all give a reasonable overall agreement to the DFT reference data, with quantitative deviations for the simpler one-layer models (left panel).

In contrast, none of the models trained on the OMAT dataset or the combination of OAM datasets provides a physically or chemically sensible extrapolation of the energy distances that differ significantly from the equilibrium distance (middle panel of Fig.~\ref{fig:Ni_NiO_Defoliation_Plots}). 
This failure is a direct manifestation of the "Hubbard $U$ pathology" also recently identified in foundational datasets \cite{warford2026better}. Because datasets like OMat24 and Alexandria inherit the Materials Project's selective application of the $+U$ correction-applying it only when oxygen or fluorine are present-they effectively force the MLIP to interpolate between two incompatible potential energy surfaces (PES). As the NiO layer is pulled away from the metallic Ni slab, the models encounter a discontinuous energy landscape, leading to the spurious repulsion and severe underbinding observed in the MACE-OMAT and eSEN models~\cite{warford2026better}.

Analogous failures are evident for all other MLIPs that we tested (right panel of Fig.~\ref{fig:Ni_NiO_Defoliation_Plots}). 
Without the formal and consistent sampling of the SMAX dataset, which avoids these $+U$ inconsistencies by maintaining a uniform PES, none of the main datasets allows for the parameterization of a MLIP that predicts the energies associated with bond breaking during Ni-NiO defoliation.

% prepared DFT data in http://localhost:7777/notebooks/storage/UFF-tests/NiO/MeOxide-defoliation.ipynb
% submit DFT to ZGH-/home/users/lysogy36/tools/VASP/Ni-NiO-slab
% http://localhost:7777/notebooks/storage/UFF-tests/NiO/PLOT-NiO-defoliation-test.ipynb

% \begin{figure*}
%     \centering
%     \includegraphics[width=\linewidth]{figures/BCC-HEA-GB-seg-profile-2L.pdf}
%     \caption{Segregation in BCC-HEA GB, as computed with GRACE-2L-UEA-OMAT-large}
%     \label{fig:BCC_HEA_GB_seg_profile_2L}
% \end{figure*}

% http://localhost:7777/lab/workspaces/auto-d/tree/acefit/GRACE/UEA/FOR_PAPER_Feature_map_1L-ALOE-MATPES-UEA-OMAT-global-per-element-resolution.ipynb

% http://localhost:7777/lab/workspaces/auto-d/tree/acefit/GRACE/UEA/FOR_PAPER_Feature_map_1L-per-element.ipynb
% \begin{figure*}
%     \centering
%     \includegraphics[width=\linewidth, height=0.9\textheight, keepaspectratio]{figures/elementwise_pca_comparison_1L.pdf}
%     \caption{Elementwise PCA projections comparison for UEA and OMAT subsets}
%     \label{fig:element_wise_pca}
% \end{figure*}

\subsection{Simulation details}

\subsubsection{Combining SMAX and OMAT dataset}\label{sec:data_comb}

To merge the OMAT and SMAX datasets effectively, we applied a filtering process to exclude incompatible entries and statistical outliers. The filtering criteria were as follows:
\begin{itemize}
    \item The OMAT dataset utilizes PBE+U for configurations containing O or F in combination with specific elements~\cite{barroso2024open}. We removed these entries to ensure energy consistency with the SMAX dataset, which relies exclusively on pure PBE calculations specifically to avoid conflict between incompatible PESs.
    \item Configurations containing Gd, Eu, and Yb were excluded because OMAT uses different pseudopotentials for these elements compared to SMAX.
    \item We identified outliers in the OMAT dataset characterized by a lack of correlation between GRACE model stress predictions and reference values. To validate this discrepancy, we recomputed 100 configurations with the largest stress mismatch using DFT. The model predictions showed significantly better agreement with the recomputed values than with the original labels, suggesting the original discrepancies were likely due to unconverged calculations. Notably, many of these problematic configurations contained Ce atoms. Consequently, we removed configurations where the maximum absolute stress prediction error exceeded the error distribution's standard deviation of 0.084 eV/$\mathrm{\AA}^3$.
\end{itemize}

The final combined dataset amounts to 98,700,740 configurations and includes 94 elements. It is worth noting that the elements Ac, Np, Pa, Pu, Th, and U are present only in OMAT, while At, Fr, Po, Ra, and Rn appear unique to SMAX.

For model parametrization we utilize a loss function that consists of three parts,
\begin{equation}
\mathcal{L} = \alpha_\mathrm{E} \mathcal{L}_\mathrm{E} + \alpha_\mathrm{F} \mathcal{L}_\mathrm{F} + \alpha_\mathrm{S} \mathcal{L}_\mathrm{S},
\end{equation}
where $\mathcal{L}_\mathrm{E}$, $\mathcal{L}_\mathrm{F}$, and $\mathcal{L}_\mathrm{S}$ correspond to losses of energy per atom, force components, and stress components, respectively.
We use $\alpha_\mathrm{E}:\alpha_\mathrm{F}:\alpha_\mathrm{S} = 48:128:128$ for all SMAX models and $32:128:128$ for combined SMAX-OMAT models. All SMAX models were trained for 1000 epochs, while combined models were trained for 12 and 8 epochs for 1L and 2L models, respectively.
We utilize the Huber loss for $\mathcal{L}$ with parameter $\delta=0.01$ for all components.
For minimizing the loss function, we employed the Adam~\cite{kingma2014adam} optimizer with cosine learning rate reduction scheme, an initial learning rate of 8$\times10^{-3}$ and a minimum learning rate of 5$\times10^{-4}$.

\begin{table*}
\centering
\caption{Cutoff values in $\AA$ for each element used for parametrizing SMAX and SMAT-OMAT models 
}
\label{tab:element_cut}
\vspace{0.2cm}
\begin{tabular}{lc| @{\hspace{1cm}} lc| @{\hspace{1cm}} lc| @{\hspace{1cm}} lc}
\hline
Element & Cutoff & Element & Cutoff & Element & Cutoff & Element & Cutoff \\
\hline
Ag & 6.0 & F  & 6.0 & N  & 6.0 & Sb & 5.5 \\
Al & 6.0 & Fe & 6.0 & Na & 6.0 & Sc & 6.0 \\
Ar & 6.2 & Fr & 7.5 & Nb & 6.0 & Se & 5.5 \\
As & 6.0 & Ga & 6.0 & Nd & 6.0 & Si & 5.5 \\
Au & 6.0 & Gd & 6.0 & Ne & 6.0 & Sm & 6.0 \\
B  & 5.5 & Ge & 6.0 & Ni & 6.0 & Sn & 6.0 \\
Ba & 6.6 & H  & 6.0 & O  & 6.0 & Sr & 6.5 \\
Be & 5.0 & He & 6.0 & Os & 6.0 & Ta & 6.0 \\
Bi & 6.0 & Hf & 6.0 & P  & 5.5 & Tb & 6.0 \\
Br & 6.0 & Hg & 6.0 & Pb & 6.0 & Tc & 6.0 \\
C  & 5.0 & Ho & 6.0 & Pd & 6.0 & Te & 6.0 \\
Ca & 6.1 & I  & 6.0 & Pm & 6.0 & Ti & 6.0 \\
Cd & 6.0 & In & 6.0 & Po & 6.0 & Tl & 6.0 \\
Ce & 6.0 & Ir & 6.0 & Pr & 6.0 & Tm & 6.0 \\
Cl & 6.0 & K  & 7.0 & Pt & 6.0 & V  & 6.0 \\
Co & 6.0 & Kr & 6.8 & Ra & 6.8 & W  & 6.0 \\
Cr & 6.0 & La & 6.0 & Rb & 7.3 & Xe & 7.1 \\
Cs & 7.5 & Li & 6.0 & Re & 6.0 & Y  & 6.0 \\
Cu & 6.0 & Lu & 6.0 & Rh & 6.0 & Yb & 6.1 \\
Dy & 6.0 & Mg & 6.0 & Rn & 7.5 & Zn & 6.0 \\
Er & 6.0 & Mn & 6.0 & Ru & 6.0 & Zr & 6.0 \\
Eu & 6.2 & Mo & 6.0 & S  & 5.5 &    &     \\
\bottomrule
\end{tabular}
\end{table*}

\subsubsection{Accurate and efficient atomic interactions}\label{sec:mod_detail}

We use three distinct model categories, based on the training datasets combination: SMAX, OMAT and combination of SMAX with OMAT.
For these three categories we further trained the following models:

\paragraph{SMAX-trained}
\begin{itemize}
    \item GRACE-1L-SMAX-L: one layer, large
    \item GRACE-2L-SMAX: two layers
    \item GRACE-2L-SMAX-L: two layers, large 
\end{itemize}

\paragraph{OMAT-trained}
(OAM models were finetuned on sAlex and MPtrj. All models as published in \cite{lysogorskiyGraphAtomicCluster2025a}.)
\begin{itemize}
    \item GRACE-2L-OMAT-M-base
    \item GRACE-2L-OMAT
    \item GRACE-2L-OMAT-M-ft-E
    \item GRACE-1L-OAM
    \item GRACE-1L-OAM-L
    \item GRACE-2L-OAM
    \item GRACE-2L-OAM-M
    \item GRACE-2L-OAM-L
    \item GRACE-1L-OMAT
    \item GRACE-1L-OMAT-L-base
    \item GRACE-1L-OMAT-L-ft-E
    \item GRACE-2L-OMAT-L-base
    \item GRACE-2L-OMAT-L-ft-E
\end{itemize}

\paragraph{SMAX and OMAT combined}
\begin{itemize}
    \item GRACE-1L-SMAX-OMAT-L: 1 layer, large
    \item GRACE-2L-SMAX-OMAT-M: 2 layers, medium
    \item GRACE-2L-SMAX-OMAT-L: 2 layers, large
\end{itemize}

\subsubsection{Tin large deformation maps}

Distorted crystal structures were manually created by uniformly scaling both the simulation cells and the atoms contained within it according to a range of volumes and uniaxial compressions.  Single point DFT calculations were then carried out for each strained configurations. Comparisons with the DFT reference results were carried out with the GRACE-2L-SMAX-L, GRACE-2L-OMAT-L and GRACE-2L-SMAX-OMAT-L models.

\subsubsection{Interdiffusion in the W/Be system}

We extracted binary prototypical structures from the AFLOW database \cite{eckert2024aflow}
and randomly decorated the lattice sites with W and Be atoms. Stoichiometry switches (for e.g., AB to BA, A3B2 to B3A2, etc) for each prototype  were performed but permutations of the chemical species on different Wyckoff sites was not performed here. In total, 1,224 prototype structures belonging to different space groups were fully relaxed using DFT. Defects (vacancies on the Be site, W site and 1NN W and Be sites) were then introduced into three of the most stable structures: WBe$_2$ (96 atom, 2x2x2 supercell), WBe$_{12}$ (104 atom 2x2x2 supercell) and WBe$_{22}$ (46 atom 1x1x1 unitcell). In addition, Be interstitials in BCC W (4x4x4 128 atom supercell) and W interstitials in HCP Be (4x4x3 96 atom supercell) were also generated. 
Comparisons with the DFT reference results were carried out with the GRACE-2L-SMAX-L, GRACE-2L-OMAT-L and GRACE-2L-SMAX-OMAT-L models.

\subsubsection{Vacancies in W/Ta/Cr/V high-entropy alloys}

A 3x3x3 54 atom BCC supercell was used to generate 20 HEA configurations each for the WTaCrV HEA \cite{el2019outstanding}   and WTaNbTiCr HEA \cite{tunes2023high}. In WTaCrV alloy, W, Ta, Cr and V are at 38.88 at.\%, 35.2 at.\%, 14.8 at.\% and 11.1 at.\% respectively. In the WTaNbTiCr alloy, the concentrations of W, Ta, Ti, Cr and Nb are at 40.7 at.\%, 40.7 at.\%, 3.7 at.\%, 3.7 at.\% and 11.1 at.\% respectively. Chemical species are assigned randomly in both the alloys for all configurations. 
For each alloy and for each of the 20 configurations, mono-vacancies are introduced on all chemically unique sites (chosen randomly) i.e., on W, Ta, Cr and V sites for the WTaCrV alloy and on W, Ta, Nb, Ti and Cr sites for the WTaNbTiCr alloy. This means we study 80 defect configurations for the WTaCrV alloy and 100 defect configurations for the WTaNbTiCr alloy. All structures were then completely relaxed with DFT. 
Comparisons with the DFT reference results were carried out with the GRACE-2L-SMAX-L, GRACE-2L-OMAT-L and GRACE-2L-SMAX-OMAT-L models.

\begin{comment}
\subsubsection{Metal Organic Frameworks}

The models were tested to more than 20,000 DFT relaxed MOF structures from the QMOF database \cite{rosen2021machine}. A total of 79 unique elements occur in the QMOF database, and the GRACE models were tested to MOFs that contain no actinide elements (Th, U, Np, and Pu). 

In the case of the Zn-MOF database \cite{yue2024toward}, a subset of structures from the QMOF database focusing on Zn-based MOFs with N, O, C, Zn and H as the only unique elements occurring in the database. The Zn-based MOFs extracted from the QMOF database were then subjected to homogeneous strain and strong perturbations. 
We note that these datasets were obtained with a slightly different DFT flavor than the one used to obtain SMAX (e.g., they include D3 dispersion corrections \cite{grimme2010consistent}).

Comparisons with the DFT reference results were carried out with the GRACE-2L-SMAX-L, GRACE-2L-OMAT-L and GRACE-2L-SMAX-OMAT-L models.
\end{comment}

\subsubsection{Segregation in W/Nb/Mo/Ta/V alloy}

The GRACE-1L-SMAX-OMAT-L model was employed for the segregation studies of refractory high-entropy alloys (HEAs). 
The simulation systems comprised 5,838 atoms for the quaternary Mo-Nb-Ta-W system and 20,416 atoms for the quinary W-Nb-Mo-Ta-V system containing impurities.

\subsubsection{Reaction barriers in catalytic systems}

The OC20NEB-OOD dataset \cite{wander2025cattsunami} comprises 460 nudged elastic band (NEB) calculations for heterogeneous catalytic reactions. It has recently been proposed as a standard benchmark for evaluating the transferability of universal MLIPs \cite{peng2025lambench}.
Following the LAMBench/OC20NEB-OOD evaluation protocol \cite{OpenLAM_LAMBench_2025}, NEBs were not recomputed using the MLIP models. Instead, single-point energy predictions were performed on the DFT-relaxed initial, saddle-point, and final configurations\footnote{We excluded the dissociation 208-9669 NEB due to unphysical reference DFT energies, likely arising from a convergence issue in the NEB or underlying DFT calculation.}. Reaction barrier errors were then evaluated relative to the DFT reference.

\subsubsection{Structure formation in multi-elemental mixtures}
The ``Lava'' simulations were performed on a system of 4,983 atoms under NPT conditions using the GRACE-1L-SMAX-OMAT-large model. 
Toward the end of the simulation, the structure formed a thin slab with a cell size of 124 $\mathrm{\AA}$ $\times$ 110 $\mathrm{\AA}$ $\times$ 5.2 $\mathrm{\AA}$.
To check the accuracy of the model, we randomly cut out small cylindrical clusters with a 7 $\mathrm{\AA}$ radius from two snapshots in the simulation: the middle (out-of-equilibrium) and the final (equilibrium) snapshots.
These clusters were placed in a periodic cell with 10 $\mathrm{\AA}$ of vacuum in the x and y directions. 
We successfully calculated the forces for twelve clusters from the last snapshot and twenty-six clusters from the middle snapshot using DFT.
Table~\ref{tab:croppedforceerrors} shows the errors for different models compared to these DFT results. 
The ``Bulk forces MAE'' only includes atoms that are further than 6\,$\mathrm{\AA}$ from the open surface of the cluster, while the ``Total Force MAE`` includes all atoms.
The results show that the two-layer (2L) models had the lowest errors, especially the ones trained on both SMAX and OMAT data. 
The single-layer (1L) models had larger errors. 
This is likely because the 1L models can only ``see'' within a single cutoff radius. 
Because of this, they cannot easily account for the complex atomic environment at the cluster surface, where broken bonds make the potential energy surface more difficult. 
The 2L models perform significantly better here because their multi-layer structure allows them to capture information from a wider neighborhood, effectively accounting for long-range environmental effects.
We also performed similar simulations with the GRACE-2L-SMAX-OMAT-large model and observed the same formation of Fe-Si clusters embedded in an oxide matrix. However, the final structures in those cases were less illustrative than those produced by the 1L model.

% source: http://localhost:7777/lab/tree/acefit/GRACE/UEA/MENDELEEV_ALLOY/compare_lava_to_DFT.ipynb
\begin{table*}[]
    \centering
\begin{tabular}{|l|c|c|c|}
\hline
 & Energy MAE (meV/atom) & Total Force MAE (meV/$\mathrm{\AA}$) & Bulk Force MAE (meV/$\mathrm{\AA}$) \\
 \hline
2L-SMAX-OMAT-medium & 19.0 & \textbf{131.7} & \textbf{110.3} \\
2L-SMAX-OMAT-large & \textbf{16.0} & 134.1 & 111.8 \\
2L-OMAT-medium-ft-E & 31.2 & 144.3 & 116.6 \\
2L-OMAT-large-ft-E & 29.2 & 140.1 & 126.4 \\
2L-OAM & 23.4 & 176.6 & 134.8 \\
2L-OMAT & 30.5 & 167.8 & 149.9 \\
1L-OMAT-large-ft-E & 42.9 & 277.8 & 214.6 \\
2L-SMAX-large & 73.4 & 310.6 & 219.0 \\
1L-OMAT & 49.7 & 295.2 & 227.2 \\
1L-SMAX-OMAT-large & 49.8 & 341.4 & 235.5 \\
2L-SMAX-medium & 97.7 & 339.6 & 249.9 \\
1L-OAM & 59.3 & 367.0 & 260.0 \\
1L-SMAX-large & 147.5 & 552.0 & 436.9 \\
\hline
\end{tabular}
    \caption{MAE for energies and forces from finite size clusters (cylinders), cropped from the simulation.
    Bulk force MAE were computed only on the atoms that are not within a 6\AA radius to a surface.}
    \label{tab:croppedforceerrors}
\end{table*}

\begin{figure*}
    \centering
    \includegraphics[width=\linewidth]{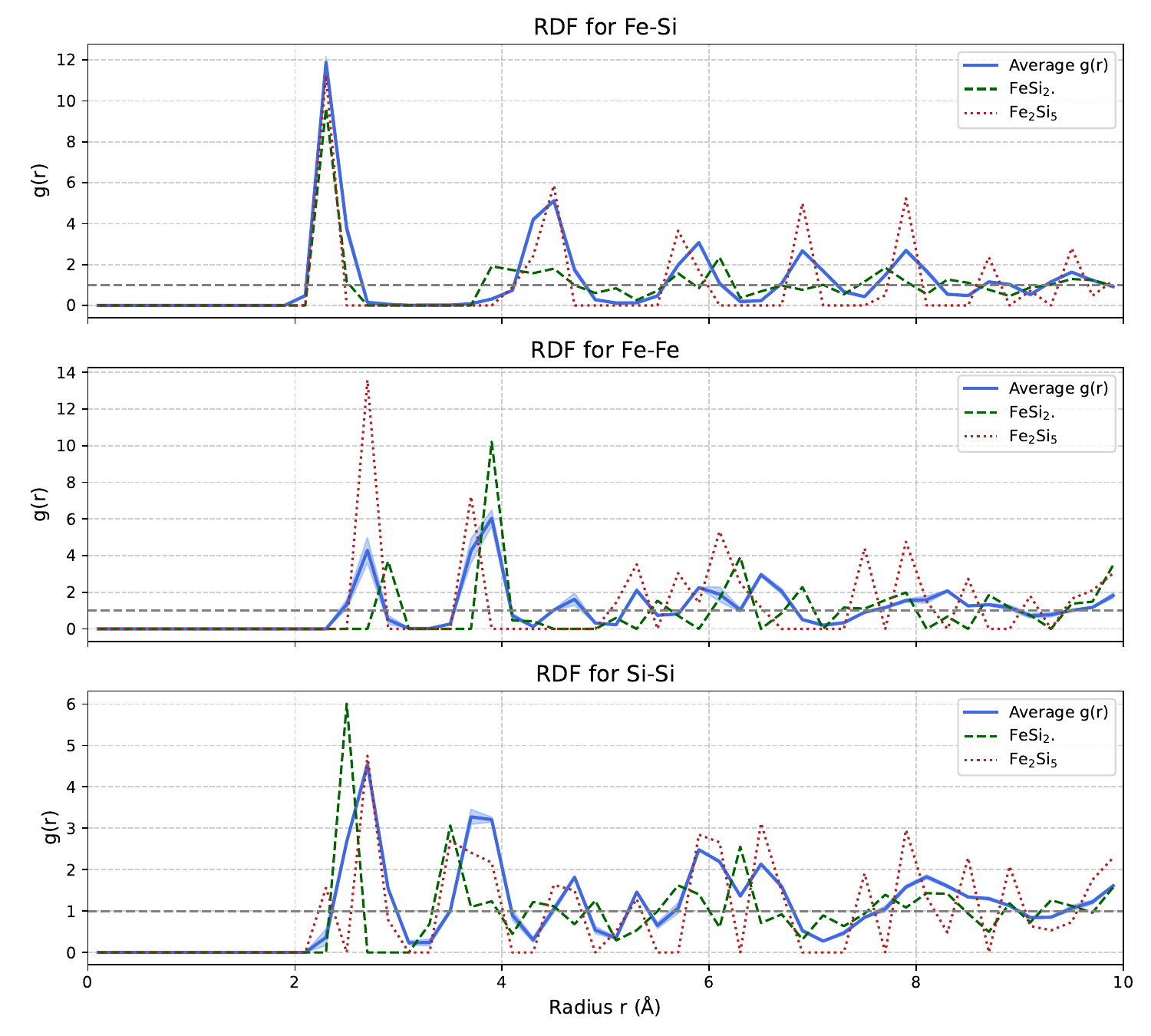}
    \caption{Comparison of RDF for FeSi region from ``Lava'' simulations with reference for FeSi2 (mp-1714) and Fe2Si5(mp-1225170)}
    \label{fig:FeSi_RDF}
\end{figure*}

We performed a structural analysis of the Fe-Si phase observed in the simulation by computing its radial distribution function (RDF), as shown in Fig.~\ref{fig:FeSi_RDF}.
By comparing this result to the RDFs of known FeSi$_2$ and Fe$_2$Si$_5$ phases, we find that the newly formed iron silicide is structurally most similar to the FeSi$_2$ phase.

 % source: http://localhost:7777/lab/tree/acefit/GRACE/UEA/MENDELEEV_ALLOY/analyze_lava_RDF.ipynb

% http://localhost:8888/lab/tree/acefit/GRACE/UEA/Mendeleev_alloy_simulation_setup-MANY_ELEMENTS-Many-2L.ipynb

\begin{figure*}
    \centering
    \includegraphics[width=0.7\linewidth]{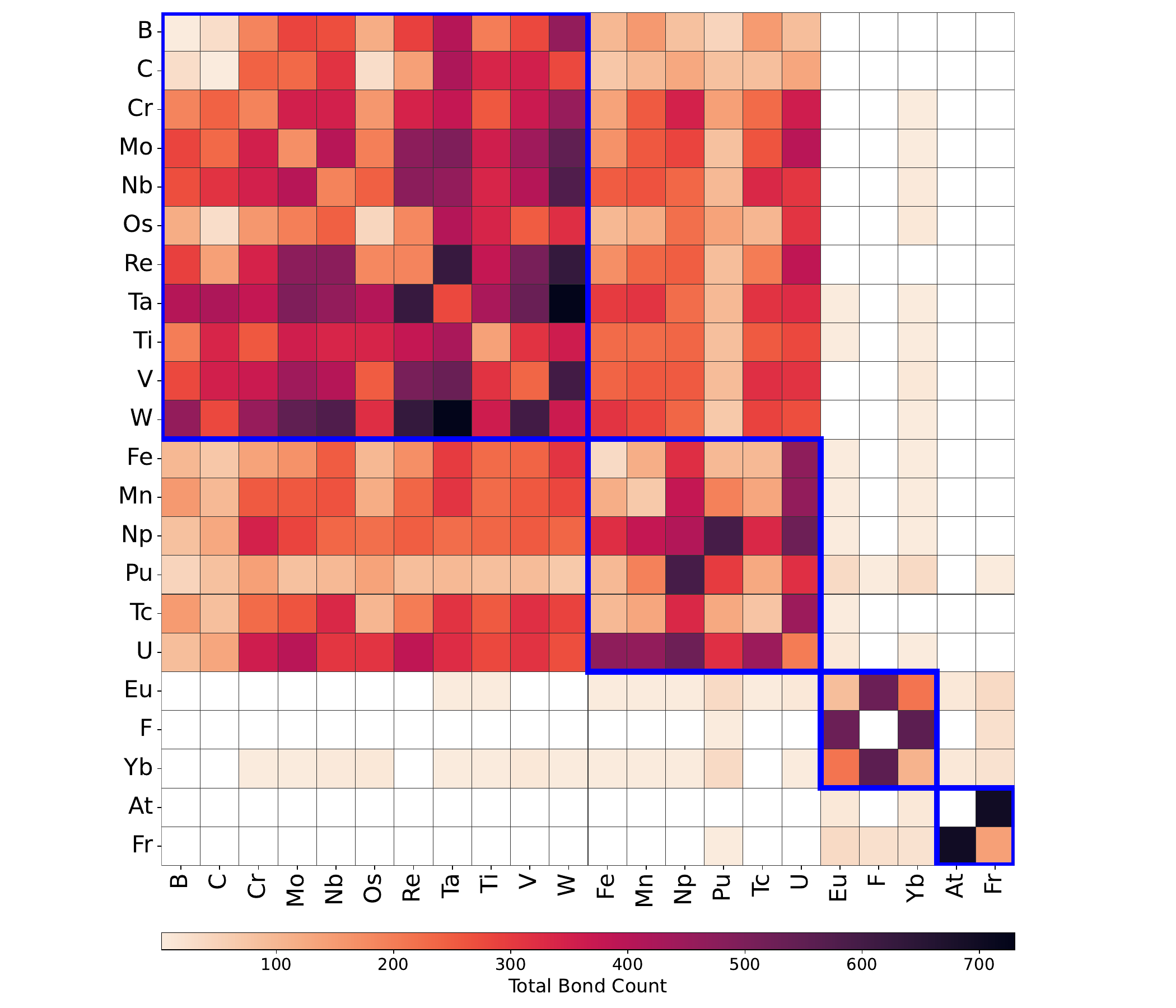}
    \caption{Heat map of the total bond counts between elements in the Mendeleev material simulation. The matrix reveals four distinct chemical clusters (highlighted by blue borders) that emerged spontaneously during the quenching process. }
    \label{fig:mend_phase_sep}
\end{figure*}

To identify chemical groupings in the Mendeleev material, we used a graph-theoretical approach where elements act as nodes and the number of bonds between them determines the edge weights. 
The resulting connectivity is visualized as a bond-count heat map in Fig.~\ref{fig:mend_phase_sep}, where strong intra-cluster correlations are highlighted by blue frames.

We applied a cluster detection algorithm to this connectivity graph to extract the four most significant clusters: refractory carbides/borides, actinide-iron mixtures, stable fluorides, and radioactive salts. 
This method allows for an objective analysis of phase separation and local bonding preferences across the entire periodic table.

\begin{comment}
\subsubsection{Metal-Organic Frameworks}
\rd{Still needs to be written}
\end{comment}

\subsection{VASP settings}
\label{supp:vasp}
The following settings were used for the VASP calculations.

%\rd{I guess ISPIN = 2 ?. Should we also mention how magnetism was initialized?}

\begin{verbatim}
 ENCUT = 500.000000
 SIGMA = 0.100000
 EDIFF = 1.00e-06
 EDIFFG = -5.00e-03
 GGA = PE
 PREC = Accurate
 IBRION = -1
 ISIF = 2
 ISMEAR = 0
 ISPIN = 2
 NELM = 120
 NSW = 0
 ADDGRID = .TRUE.
 LCHARG = .FALSE.
 LWAVE = .FALSE.
 LREAL = .FALSE.
 KSPACING = 0.125
\end{verbatim}

Magnetic moments were initialized as shown in Table.\ \ref{tab:initial_magmoms}.

\begin{table}[h]
\centering
\caption{Initial magnetic moments used for spin-polarized calculations.}
\label{tab:initial_magmoms}
\begin{tabular}{lc}
\toprule
Element & Initial magnetic moment ($\mu_B$) \\
\midrule
Default & 1.00 \\
Ce & 5.00 \\
Co & 2.00 \\
Cr & 2.00 \\
Dy & 10.65 \\
Er & 9.58 \\
Eu & 10.00 \\
Fe & 2.50 \\
Gd & 7.94 \\
Ho & 10.60 \\
Mn & 5.00 \\
Nd & 3.62 \\
Ni & 1.50 \\
Pm & 2.68 \\
Pr & 3.58 \\
Sm & 0.85 \\
Tb & 9.72 \\
Tm & 7.56 \\
Yb & 4.54 \\
\bottomrule
\end{tabular}
\end{table}

\section{References}

\bibliography{main} %You need to replace "rsc" on this line with the name of your .bib file

@article{el2023quinary,
  title={A quinary WTaCrVHf nanocrystalline refractory high-entropy alloy withholding extreme irradiation environments},
  author={El Atwani, Osman and Vo, Hi T and Tunes, Matheus Araujo and Lee, Chanho and Alvarado, A and Krienke, N and Poplawsky, Jonathan D and Kohnert, Aaron A and Gigax, Jonathan and Chen, W-Y and others},
  journal={Nature communications},
  volume={14},
  number={1},
  pages={2516},
  year={2023},
  publisher={Nature Publishing Group UK London}
}

@article{el2019outstanding,
  title={Outstanding radiation resistance of tungsten-based high-entropy alloys},
  author={El-Atwani, Osman and Li, Nan and Li, Meimei and Devaraj, Arun and Baldwin, Jon Kevin Scott and Schneider, Matthew M and Sobieraj, Damian and Wr{\'o}bel, Jan S and Nguyen-Manh, Duc and Maloy, Stuart A and others},
  journal={Science advances},
  volume={5},
  number={3},
  pages={eaav2002},
  year={2019},
  publisher={American Association for the Advancement of Science}
}

@article{qin2024influence,
  title={Influence of alloying on ductility and mechanical properties of W--Ta--Cr--V high-entropy alloys},
  author={Qin, Chenglong and Liu, Jinde and Ma, Shiyin and Du, Jiguang and Jiang, Gang and Zhao, Liang},
  journal={Intermetallics},
  volume={172},
  pages={108384},
  year={2024},
  publisher={Elsevier}
}

@inproceedings{cusentino2023molecular,
  title={Molecular dynamics of high pressure tin phases: Empirical and machine learned interatomic potentials},
  author={Cusentino, Mary Alice and Nebgen, Ben and Barros, Kipton M and Smith, Justin S and Shimanek, John D and Allen, Alice and Thompson, Aidan P and Fensin, Saryu J and Lane, J Matthew D},
  booktitle={AIP conference proceedings},
  volume={2844},
  number={1},
  pages={320002},
  year={2023},
  organization={AIP Publishing LLC}
}

@article{ChenPRM2023,
  author  = {Chen, T. and Yuan, F. and Liu, J. and Geng, H. and Zhang, L. and Wang, H. and Chen, M.},
  title   = {Modeling the high-pressure solid and liquid phases of tin from deep potentials with \emph{ab initio} accuracy},
  journal = {Phys. Rev. Materials},
  volume  = {7},
  pages   = {053603},
  year    = {2023},
  doi     = {10.1103/PhysRevMaterials.7.053603}
}

@article{Aguado2003,
  author  = {Aguado, A.},
  title   = {First-principles study of elastic properties and pressure-induced phase transitions of Sn: LDA versus GGA results},
  journal = {Phys. Rev. B},
  volume  = {67},
  pages   = {212104},
  year    = {2003},
  doi     = {10.1103/PhysRevB.67.212104}
}

@article{Matthews2013JETILW,
  author  = {Matthews, G. F. and JET EFDA Contributors and ASDEX Upgrade Team},
  title   = {Plasma operation with an all metal first-wall: Comparison of an ITER-like wall with a carbon wall in JET},
  journal = {Journal of Nuclear Materials},
  volume  = {438},
  pages   = {S2--S10},
  year    = {2013},
  doi     = {10.1016/j.jnucmat.2013.01.282}
}

@article{Brezinsek2015BeW_PSI,
  author  = {Brezinsek, S. and et al.},
  title   = {Plasma-surface interaction in the Be/W environment: Conclusions drawn from the JET-ILW},
  journal = {Journal of Nuclear Materials},
  volume  = {463},
  pages   = {11--21},
  year    = {2015},
  doi     = {10.1016/j.jnucmat.2014.12.007}
}

@article{Okamoto1986BeW,
  author  = {Okamoto, H. and Tanner, L. E.},
  title   = {The Be--W (Beryllium--Tungsten) system},
  journal = {Bulletin of Alloy Phase Diagrams},
  volume  = {7},
  pages   = {356--358},
  year    = {1986},
  doi     = {10.1007/BF02873019}
}

@article{Wiltner2021WBeInterface,
  author  = {Wiltner, A. and coauthors},
  title   = {Interface formation between Be and W layers depending on thickness and ordering},
  journal = {Applied Surface Science},
  volume  = {551},
  pages   = {149427},
  year    = {2021},
  doi     = {10.1016/j.apsusc.2020.147636}
}

@article{Doerner2005BeWmixed,
  author  = {Doerner, R. P. and Baldwin, M. J. and Nishijima, D. and others},
  title   = {Beryllium–tungsten mixed-material interactions},
  journal = {Journal of Nuclear Materials},
  volume  = {337--339},
  pages   = {970--974},
  year    = {2005},
  doi     = {10.1016/j.jnucmat.2004.10.105}
}

@article{li2020complex,
  title={Complex strengthening mechanisms in the NbMoTaW multi-principal element alloy},
  author={Li, Xiang-Guo and Chen, Chi and Zheng, Hui and Zuo, Yunxing and Ong, Shyue Ping},
  journal={npj Computational Materials},
  volume={6},
  number={1},
  pages={70},
  year={2020},
  publisher={Nature Publishing Group UK London}
}

@article{karabin2020entropy,
  title={An entropy-maximization approach to automated training set generation for interatomic potentials},
  author={Karabin, Mariia and Perez, Danny},
  journal={The Journal of Chemical Physics},
  volume={153},
  number={9},
  year={2020},
  publisher={AIP Publishing}
}

@article{montes2022training,
  title={Training data selection for accuracy and transferability of interatomic potentials},
  author={Montes de Oca Zapiain, David and Wood, Mitchell A and Lubbers, Nicholas and Pereyra, Carlos Z and Thompson, Aidan P and Perez, Danny},
  journal={npj Computational Materials},
  volume={8},
  number={1},
  pages={189},
  year={2022},
  publisher={Nature Publishing Group UK London}
}

@article{pa2025information,
  title={Information-entropy-driven generation of material-agnostic datasets for machine-learning interatomic potentials},
  author={PA Subramanyam, Aparna and Perez, Danny},
  journal={npj Computational Materials},
  volume={11},
  number={1},
  pages={218},
  year={2025},
  publisher={Nature Publishing Group UK London}
}

@article{fu2022forces,
  title={Forces are not enough: Benchmark and critical evaluation for machine learning force fields with molecular simulations},
  author={Fu, Xiang and Wu, Zhenghao and Wang, Wujie and Xie, Tian and Keten, Sinan and Gomez-Bombarelli, Rafael and Jaakkola, Tommi},
  journal={arXiv preprint arXiv:2210.07237},
  year={2022}
}

@article{wood2018extending,
  title={Extending the accuracy of the SNAP interatomic potential form},
  author={Wood, Mitchell A and Thompson, Aidan P},
  journal={The Journal of chemical physics},
  volume={148},
  number={24},
  year={2018},
  publisher={AIP Publishing}
}

@article{drautz2019atomic,
  title={Atomic cluster expansion for accurate and transferable interatomic potentials},
  author={Drautz, Ralf},
  journal={Physical Review B},
  volume={99},
  number={1},
  pages={014104},
  year={2019},
  publisher={APS}
}

@article{ramakrishnan2014quantum,
  title={Quantum chemistry structures and properties of 134 kilo molecules},
  author={Ramakrishnan, Raghunathan and Dral, Pavlo O and Rupp, Matthias and Von Lilienfeld, O Anatole},
  journal={Scientific data},
  volume={1},
  number={1},
  pages={1--7},
  year={2014},
  publisher={Nature Publishing Group}
}

@article{schreiner2022transition1x,
  title={Transition1x-a dataset for building generalizable reactive machine learning potentials},
  author={Schreiner, Mathias and Bhowmik, Arghya and Vegge, Tejs and Busk, Jonas and Winther, Ole},
  journal={Scientific Data},
  volume={9},
  number={1},
  pages={779},
  year={2022},
  publisher={Nature Publishing Group UK London}
}

@inproceedings{ahn2014flux,
  title={Flux: A next-generation resource management framework for large HPC centers},
  author={Ahn, Dong H and Garlick, Jim and Grondona, Mark and Lipari, Don and Springmeyer, Becky and Schulz, Martin},
  booktitle={2014 43rd International Conference on Parallel Processing Workshops},
  pages={9--17},
  year={2014},
  organization={IEEE}
}

@article{hafner2008ab,
  title={Ab-initio simulations of materials using VASP: Density-functional theory and beyond},
  author={Hafner, J{\"u}rgen},
  journal={Journal of computational chemistry},
  volume={29},
  number={13},
  pages={2044--2078},
  year={2008},
  publisher={Wiley Online Library}
}

@article{barroso2024open,
  title={Open materials 2024 (omat24) inorganic materials dataset and models},
  author={Barroso-Luque, Luis and Shuaibi, Muhammed and Fu, Xiang and Wood, Brandon M and Dzamba, Misko and Gao, Meng and Rizvi, Ammar and Zitnick, C Lawrence and Ulissi, Zachary W},
  journal={arXiv preprint arXiv:2410.12771},
  year={2024}
}

@article{kuner2025mp,
  title={MP-ALOE: an r2SCAN dataset for universal machine learning interatomic potentials},
  author={Kuner, Matthew C and Kaplan, Aaron D and Persson, Kristin A and Asta, Mark and Chrzan, Daryl C},
  journal={npj Computational Materials},
  volume={11},
  number={1},
  pages={352},
  year={2025},
  publisher={Nature Publishing Group UK London}
}

@article{kaplan2025foundational,
  title={A foundational potential energy surface dataset for materials},
  author={Kaplan, Aaron D and Liu, Runze and Qi, Ji and Ko, Tsz Wai and Deng, Bowen and Riebesell, Janosh and Ceder, Gerbrand and Persson, Kristin A and Ong, Shyue Ping},
  journal={arXiv preprint arXiv:2503.04070},
  year={2025}
}

@article{mazitov2025massive,
  title={Massive Atomic Diversity: a compact universal dataset for atomistic machine learning},
  author={Mazitov, Arslan and Chorna, Sofiia and Fraux, Guillaume and Bercx, Marnik and Pizzi, Giovanni and De, Sandip and Ceriotti, Michele},
  journal={Scientific Data},
  volume={12},
  number={1},
  pages={1857},
  year={2025},
  publisher={Nature Publishing Group UK London}
}

@article{wander2025cattsunami,
  title={CatTSunami: Accelerating transition state energy calculations with pretrained graph neural networks},
  author={Wander, Brook and Shuaibi, Muhammed and Kitchin, John R and Ulissi, Zachary W and Zitnick, C Lawrence},
  journal={ACS Catalysis},
  volume={15},
  number={7},
  pages={5283--5294},
  year={2025},
  publisher={ACS Publications}
}

@article{peng2025lambench,
  title={LAMBench: A Benchmark for Large Atomic Models},
  author={Peng, Anyang and Cai, Chun and Guo, Mingyu and Zhang, Duo and Zhang, Chengqian and Loew, Antoine and Zhang, Linfeng and Wang, Han},
  journal={arXiv preprint arXiv:2504.19578},
  year={2025}
}

@misc{OpenLAM_LAMBench_2025,
  title        = {{LAMBench}: A benchmark for Large Atomistic Models},
  howpublished = {\url{https://www.aissquare.com/openlam?tab=Benchmark}},
  year         = {2025},
  note         = {Accessed: 2025-12-04}
}

@book{bhadeshiaSTEELSStructureProperties2024,
  title = {{{STEELS}}: Structure, Properties and Design},
  shorttitle = {{{STEELS}}},
  author = {Bhadeshia, H. K. D. H. and Honeycombe, Robert W. K.},
  year = 2024,
  edition = {5th edition},
  publisher = {BUTTERWORTH-HEINEMANN INC},
  address = {London},
  isbn = {978-0-443-18491-8 978-0-443-18490-1},
  langid = {english}
}

@book{reedSuperalloysFundamentalsApplications2006,
  title = {The {{Superalloys}}: {{Fundamentals}} and {{Applications}}},
  shorttitle = {The {{Superalloys}}},
  author = {Reed, Roger C.},
  year = 2006,
  month = sep,
  edition = {1},
  publisher = {Cambridge University Press},
  doi = {10.1017/CBO9780511541285},
  urldate = {2025-12-07},
  copyright = {https://www.cambridge.org/core/terms},
  isbn = {978-0-521-85904-2 978-0-521-07011-9 978-0-511-54128-5}
}

@book{youngIntroductionPolymers2011,
  title = {Introduction to Polymers},
  author = {Young, Robert Joseph and Lovell, Peter A.},
  year = 2011,
  edition = {3rd ed},
  publisher = {CRC press},
  address = {Boca Raton},
  isbn = {978-0-8493-3929-5},
  langid = {english},
  lccn = {547.7}
}

@book{urryCampbellBiology2017,
  title = {Campbell Biology},
  author = {Urry, Lisa A. and Cain, Michael L. and Wasserman, Steven Alexander and Minorsky, Peter V. and Reece, Jane B. and Campbell, Neil A.},
  year = 2017,
  edition = {Eleventh edition},
  publisher = {Pearson Education, Inc.},
  address = {New York, NY},
  isbn = {978-0-13-409341-3},
  langid = {english}
}

@article{WEEEManagementCircular2018,
  title = {{{WEEE}} Management in a Circular Economy Perspective: An Overview},
  shorttitle = {{{WEEE}} Management in a Circular Economy Perspective},
  year = 2018,
  month = nov,
  journal = {Global NEST Journal},
  volume = {20},
  number = {4},
  pages = {743--750},
  issn = {1790-7632, 2241-777X},
  doi = {10.30955/gnj.002623},
  urldate = {2025-12-10}
}

@article{bressanelliCircularEconomyWEEE2020,
  title = {Circular {{Economy}} in the {{WEEE}} Industry: A Systematic Literature Review and a Research Agenda},
  shorttitle = {Circular {{Economy}} in the {{WEEE}} Industry},
  author = {Bressanelli, Gianmarco and Saccani, Nicola and Pigosso, Daniela C.A. and Perona, Marco},
  year = 2020,
  month = jul,
  journal = {Sustainable Production and Consumption},
  volume = {23},
  pages = {174--188},
  issn = {23525509},
  doi = {10.1016/j.spc.2020.05.007},
  urldate = {2025-12-10},
  langid = {english}
}

@article{panChemicalCharacteristicsRisk2013,
  title = {Chemical Characteristics and Risk Assessment of Typical Municipal Solid Waste Incineration ({{MSWI}}) Fly Ash in {{China}}},
  author = {Pan, Yun and Wu, Zhiming and Zhou, Jizhi and Zhao, Jun and Ruan, Xiuxiu and Liu, Jianyong and Qian, Guangren},
  year = 2013,
  month = oct,
  journal = {Journal of Hazardous Materials},
  volume = {261},
  pages = {269--276},
  issn = {03043894},
  doi = {10.1016/j.jhazmat.2013.07.038},
  urldate = {2025-12-10},
  langid = {english}
}

@article{weibelChemicalAssociationsMobilization2017,
  title = {Chemical Associations and Mobilization of Heavy Metals in Fly Ash from Municipal Solid Waste Incineration},
  author = {Weibel, Gisela and Eggenberger, Urs and Schlumberger, Stefan and M{\"a}der, Urs K.},
  year = 2017,
  month = apr,
  journal = {Waste Management},
  volume = {62},
  pages = {147--159},
  issn = {0956053X},
  doi = {10.1016/j.wasman.2016.12.004},
  urldate = {2025-12-10},
  langid = {english}
}

@article{ngulimiRadioactiveWasteManagement2025,
  title = {The Radioactive Waste Management - {{State}} of the Art and Emerging Technologies},
  author = {Ngulimi, Miguta Faustine and Asghar, Kamal and Kim, Sion and Seo, Bum Kyoung and Roh, Changhyun},
  year = 2025,
  month = nov,
  journal = {Journal of Hazardous Materials Advances},
  volume = {20},
  pages = {100932},
  issn = {27724166},
  doi = {10.1016/j.hazadv.2025.100932},
  urldate = {2025-12-10},
  langid = {english}
}

@article{hanswedepohlCompositionContinentalCrust1995,
  title = {The Composition of the Continental Crust},
  author = {Hans Wedepohl, K.},
  year = 1995,
  month = apr,
  journal = {Geochimica et Cosmochimica Acta},
  volume = {59},
  number = {7},
  pages = {1217--1232},
  issn = {00167037},
  doi = {10.1016/0016-7037(95)00038-2},
  urldate = {2025-12-10},
  copyright = {https://www.elsevier.com/tdm/userlicense/1.0/},
  langid = {english}
}

@incollection{Chapter5Composition1986,
  title = {Chapter 5 {{The Composition}} of the {{Continental Crust}}},
  booktitle = {International {{Geophysics}}},
  year = 1986,
  volume = {34},
  pages = {213--241},
  publisher = {Elsevier},
  doi = {10.1016/S0074-6142(09)60137-6},
  urldate = {2025-12-10},
  isbn = {978-0-12-488950-7},
  langid = {english}
}

@incollection{rudnickCompositionContinentalCrust2003,
  title = {Composition of the {{Continental Crust}}},
  booktitle = {Treatise on {{Geochemistry}}},
  author = {Rudnick, R.L. and Gao, S.},
  year = 2003,
  pages = {1--64},
  publisher = {Elsevier},
  doi = {10.1016/B0-08-043751-6/03016-4},
  urldate = {2025-12-10},
  copyright = {https://www.elsevier.com/tdm/userlicense/1.0/},
  isbn = {978-0-08-043751-4},
  langid = {english}
}

@article{georgeHighentropyAlloys2019,
  title = {High-Entropy Alloys},
  author = {George, Easo P. and Raabe, Dierk and Ritchie, Robert O.},
  year = 2019,
  month = jun,
  journal = {Nature Reviews Materials},
  volume = {4},
  number = {8},
  pages = {515--534},
  issn = {2058-8437},
  doi = {10.1038/s41578-019-0121-4},
  urldate = {2025-12-10},
  langid = {english}
}

@article{miracleCriticalReviewHigh2017,
  title = {A Critical Review of High Entropy Alloys and Related Concepts},
  author = {Miracle, D.B. and Senkov, O.N.},
  year = 2017,
  month = jan,
  journal = {Acta Materialia},
  volume = {122},
  pages = {448--511},
  issn = {13596454},
  doi = {10.1016/j.actamat.2016.08.081},
  urldate = {2025-12-10},
  langid = {english}
}

@article{pettiforStructuresBinaryCompounds1986,
  title = {The Structures of Binary Compounds. {{I}}. {{Phenomenological}} Structure Maps},
  author = {Pettifor, D G},
  year = 1986,
  month = jan,
  journal = {Journal of Physics C: Solid State Physics},
  volume = {19},
  number = {3},
  pages = {285--313},
  issn = {0022-3719},
  doi = {10.1088/0022-3719/19/3/002},
  urldate = {2025-12-12}
}

@misc{levineOpenMolecules20252025,
  title = {The {{Open Molecules}} 2025 ({{OMol25}}) {{Dataset}}, {{Evaluations}}, and {{Models}}},
  author = {Levine, Daniel S. and Shuaibi, Muhammed and {Spotte-Smith}, Evan Walter Clark and Taylor, Michael G. and Hasyim, Muhammad R. and Michel, Kyle and Batatia, Ilyes and Csanyi, Gabor and Dzamba, Misko and Eastman, Peter and Frey, Nathan C. and Fu, Xiang and Gharakhanyan, Vahe and Krishnapriyan, Aditi S. and Rackers, Joshua A. and Raja, Sanjeev and Rizvi, Ammar and Rosen, Andrew S. and Ulissi, Zachary and Vargas, Santiago and Zitnick, C. Lawrence and Blau, Samuel M. and Wood, Brandon M.},
  year = 2025,
  publisher = {arXiv},
  doi = {10.48550/ARXIV.2505.08762},
  urldate = {2025-11-21},
  copyright = {arXiv.org perpetual, non-exclusive license}
}

@article{schmidt2023machine,
  title={Machine-learning-assisted determination of the global zero-temperature phase diagram of materials},
  author={Schmidt, Jonathan and Hoffmann, Noah and Wang, Hai-Chen and Borlido, Pedro and Carri{\c{c}}o, Pedro JMA and Cerqueira, Tiago FT and Botti, Silvana and Marques, Miguel AL},
  journal={Advanced Materials},
  volume={35},
  number={22},
  pages={2210788},
  year={2023},
  publisher={Wiley Online Library}
}

@article{wang2023symmetry,
  title={Symmetry-based computational search for novel binary and ternary 2D materials},
  author={Wang, Hai-Chen and Schmidt, Jonathan and Marques, Miguel AL and Wirtz, Ludger and Romero, Aldo H},
  journal={2D Materials},
  volume={10},
  number={3},
  pages={035007},
  year={2023},
  publisher={IOP Publishing}
}

@article{jainCommentaryMaterialsProject2013,
  title = {Commentary: {{The Materials Project}}: {{A}} Materials Genome Approach to Accelerating Materials Innovation},
  shorttitle = {Commentary},
  author = {Jain, Anubhav and Ong, Shyue Ping and Hautier, Geoffroy and Chen, Wei and Richards, William Davidson and Dacek, Stephen and Cholia, Shreyas and Gunter, Dan and Skinner, David and Ceder, Gerbrand and Persson, Kristin A.},
  year = 2013,
  month = jul,
  journal = {APL Materials},
  volume = {1},
  number = {1},
  pages = {011002},
  issn = {2166-532X},
  doi = {10.1063/1.4812323},
  urldate = {2025-03-13},
  langid = {english}
}

@article{bochkarev2024,
  title = {Graph Atomic Cluster Expansion for Semilocal Interactions beyond Equivariant Message Passing},
  author = {Bochkarev, Anton and Lysogorskiy, Yury and Drautz, Ralf},
  journal = {Phys. Rev. X},
  volume = {14},
  issue = {2},
  pages = {021036},
  numpages = {28},
  year = {2024},
  month = {Jun},
  publisher = {American Physical Society},
  doi = {10.1103/PhysRevX.14.021036},
  url = {https://link.aps.org/doi/10.1103/PhysRevX.14.021036}
}

@article{drautzAtomicClusterExpansion2019,
  title = {Atomic Cluster Expansion for Accurate and Transferable Interatomic Potentials},
  author = {Drautz, Ralf},
  year = 2019,
  month = jan,
  journal = {Phys. Rev. B},
  volume = {99},
  number = {1},
  pages = {014104},
  publisher = {American Physical Society},
  doi = {10.1103/PhysRevB.99.014104}
}

@misc{lysogorskiyGraphAtomicCluster2025a,
  title = {Graph Atomic Cluster Expansion for Foundational Machine Learning Interatomic Potentials},
  author = {Lysogorskiy, Yury and Bochkarev, Anton and Drautz, Ralf},
  year = 2025,
  publisher = {arXiv},
  doi = {10.48550/ARXIV.2508.17936},
  urldate = {2025-11-20},
  copyright = {Creative Commons Attribution 4.0 International}
}

@article{kormannPhononBroadeningHigh2017,
  title = {Phonon Broadening in High Entropy Alloys},
  author = {K{\"o}rmann, Fritz and Ikeda, Yuji and Grabowski, Blazej and Sluiter, Marcel H. F.},
  year = 2017,
  month = sep,
  journal = {npj Computational Materials},
  volume = {3},
  number = {1},
  pages = {36},
  issn = {2057-3960},
  doi = {10.1038/s41524-017-0037-8},
  urldate = {2025-12-13},
  langid = {english}
}

@article{senkovRefractoryHighentropyAlloys2010,
  title = {Refractory High-Entropy Alloys},
  author = {Senkov, O.N. and Wilks, G.B. and Miracle, D.B. and Chuang, C.P. and Liaw, P.K.},
  year = 2010,
  month = sep,
  journal = {Intermetallics},
  volume = {18},
  number = {9},
  pages = {1758--1765},
  issn = {09669795},
  doi = {10.1016/j.intermet.2010.05.014},
  urldate = {2025-12-13},
  copyright = {https://www.elsevier.com/tdm/userlicense/1.0/},
  langid = {english}
}

@article{liuRecentProgressBCCStructured2022,
  title = {Recent {{Progress}} with {{BCC-Structured High-Entropy Alloys}}},
  author = {Liu, Fangfei and Liaw, Peter and Zhang, Yong},
  year = 2022,
  month = mar,
  journal = {Metals},
  volume = {12},
  number = {3},
  pages = {501},
  issn = {2075-4701},
  doi = {10.3390/met12030501},
  urldate = {2025-12-13},
  langid = {english}
}

@article{podryabinkin2017active,
  title={Active learning of linearly parametrized interatomic potentials},
  author={Podryabinkin, Evgeny V and Shapeev, Alexander V},
  journal={Computational Materials Science},
  volume={140},
  pages={171--180},
  year={2017},
  publisher={Elsevier}
}

@article{kingma2014adam,
  title={Adam: A method for stochastic optimization},
  author={Kingma, Diederik P and Ba, Jimmy},
  journal={arXiv preprint arXiv:1412.6980},
  year={2014}
}

@article{LAMMPS,
  title={LAMMPS-a flexible simulation tool for particle-based materials modeling at the atomic, meso, and continuum scales},
  author={Thompson, Aidan P and Aktulga, H Metin and Berger, Richard and Bolintineanu, Dan S and Brown, W Michael and Crozier, Paul S and In't Veld, Pieter J and Kohlmeyer, Axel and Moore, Stan G and Nguyen, Trung Dac and others},
  journal={Computer physics communications},
  volume={271},
  pages={108171},
  year={2022},
  publisher={Elsevier}
}

@article{de2015charting,
  title={Charting the complete elastic properties of inorganic crystalline compounds},
  author={De Jong, Maarten and Chen, Wei and Angsten, Thomas and Jain, Anubhav and Notestine, Randy and Gamst, Anthony and Sluiter, Marcel and Krishna Ande, Chaitanya and Van Der Zwaag, Sybrand and Plata, Jose J and others},
  journal={Scientific data},
  volume={2},
  number={1},
  pages={1--13},
  year={2015},
  publisher={Nature Publishing Group}
}

@article{golesorkhtabar2013elastic,
  title={ElaStic: A tool for calculating second-order elastic constants from first principles},
  author={Golesorkhtabar, Rostam and Pavone, Pasquale and Spitaler, J{\"u}rgen and Puschnig, Peter and Draxl, Claudia},
  journal={Computer Physics Communications},
  volume={184},
  number={8},
  pages={1861--1873},
  year={2013},
  publisher={Elsevier}
}

@article{onwuliElementSimilarityHighdimensional2023,
  title = {Element Similarity in High-Dimensional Materials Representations},
  author = {Onwuli, Anthony and Hegde, Ashish V. and Nguyen, Kevin V. T. and Butler, Keith T. and Walsh, Aron},
  year = 2023,
  journal = {Digital Discovery},
  volume = {2},
  number = {5},
  pages = {1558--1564},
  issn = {2635-098X},
  doi = {10.1039/D3DD00121K},
  urldate = {2026-02-01},
  langid = {english}
}

@article{zhouLearningAtomsMaterials2018,
  title = {Learning Atoms for Materials Discovery},
  author = {Zhou, Quan and Tang, Peizhe and Liu, Shenxiu and Pan, Jinbo and Yan, Qimin and Zhang, Shou-Cheng},
  year = 2018,
  month = jul,
  journal = {Proceedings of the National Academy of Sciences},
  volume = {115},
  number = {28},
  issn = {0027-8424, 1091-6490},
  doi = {10.1073/pnas.1801181115},
  urldate = {2026-02-01},
  langid = {english}
}

@misc{cerqueiraNonorthogonalRepresentationChemical2024,
  title = {A Non-Orthogonal Representation of the Chemical Space},
  author = {Cerqueira, Tiago F. T. and Wang, Haichen and Botti, Silvana and Marques, Miguel A. L.},
  year = 2024,
  publisher = {arXiv},
  doi = {10.48550/ARXIV.2406.19761},
  urldate = {2026-02-01},
  copyright = {arXiv.org perpetual, non-exclusive license}
}

@article{batatiaDesignSpaceE3equivariant2025,
  title = {The Design Space of {{E}}(3)-Equivariant Atom-Centred Interatomic Potentials},
  author = {Batatia, Ilyes and Batzner, Simon and Kov{\'a}cs, D{\'a}vid P{\'e}ter and Musaelian, Albert and Simm, Gregor N. C. and Drautz, Ralf and Ortner, Christoph and Kozinsky, Boris and Cs{\'a}nyi, G{\'a}bor},
  year = 2025,
  month = jan,
  journal = {Nature Machine Intelligence},
  volume = {7},
  number = {1},
  pages = {56--67},
  issn = {2522-5839},
  doi = {10.1038/s42256-024-00956-x},
  urldate = {2025-02-23},
  langid = {english}
}

@inproceedings{NEURIPS2022_4a36c3c5,
  title = {{{MACE}}: {{Higher}} Order Equivariant Message Passing Neural Networks for Fast and Accurate Force Fields},
  booktitle = {Advances in Neural Information Processing Systems},
  author = {Batatia, Ilyes and Kovacs, David P and Simm, Gregor and Ortner, Christoph and Csanyi, Gabor},
  editor = {Koyejo, S. and Mohamed, S. and Agarwal, A. and Belgrave, D. and Cho, K. and Oh, A.},
  year = 2022,
  volume = {35},
  pages = {11423--11436},
  publisher = {Curran Associates, Inc.}
}

@article{eckert2024aflow,
  title={The AFLOW library of crystallographic prototypes: Part 4},
  author={Eckert, Hagen and Divilov, Simon and Mehl, Michael J and Hicks, David and Zettel, Adam C and Esters, Marco and Campilongo, Xiomara and Curtarolo, Stefano},
  journal={Computational Materials Science},
  volume={240},
  pages={112988},
  year={2024},
  publisher={Elsevier}
}

@article{tunes2023high,
  title={From high-entropy alloys to high-entropy ceramics: The radiation-resistant highly concentrated refractory carbide (CrNbTaTiW) C},
  author={Tunes, Matheus A and Fritze, Stefan and Osinger, Barbara and Willenshofer, Patrick and Alvarado, Andrew M and Martinez, Enrique and Menon, Ashok S and Str{\"o}m, Petter and Greaves, Graeme and Lewin, Erik and others},
  journal={Acta Materialia},
  volume={250},
  pages={118856},
  year={2023},
  publisher={Elsevier}
}

@article{rosen2021machine,
  title={Machine learning the quantum-chemical properties of metal--organic frameworks for accelerated materials discovery},
  author={Rosen, Andrew S and Iyer, Shaelyn M and Ray, Debmalya and Yao, Zhenpeng and Aspuru-Guzik, Al{\'a}n and Gagliardi, Laura and Notestein, Justin M and Snurr, Randall Q},
  journal={Matter},
  volume={4},
  number={5},
  pages={1578--1597},
  year={2021},
  publisher={Elsevier}
}

@article{yue2024toward,
  title={Toward a generalizable machine-learned potential for metal--organic frameworks},
  author={Yue, Yifei and Mohamed, Saad Aldin and Loh, N Duane and Jiang, Jianwen},
  journal={ACS nano},
  volume={19},
  number={1},
  pages={933--949},
  year={2024},
  publisher={ACS Publications}
}

@article{grimme2010consistent,
  title={A consistent and accurate ab initio parametrization of density functional dispersion correction (DFT-D) for the 94 elements H-Pu},
  author={Grimme, Stefan and Antony, Jens and Ehrlich, Stephan and Krieg, Helge},
  journal={The Journal of chemical physics},
  volume={132},
  number={15},
  year={2010},
  publisher={AIP Publishing}
}

@article{warford2026better,
  title={Better without U: Impact of Selective Hubbard U Correction on Foundational MLIPs},
  author={Warford, Thomas and Thiemann, Fabian L and Csanyi, Gabor},
  journal={arXiv preprint arXiv:2601.21056},
  year={2026}
}

\end{document}